%% file: paper.tex
\documentclass[10pt,journal]{IEEEtran} %
\input{config}

\input{latex-commands}

\makeatletter
\patchcmd{\@makecaption}
  {\scshape}
  {}
  {}
  {}
\makeatother

\usepackage{comment}

\allowdisplaybreaks
\usetikzlibrary{external}
\begin{document}
\bstctlcite{MyBSTcontrol}

\title{Adaptive Network Security Policies\\via Belief Aggregation and Rollout}

\author{\IEEEauthorblockN{Kim Hammar\IEEEauthorrefmark{2}, Yuchao Li\IEEEauthorrefmark{2}, Tansu Alpcan\IEEEauthorrefmark{3}, Emil C. Lupu\IEEEauthorrefmark{4}, and Dimitri Bertsekas\IEEEauthorrefmark{2}}\\
 \IEEEauthorblockA{\IEEEauthorrefmark{2}
   School of Computing and Augmented Intelligence, Arizona State University, USA\\
 }
 \IEEEauthorblockA{\IEEEauthorrefmark{3}
   Department of Electrical and Electronic Engineering, University of Melbourne, Australia\\
 } 
 \IEEEauthorblockA{\IEEEauthorrefmark{4}
   Department of Computing, Imperial College London, United Kingdom\\
 }
 Email: \{khammar1,yuchaoli,dbertsek\}@asu.edu, tansu.alpcan@unimelb.edu.au, e.c.lupu@imperial.ac.uk
}
\maketitle
 
\begin{abstract}
Evolving security vulnerabilities and shifting operational conditions require frequent updates to network security policies. These updates include adjustments to incident response procedures and modifications to access controls, among others. Reinforcement learning methods have been proposed for automating such policy adaptations, but most methods in the research literature lack performance guarantees and adapt slowly to changes. In this paper, we address these limitations and present a method for computing security policies that is scalable, offers theoretical guarantees, and adapts quickly to changes. The method uses a model or simulator of the system, which is updated when changes occur, and combines three components: belief estimation through particle filtering, offline policy computation through feature-based aggregation, and online policy adaptation through rollout. In particular, feature-based aggregation enables scalable offline optimization of a policy, while rollout adapts the policy online to changes in the system model without repeating the offline optimization. We analyze the approximation error of the aggregation and show that the rollout efficiently adapts policies to changes under certain conditions. Simulations and testbed results demonstrate that our method outperforms state-of-the-art methods on several benchmarks, including CAGE-2.
\end{abstract}
\section{Introduction}
\lettrine[lines=2]{\textbf{N}}{etwork} security policies dictate how security measures are implemented and applied to protect a networked system against attacks. Such policies can be enforced at the physical, network, and service layers and include access control, flow control, and intrusion response policies, among others. Traditionally, such policies have been defined, implemented, and updated by domain experts \cite{denning_1}. Although this approach can provide effective security policies for systems that change infrequently, it becomes impractical for systems with frequent changes, such as those caused by shifting operational requirements, fluctuating workloads, component failures, or software updates. These dynamic changes make policy adjustments and reconfigurations an inherent part of operations, necessitating adaptive approaches to managing security policies \cite{8453023}.

\tikzexternaldisable
\begin{figure}
  \centering
\scalebox{0.83}{
  \input{tikz/rollout_framework11.tex}
  }
  \caption{Our method for computing adaptive network security policies. A base policy and cost function are computed offline via dynamic programming in an aggregate belief space, where beliefs represent uncertainty about the system's security state. At runtime, the belief is estimated via particle filtering and the base policy is adapted via rollout simulations and lookahead optimization guided by the cost function. This lookahead allows the system to anticipate possible threats and assess the impact of various security controls.}
  \label{fig:framework}
\end{figure}
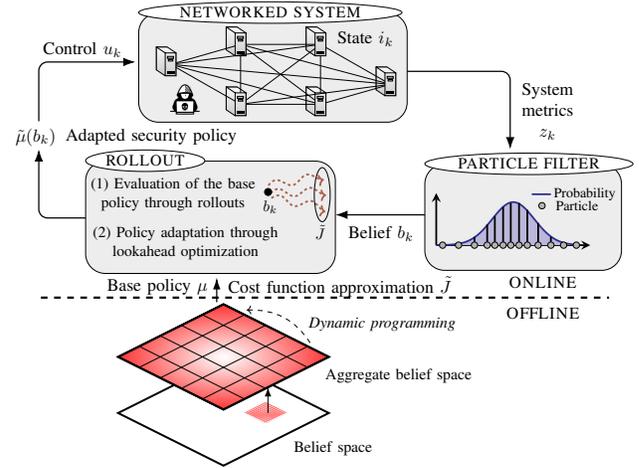
\tikzexternalenable

A promising solution is to frame the problem of obtaining an effective security policy as a sequential decision-making problem, which enables automatic policy adaptation via reinforcement learning (RL) \cite{deep_rl_cyber_sec}. In this formulation, a security policy is defined as a function that prescribes security controls (e.g., access or flow controls) based on a \textit{belief} about the system's security state (e.g., whether it is compromised or not). This belief is defined as a probability distribution over possible system states and is updated sequentially based on system metrics (e.g., logs and alerts). These metrics are also used to track changes to the system and iteratively adapt the security policy to meet a given objective quantified by a cost function. Recent work shows the potential of this approach to adapt a broad range of security policies, including intrusion response \cite{8328972,5270307}, penetration testing \cite{299699}, deception \cite{10234402}, replication \cite{dsn24_hammar_stadler}, and fuzzing \cite{274553} policies. Despite these advances, important limitations remain. In particular, most methods for policy adaptation proposed in the literature use \textit{deep} RL, which has limited performance guarantees and requires extensive offline retraining to adapt. Moreover, most methods have only been tested in simulation, leaving their practical utility unvalidated.

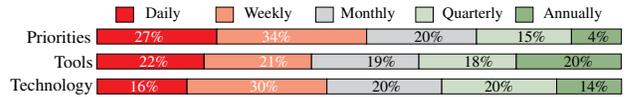
\begin{figure}
  \centering
\scalebox{0.83}{
   \input{tikz/stats.tex}
  }
  \caption{Frequency of change in networked systems \cite{devops_trends}.}
  \label{fig:stats}
\end{figure}

\begin{figure}
  \centering
\scalebox{0.9}{
    \input{tikz/outages2.tex}
  }
  \caption{Most common causes of outages in networked systems \cite{observability_trends}.}
  \label{fig:outages}
\end{figure}
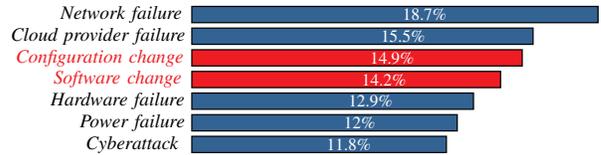

In this paper, we address these limitations by presenting and validating a method for computing security policies that is scalable, offers performance guarantees, and adapts quickly to changes. Our method includes three main components, as shown in Fig.~\ref{fig:framework}. First, we estimate a probabilistic belief about the system state through \textit{particle filtering}. This belief quantifies the likelihood of potential system compromises and enables the security policy to account for uncertainty. Second, we aggregate the space of such beliefs into a finite set of representative ones, enabling efficient (offline) computation of a \textit{base policy}, as well as approximation of the optimal cost function through dynamic programming (DP) \cite{bertsekas2012dynamic}. We show that the error of this approximation is bounded. Third, we use (online) \textit{rollout} \cite{bertsekas2021rollout} and lookahead optimization to adapt the base policy to changes in a given system model, such as changing operational conditions or security objectives; see Figs.~\ref{fig:stats} and \ref{fig:outages}. We show experimentally that this adaptation can be completed in seconds using commodity computing hardware and show theoretically that it is guaranteed to improve the base policy under general conditions.

Although our method is designed for security policies, it can be used more generally for adaptive control of partially observable dynamic systems. Compared to other approximation schemes for such systems \cite{bertsekas2021rollout,NIPS1996_996009f2,10.5555/1036843.1036918,saldi2016,bertsekas2018featurebasedaggregationdeepreinforcement,bertsekas2024reinforcement,bertsekas2019biasedaggregationrolloutenhanced,noms24_rollout,liu2020reinforcement,liu2022rollout,10955193,automotive_rollout,24227}, our method introduces a novel \textit{feature-based} aggregation technique, which improves scalability and flexibility. Instead of aggregating beliefs over the state space directly, we first aggregate states into a small set of feature states and then aggregate beliefs over this set.

In this paper, we focus on control-theoretic system models where the attacker’s strategy is fixed. If the attacker’s strategy changes, our method adapts the policy to the change through rollout with an updated model. Extending our method to game-theoretic models where the attacker's strategy is coupled with the security policy is beyond the scope of this paper. Related aggregation methods for such models are discussed in \cite{Hammar_Alpcan_2026}.

We summarize our contributions as follows:
\begin{itemize}
\item We develop a scalable method for computing adaptive network security policies, which involves a novel combination of particle filtering, aggregation, and rollout.
\item We establish a bound on the approximation error of the cost function obtained through our aggregation method.
\item We show conditions under which our rollout method for policy adaptation improves the security policy.
\item We evaluate our method through simulations and testbed experiments. The results show state-of-the-art performance on several benchmarks, including CAGE-2.
\end{itemize}

\section{Related Work}\label{sec:related_work}
The problem of computing security policies has engaged security experts, control engineers, and game theorists for over two decades \cite{nework_security_alpcan}. Surveys \cite{kim_phd_thesis, 10.1145/3729213}, and \cite{deep_rl_cyber_sec} give an extensive account of these efforts and an appraisal of the state of the art. A driving factor behind this research is the development of evaluation benchmarks, which allow researchers to compare different methods. Currently, the most popular benchmark is the cyber autonomy gym for experimentation 2 (CAGE-2) \cite{cage_challenge_2_announcement}, which involves computing an intrusion response policy.

More than $35$ methods have been evaluated against CAGE-2 \cite{cage_challenge_2_announcement}. Detailed descriptions of some methods can be found in \cite{vyas2023automated,alan_turing_1,beyond_cage,foley_cage_1,foley2023inroads,10476122,cheng2024rice,palmer2025empiricalgametheoreticanalysisautonomous,sussex_1,TANG2024103871,wiebe2023learning,singh2024hierarchicalmultiagentreinforcementlearning,yan2024depending,llm_cage_2_5,hammar2024optimaldefenderstrategiescage2,ramamurthy2025generalautonomouscybersecuritydefense,10991969}. Although good results have been obtained, key aspects remain unexplored. For example, current methods focus narrowly on offline (deep) RL, which adapts slowly to changes and lacks performance guarantees. One exception is the method based on tree search presented in \cite[Alg. 1]{hammar2024optimaldefenderstrategiescage2}. However, this method is customized for a specific system and not generalizable. None of the current methods considers aggregation and rollout, which we introduce in this paper. The benefit of our approach is that it provides performance guarantees and adapts policies quickly to changes after the system model is updated. Another difference is that the referenced methods are evaluated only in simulation, whereas we evaluate our method both in simulation and in a testbed; see Table~\ref{tab:related_work}.
\tikzexternaldisable
\tikzexternalenable

\tikzexternaldisable
\tikzexternalenable
\begin{table}[H]
  \centering
  \scalebox{0.84}{
    \begin{tabular}{lllll} \toprule
\rowcolor{lightgray}
      {\textit{Method}} & {\textit{CAGE-2 sota}} & {\textit{Adaptive}} & {\textit{Testbed}} & {\textit{Formal guarantees}}\\ \midrule
    \rowcolor{lightgreen}
      Ours (Fig.~\ref{fig:framework}) & \cmark & \cmark & \cmark & \cmark \\
      Deep RL \cite{vyas2023automated,alan_turing_1} & \cmark & \xmark & \xmark & \xmark \\
      Deep RL \cite{beyond_cage,foley_cage_1,foley2023inroads,10476122,cheng2024rice,palmer2025empiricalgametheoreticanalysisautonomous} & \xmark & \xmark & \xmark & \xmark \\
      Evolution \cite{sussex_1} & \xmark & \xmark & \xmark & \xmark \\
      Deep MARL \cite{TANG2024103871,wiebe2023learning,singh2024hierarchicalmultiagentreinforcementlearning} & \xmark & \xmark & \xmark & \xmark \\
      LLM \cite{yan2024depending,llm_cage_2_5,10991969} & \xmark & \xmark & \xmark & \xmark \\
      Tree search \cite{hammar2024optimaldefenderstrategiescage2} & \cmark & \cmark & \xmark & \cmark \\
    \bottomrule\\
  \end{tabular}}
  \caption{Comparison with related work. Our method is the first to be validated on a testbed and achieve state-of-the-art (SOTA) results on CAGE-2 \cite{cage_challenge_2_announcement} while providing theoretical guarantees and adapting quickly to changes.}\label{tab:related_work}
\end{table}
\vspace{-5pt}

Beyond methods evaluated on CAGE-2, our method is related to other control-theoretic approaches in cybersecurity. For example, Aydeger et al. \cite{9328143} and Zhu et. al. \cite{10.1007/978-3-319-02786-9_15} use adaptive control techniques for moving target defense; Hammar et al. \cite{hammar_stadler_tnsm,10955193} employ model-predictive control for incident response; Miehling et al. \cite{8325528} and Zonouz et al. \cite{5270307} use tree search for attack mitigation; and Yue et al. \cite{11187396} use game theory for security decision-making in control systems.

Our method differs from these methods in three main ways. First, the referenced methods are each designed for a specific security application. By contrast, our method is applicable to any security problem that can be cast as a finite-state partially observable Markov decision problem. Second, we introduce a novel feature-based aggregation technique, which improves scalability. Third, we demonstrate this scalability by evaluating our method on CAGE-2, a system of considerably greater complexity than the systems considered in the referenced studies. Whether the methods proposed in those studies could scale to systems of comparable size remains an open question.

Lastly, we note that our method draws on ideas from model-predictive control (see e.g., Borrelli et al. \cite{Borrellietal2017}) and adaptive control (see e.g., Åström and Wittenmark \cite{69dd8cb6cfee477185172ca75d58cd35}), both of which are well-developed research areas with a long history.
\section{Example Use Case: Intrusion Recovery}\label{sec:use_case}
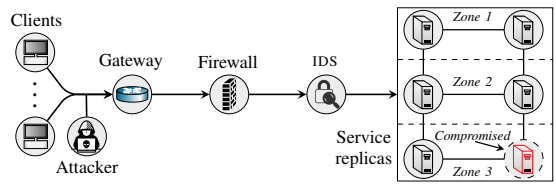
\begin{figure}[H]
  \centering
\scalebox{0.86}{
\input{tikz/service_chain.tex}
  }
  \caption{Architecture of the networked system in the example use case.}
  \label{fig:service_chain}
\end{figure}

To illustrate the need for adaptive security policies, consider the networked system in Fig.~\ref{fig:service_chain}. This system consists of service replicas that collectively provide services to a client population through a public gateway. Though intended for service delivery, this gateway is also accessible to a potential attacker who may compromise replicas. To maintain service to the clients even in the face of such attacks, the system can recover a replica suspected of being compromised by restarting it from a new virtual machine image. The decision whether to recover a replica or not is governed by a \textit{security policy} based on alerts generated by an intrusion detection system (IDS).

A key challenge when designing this policy is that the distribution of alerts depends on many factors that change over time, such as the service load and the system configuration. As a result, the policy must be frequently adapted to remain effective. Without adaptation, the policy risks either overlooking attacks or overreacting to benign alerts, both of which degrade system performance and incur operational costs.

We formalize the problem of adapting such policies in the next section, after which we present our approach.

\tikzexternaldisable
\tikzexternalenable

\section{Problem Formulation}\label{sec:preliminaries:model}
We formulate the problem of obtaining an effective security policy for a networked system as a partially observable Markov decision problem (POMDP). Following this formalism, a security policy is a function that sequentially prescribes \textit{controls} (i.e., security measures) based on a series of \textit{observations} (e.g., system metrics). These controls stochastically influence the evolution of the system's \textit{state}, which captures its security and service status. Due to limited monitoring capabilities or intentional concealment by a potential attacker, the state of the system cannot be observed directly. Therefore, controls are selected based on a state of \textit{belief}, which represents the conditional probability distribution over possible states of the system given observations. The effectiveness of these controls is measured with respect to a specified objective, which is quantified through a \textit{cost function} that should be minimized.

We denote the set of controls by $U$, the set of observations by $Z$, and the set of states by $X=\{1,\hdots, n\}$, all of which are finite. State transitions $i \rightarrow j$ under control $u$ occur at discrete times $k$ according to transition probabilities $p_{ij}(u)$. Each transition is associated with a cost $g(i,u,j) \in \Re$ and an observation $z$, which is generated with probability $p(z \mid j, u)$.

While the POMDP involves imperfect state information, it can be formulated as an equivalent problem with perfect state information \cite{ASTROM1965174}. In this formulation, the system is described by the belief state $b = \big(b(1),\dots,b(n)\big)$, where $b(i)$ is the conditional probability that the state is $i$, given the history of controls and observations. This vector belongs to the belief space $B$ and is updated through a belief estimator $F$ as
\begin{align}
b_{k}&= F(b_{k-1}, u_{k-1}, z_k). \label{eq:belief_estimator}
\end{align}
We adopt the belief-space formulation and consider security policies $\mu$ that map the belief space $B$ to the control space $U$. The cost function of such a policy is defined as
\begin{align}
J_{\mu}(b_0)=\lim_{N\rightarrow \infty}E\left\{\sum_{k=0}^{N-1}\alpha^k\hat{g}\big(b_k, \mu(b_k)\big)\right\},\label{eq:pomdp_minimization}
\end{align}
where $E\{\cdot\}$ denotes the expected value, $\alpha \in (0,1)$ is a discount factor, and the stage cost $\hat{g}(b,u)$ is defined as
\begin{align}
\hat{g}(b,u)=&\sum_{i=1}^nb(i)\sum_{j=1}^np_{ij}(u)g(i,u,j)\label{eq:hat_g}.
\end{align}
The optimal cost function $J^{\star}$, derived by optimizing over all possible policies $\mu$, uniquely satisfies the Bellman equation
\begin{align}
J^{\star}(b) &= \min_{u \in U}\left[\hat{g}(b, u) + \alpha \sum_{z \in Z}\hat{p}(z \mid b, u)J^{\star}(F(b,u,z))\right],\label{eq:optimal_cost}
\end{align}
where the probability $\hat{p}(z \mid b, u)$ is defined as
\begin{align}
\hat{p}(z \mid b,u)=&\sum_{i=1}^nb(i)\sum_{j=1}^np_{ij}(u)p(z \mid j,u).\label{eq:hat_p}
\end{align}
We say that a policy $\mu^{\star}$ is optimal if $J_{\mu^\star}=J^{\star}$. Although such a policy exists (see e.g., \cite[Thm. 7.6.1]{krishnamurthy_2016} or \cite[§ 5.6]{bertsekas2012dynamic}), there are no efficient algorithms to obtain it. Consequently, approximations are required in practice. A further complication is that the POMDP model may change over time due to changes in the networked system. When such changes occur, the policy must be adapted to remain effective.

We use the following POMDP as a running example.
\tikzexternaldisable
\begin{examplebox}
\setword{Consider}{ex:model_1} the recovery use case described in \S\ref{sec:use_case}, which involves a networked system with $K$ service replicas. Each replica has two states: $1$ (compromised) or $0$ (safe), i.e., $i=(i^1,\hdots,i^K)$ where $i^l \in \{0,1\}$. Compromises occur randomly over time and incur operational costs. Intrusion detection systems generate security alerts $z = (z^1, \ldots, z^K)$ that provide partial indications of the replicas' states. The security policy $\mu$ prescribes the control vector $u=(u^1,\hdots,u^K)$, where each $u^l$ determines whether to recover component $l$ ($u^l = 1$) or take no action ($u^l = 0$). The goal is to determine an optimal recovery policy $\mu^\star$ that balances security requirements against recovery costs. (Further details about this POMDP are provided in \S\ref{sec:evaluation}.)
\begin{figure}[H]
  \centering
  \vspace{-0.38cm}
\scalebox{0.85}{
    \input{tikz/example.tex}
  }
\end{figure}
\end{examplebox}
\tikzexternalenable
\section{Our Method for Computing Security Policies}\label{sec:framework}
Building on the above problem formulation, we develop a method for approximating optimal security policies. It consists of three components: (\textit{i}) belief estimation through particle filtering; (\textit{ii}) offline policy computation through aggregation \cite{bertsekas2012dynamic}; and (\textit{iii}) online policy adaptation through rollout \cite{bertsekas2021rollout}.

\subsection{Belief Estimation through Particle Filtering}
In a security context, the belief state represents a probabilistic estimate of the system's security state. Consequently, accurate belief estimation is key to making informed security decisions amidst uncertainty about potential attacks.

The belief state can be computed via the following recursion
\begin{align}
b_{k}(j)&= \frac{p(z_{k} \mid j, u_{k-1})\sum^{n}_{i=1}b_{k-1}(i)p_{ij}(u_{k-1})}{\sum_{i^{\prime}=1}^{n}\sum^{n}_{j^{\prime}=1}p(z_{k} \mid j^{\prime}, u_{k-1})b_{k-1}(i^{\prime})p_{i^{\prime}j^{\prime}}(u_{k-1})}.\label{eq:bayes_belief}
\end{align}
However, the complexity of this calculation is quadratic in the number of states $n$, which typically grows exponentially with the number of system components; see, e.g., the example POMDP. For this reason, we estimate the belief state by
\begin{align}
\hat{b}_k(j) = \frac{1}{M}\sum_{s=1}^M\delta_{j\hat{j}_k^{s}}, && \text{for all }j \in X,\label{eq:estimate_belief}
\end{align}
where $\delta_{ij}=1$ if $i=j$ and $\delta_{ij}=0$ if $i\neq j$. The states (particles) $\hat{j}_k^{1},\hdots,\hat{j}_k^{M}$ are sampled with probability proportional to the numerator in Eq.~\eqref{eq:bayes_belief}, and $M$ is the number of particles. Such sampling ensures that the estimate $\hat{b}_k$ converges (almost surely) to $b_k$ when $M \rightarrow \infty$; see e.g., \cite{particle_filter_survey}. Hence, Eq.~\eqref{eq:estimate_belief} provides a consistent way to estimate beliefs while allowing computational cost to be adjusted by tuning $M$.

\subsection{Offline Policy Computation through Belief Aggregation}\label{sec:belief_aggregation}
While the particle filter enables efficient estimation of beliefs [cf.~Eq.~\eqref{eq:belief_estimator}], the problem of computing an optimal policy remains intractable. We address this challenge by aggregating the belief space into a finite set of \textit{representative beliefs}. Through this aggregation, we construct an aggregate Markov decision problem (MDP) with a finite state space, whose solution can be used to approximate that of the POMDP.

The aggregation consists of two stages. First, we aggregate the states $i \in X$ into \textit{feature states} $y\in \mathcal{F}$, where the feature space $\mathcal{F}$ is smaller than $X$. Our method does not restrict how $\mathcal{F}$ is constructed; it accommodates both hand-crafted features based on engineering intuition and automatic feature extraction through data-driven approaches. We illustrate the construction of the feature space $\mathcal{F}$ in the gray box to the right.

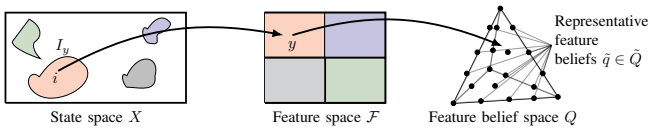
\begin{figure}
  \centering
\scalebox{0.5}{
    \input{tikz/aggregation_3.tex}
  }
  \caption{Feature-based belief aggregation: we map the state space $X$ into a feature space $\mathcal{F}$, over which beliefs are aggregated via discretization. In this illustration, a subset of states is mapped to a feature space with $4$ elements, where $I_y$ denotes the set of states that aggregate to feature state $y\in \mathcal{F}$. The resulting feature belief space $Q$ is the 3-dimensional unit-simplex.}
  \label{fig:aggregation_4}
\end{figure}

\tikzexternaldisable
\begin{examplefeatures}
In the context of our running example, the feature space $\mathcal{F}$ can be obtained by grouping the $K$ service replicas based on their network zone. Specifically, we can define a feature state as $y=(y^1,\hdots,y^V)$, where $V < K$ is the number of zones and $y^v=1$ if any replica in zone $v$ is compromised and $y^v=0$ otherwise. This yields a feature space of size $2^{V}$, which can be substantially smaller than the number of states, which is $n=2^K$ in this example.
\begin{figure}[H]
  \centering
  \vspace{-0.38cm}
\scalebox{0.85}{
    \input{tikz/example2.tex}
  }
\end{figure}
  \vspace{-0.55cm}
\end{examplefeatures}
\tikzexternalenable
After aggregating states into feature states, the second stage of our aggregation method involves grouping beliefs over the feature space $\mathcal{F}$. We denote these beliefs by $q$ and the corresponding feature belief space by $Q$. We aggregate them via discretization into a finite set of \textit{representative feature beliefs} $\Tilde{Q} \subset Q$, whose elements are written as $\tilde{q}$. This two-stage aggregation is illustrated conceptually in Fig.~\ref{fig:aggregation_4}.

\tikzexternaldisable
\tikzexternalenable

The first aggregation stage (the left arrow in Fig.~\ref{fig:aggregation_4}) involves connecting states $i \in X$ with feature states $y \in \mathcal{F}$. We specify this connection as follows.
\begin{itemize}
\item With every feature state $y\in \mathcal{F}$, we associate a subset $I_y \subset X$. We require that the sets $(I_y)_{y\in\mathcal{F}}$ are disjoint.
\item With every feature state $y\in \mathcal{F}$, we associate its \textit{disaggregation probability distribution} $\{d_{yi} \mid i \in X\}$. We require that $d_{yi}=0$ for all $i \not\in I_y$.
\item With every state $j\in X$, we associate its \textit{aggregation probability distribution} $\{\phi_{jy} \mid y \in \mathcal{F}\}$. We require that $\phi_{jy} = 1$ for all $j \in I_y$ and $y\in \mathcal{F}$.
\end{itemize}
The second aggregation stage (the right arrow in Fig.~\ref{fig:aggregation_4}) involves specifying a finite set of beliefs over the feature space $\mathcal{F}$. We construct such a set via uniform discretization as 
\begin{align}
&\Tilde{Q}\!\!=\!\!\left\{\tilde{q} \,\bigg|\, \tilde{q} \in Q, \tilde{q}(y)\!\!=\!\!\frac{\beta_y}{\rho}, \sum_{y\in \mathcal{F}}\beta_y = \rho,\beta_y\!\!\in\!\!\{0,\hdots,\rho\}\right\}, \label{eq:aggregate_belief_space}
\end{align}
where $Q$ denotes the belief space over $\mathcal{F}$ and $\rho \in \{1,2,\hdots\}$ can be interpreted as the \textit{discretization resolution}. We refer to the elements of this subset as \textit{representative feature beliefs}.

\vspace{2mm}

\noindent\textbf{\textit{Approximation of a policy and cost function for the POMDP.}}
We now use the set of representative feature beliefs $\Tilde{Q}$ [cf.~Eq.~\eqref{eq:aggregate_belief_space}], the disaggregation probabilities $d_{yi}$, and the aggregation probabilities $\phi_{iy}$ to construct a (computationally tractable) \textit{aggregate MDP} whose solution can be used to approximate that of the original POMDP.

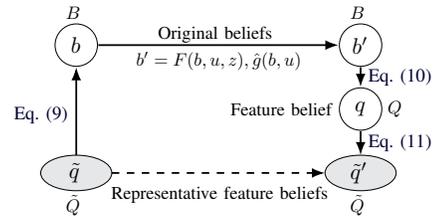
\begin{figure}[H]
  \centering
\scalebox{1.05}{
    \input{tikz/aggregation_9.tex}
  }
  \caption{One-step transition diagram of the aggregate MDP with finite state space $\Tilde{Q}$ constructed through our feature-based belief aggregation method.}\label{fig:aggregation_5}
\end{figure}

The aggregate MDP starts from a representative feature belief $\tilde{q} \in \Tilde{Q}$ and evolves as follows. First, it transitions from $\tilde{q}$ to a belief $b \in B$ via the disaggregation probabilities as
  \begin{align}
b(i) &= \sum_{y\in \mathcal{F}}\tilde{q}(y)d_{yi}, && \text{for all } i \in X. \label{eq:belief_transition_2}    
  \end{align}
Subsequently, a control $u \in U$ is applied, which generates an observation $z \in Z$ according to Eq.~\eqref{eq:hat_p} and incurs a cost $\hat{g}(b,u)$ according to Eq.~\eqref{eq:hat_g}. The belief $b$ is then updated as $b'=F(b,u,z)$ [cf.~Eq.~\eqref{eq:belief_estimator}], after which the MDP transitions to a feature belief $q \in Q$ via the aggregation probabilities as
\begin{align}
q(y) &= \sum_{i=1}^{n} b^{\prime}(i)\phi_{iy}, && \text{for all } y \in \mathcal{F}.\label{eq:belief_transition_1}
\end{align}
Finally, the resulting feature belief $q$ is mapped to a \textit{representative} feature belief $\tilde{q}' \in \Tilde{Q}$ via the nearest-neighbor mapping
\begin{align}
\tilde{q}' \in \argmin_{\tilde{q} \in \Tilde{Q}} \norm{q - \tilde{q}},\label{eq:belief_transition_3}
\end{align}
where tie-breaking is consistent and $\norm{\cdot}$ is the maximum norm. From the new representative feature belief $\tilde{q}^{\prime} \in \Tilde{Q}$, the MDP proceeds analogously by repeating the transitions in Eqs.~\eqref{eq:belief_transition_2}-\eqref{eq:belief_transition_3}. As a result, we obtain a well-defined MDP with state space $\Tilde{Q}$ whose one-step transition diagram is shown in Fig.~\ref{fig:aggregation_5}.
\tikzexternaldisable
\tikzexternalenable
Due to the finite state space, the aggregate MDP in Fig.~\ref{fig:aggregation_5} can be efficiently solved using DP or RL. Let $r^{\star}$ and $\pi^{\star}$ denote the optimal cost function and policy in this MDP, respectively. Further, let $\Phi: B\mapsto\Tilde{Q}$ denote the mapping defined by Eqs.~\eqref{eq:belief_transition_1}-\eqref{eq:belief_transition_3}, i.e.,
\begin{align}
\Phi(b) \in \argmin_{\tilde{q} \in \Tilde{Q}}\max_{y \in \mathcal{F}}\left|\tilde{q}(y)-\sum_{i=1}^{n} b(i)\phi_{iy}\right|,\label{eq:aggregation_mapping}
\end{align}
where ties in the $\argmin$ are broken using a fixed rule.

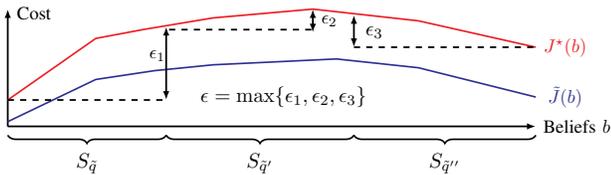
\begin{figure}[H]
  \centering
\scalebox{0.78}{
    \input{tikz/rep28.tex}
  }
  \caption{Illustration of the scalar $\epsilon$ of Prop.~\ref{prop:aggregation_bound}. The illustration is based on an approximation with three representative feature beliefs: $\Tilde{Q}=\{\tilde{q}, \tilde{q}^{\prime},\tilde{q}^{\prime\prime}\}$. The corresponding belief space partitions are: $S_{\tilde{q}}$, $S_{\tilde{q}^{\prime}}$, and $S_{\tilde{q}^{\prime\prime}}$; cf.~Eq.~\eqref{eq:partitioning}.}\label{fig:rep20}
\end{figure}

Using this mapping, we approximate the optimal cost function $J^{\star}$ and policy $\mu^{\star}$ of the original POMDP as
\begin{align}
\Tilde{J}(b) &= r^{\star}(\Phi(b)) \text{ and } \mu(b)=\pi^{\star}(\Phi(b)), \text{ for all } b \in B.\label{eq:approximation_1}
\end{align}
We refer to the difference between the cost function approximation $\tilde{J}$ obtained through Eq.~\eqref{eq:approximation_1} and the optimal cost function $J^{\star}$ as the \textit{approximation error}. To gain insight into this error, note that the aggregation mapping $\Phi$ [cf.~Eq.~\eqref{eq:aggregation_mapping}] partitions the belief space $B$ into disjoint subsets $S_{\tilde{q}}$ as
\begin{align}
B = \bigcup_{\tilde{q} \in \tilde{Q}}S_{\tilde{q}}, && \text{where } S_{\tilde{q}}=\left\{b \mid b \in B, \Phi(b) = \tilde{q}\right\}.\label{eq:partitioning}
\end{align}
In view of Eq.~\eqref{eq:approximation_1}, this partitioning means that the approximation error is determined by how much the optimal cost function $J^{\star}(b)$ varies for beliefs $b$ within the same partition $S_{\tilde{q}}$. This insight is formalized by the following proposition.
\begin{proposition}[Approximation error bound]\label{prop:aggregation_bound}
The error of the cost function approximation in Eq.~\eqref{eq:approximation_1} is bounded as
\begin{align*}
|\Tilde{J}(b) - J^{\star}(b)| \leq \frac{\epsilon}{1-\alpha}, && \text{for all } b \in S_{\tilde{q}}, \tilde{q} \in \Tilde{Q},
\end{align*}
where $\epsilon$ is a finite constant defined by
\begin{align*}
\epsilon = \max_{\tilde{q} \in \Tilde{Q}}\sup_{b,b^{\prime} \in S_{\tilde{q}}}|J^{\star}(b)-J^{\star}(b^{\prime})|.  
\end{align*}
\end{proposition}
A more general version of this proposition and additional auxiliary results are proved in \cite{lihambert}. Proposition~\ref{prop:aggregation_bound} implies that the error of the approximation $\tilde{J}$ [cf.~Eq.~\eqref{eq:approximation_1}] is small if the mapping $\Phi$ [cf.~Eq.~\eqref{eq:aggregation_mapping}] conforms to the optimal cost function $J^{\star}$ in the sense that $\Phi$ varies little in regions of the belief space where $J^{\star}$ also varies little. Hence, Prop.~\ref{prop:aggregation_bound} provides a criterion to guide feature design: we seek a feature space $\mathcal{F}$, disaggregation probabilities $d_{yi}$, and aggregation probabilities $\phi_{iy}$ that induce belief space partitions $S_{\tilde{q}}$ [cf.~Eq.~\eqref{eq:partitioning}] over which $J^{\star}$ is approximately constant; see Fig.~\ref{fig:rep20}.
\tikzexternaldisable
\tikzexternalenable
A special case of interest, which we refer to as \textit{identity aggregation}, is when each state is mapped to a unique feature state and vice versa. In this case, the cost function approximation in Eq.~\eqref{eq:approximation_1} converges to the optimal cost function $J^{\star}$ when the discretization resolution $\rho$ in Eq.~\eqref{eq:aggregate_belief_space} is increased, as stated in the following proposition.
\begin{proposition}[Asymptotic optimality]\label{thm:consistent_1}
Suppose that $X=\mathcal{F}$ and $I_i=\{i\}$ for all states $i \in X$. We have
\begin{align*}
&\lim_{\rho \rightarrow \infty} |\Tilde{J}(b) - J^{\star}(b)| = 0, && \text{for all }b \in B.
\end{align*}
\end{proposition}
This proposition implies that the approximation error vanishes under identity aggregation as the discretization resolution $\rho$ increases. We present the proof of Prop.~\ref{thm:consistent_1} in Appendix~\ref{app:proof_consistent}.

From a network security viewpoint, the benefit of the preceding propositions is that they allow a network operator to weigh the trade-off between computational expenditure and policy performance. For instance, the operator can allocate more resources and increase the discretization resolution in critical segments of the network to achieve finer security control where potential breaches carry severe consequences. 

\subsection{Online Policy Adaptation through Rollout}\label{sec:rollout}
The two-stage aggregation described above provides a scalable approach to \textit{offline} policy computation. However, it does not provide a means for \textit{online} policy adaptation. In a network security context, such adaptation is necessary as changes in networked systems occur regularly due to, e.g., evolving operational requirements and goals \cite{devops_trends}; cf.~Fig.~\ref{fig:stats}.

For this reason, we complement the \textit{base policy} $\mu$ computed offline via Eq.~\eqref{eq:approximation_1} with online lookahead optimization and \textit{rollout} \cite{bertsekas2021rollout}. Specifically, at each step of online execution, we simulate the system's evolution several steps into the future, which allows us to evaluate different controls and adapt the base policy based on their outcomes. Effectively, these simulations can be understood by a security operator as a form of `what if' analysis, where the system anticipates possible threats and assesses the impact of various security measures.

Mathematically, at each time step $k$ during online execution, we transform the (pre-computed) base policy $\mu$ [cf.~Eq.~\eqref{eq:approximation_1}] to a \textit{rollout policy} $\tilde{\mu}$ via lookahead optimization as
\begin{align}
  &\tilde{\mu}(b_k) \in \argmin_{u_k \in U}\Bigg[\hat{g}(b_k,u_k) + \min_{\mu_{k+1},\hdots,\mu_{k+\ell-1}}\mathop{E}_{b_{k+1},\hdots,b_{k+\ell}}\nonumber\\
  &\quad\quad\quad\quad\bigg\{\sum_{j=k+1}^{k+\ell-1}\alpha^{j-k}\hat{g}(b_j, \mu_j(b_j)) + \alpha^{\ell}\Tilde{J}_{\mu}(b_{k+\ell})\bigg\}\Bigg] \label{eq:rollout},
\end{align}
where $\ell \geq 1$ is the lookahead horizon, and the cost $\Tilde{J}_{\mu}(b_{k+\ell})$ is estimated based on $L$ simulations as
\begin{align}
\Tilde{J}_{\mu}(b_{k\!+\!\ell})\!\!&=\!\!\frac{1}{L}\sum_{s\!=\!1}^{L}\sum_{l\!=\!k+\ell}^{k\!+\!\ell\!+\!m\!-\!1}\!\!\alpha^{l\!-k\!-\!\ell}\hat{g}\big(b^{s}_l,\mu(b^{s}_l)\big) + \alpha^{m}\Tilde{J}(b^{s}_{k\!+\!\ell\!+\!m}),\label{eq:rollout_approximation}                                  
\end{align}
where $\Tilde{J}$ is the cost function approximation in Eq.~\eqref{eq:approximation_1}, $m$ is the rollout horizon, the cost function $\hat{g}$ is defined in Eq.~\eqref{eq:hat_g}, and $(b_{k+\ell}^s,\dots,b_{k+\ell+m}^s)$ is the belief trajectory of the $s$th simulation. Subsequently, the first control obtained through Eq.~\eqref{eq:rollout} is applied to the system, which yields an observation $z$ that is used to update the belief through Eq.~\eqref{eq:belief_estimator}. The same computation is then repeated from the updated belief.

In a security context, a key property of the lookahead optimization in Eq.~\eqref{eq:rollout} is that the computational cost can be scaled by tuning the number of lookahead steps ($\ell$) and the rollout horizon ($m$). This scalability enables our method to accommodate resource constraints that are common in operational systems. Another fundamental property of Eq.~\eqref{eq:rollout} is the \textit{policy improvement} property, which is formalized below.
\begin{proposition}[Policy improvement of the adaptation]\label{prop:improvement}
If the policy evaluation in Eq.~\eqref{eq:rollout_approximation} is exact, i.e., if $\Tilde{J}_{\mu}=J_{\mu}$, then the rollout policy $\tilde{\mu}$ improves the base policy $\mu$, i.e., $J_{\tilde{\mu}}\leq J_{\mu}$. Further, the suboptimality of $\tilde{\mu}$ is bounded as
\begin{align*}
\norm{J_{\tilde{\mu}} - J^{\star}} \leq \frac{2\alpha^{\ell}}{1-\alpha}\norm{\Tilde{J}_{\mu} - J^{\star}}.
\end{align*}
\end{proposition}
The implication of this proposition is that our method can adapt \textit{online} to changes in a given system model without repeating the offline computation. The proof follows directly from standard results by the last author; see \cite[Prop. 2.3.1]{bertsekas2021rollout} and \cite[Prop. 5.1.1]{bertsekas2019reinforcement} for details. We omit it here for brevity.

To apply the rollout method in Eq.~(15), we require a model of the system, either in the form of an analytical model or a simulator. Identifying such a model is part of the broader \textit{system identification} methodology; see e.g., Ljung \cite{Ljung1998}. When significant changes occur in the system (e.g., discovery of a new vulnerability), the identification must be repeated to update the model. The detailed updating procedure depends on the nature of the changes. In many cases, changes are structural and straightforward to incorporate. For instance, the deployment of a new service or the addition of a network link can be reflected directly in the model's structure. In other cases, changes may be more subtle, such as shifts in observation patterns, which may require statistical estimation to recalibrate the model's observation probabilities. Standard techniques for this purpose include maximum likelihood estimation and Bayesian learning; see e.g., Hammar \cite[\S 3.8]{kim_phd_thesis}. We demonstrate an approach to such estimation in \S\ref{sec:testbed_eval}.

While we require that the model be updated to reflect major changes to the system, the model does not have to be perfect for the rollout policy to be effective. Since the rollout policy is computed using DP principles, its robustness to model misspecification is related to that of other DP approaches. This robustness has been extensively analyzed in the robust control literature; see e.g., Iyengar \cite{1b127c8f-33c5-380a-bc31-862489026347}. Such an analysis is peripheral to the focus of the present paper and is therefore omitted here.

Lastly, we note that the requirement of an updated model is shared by other RL approaches. The key advantage of our method is that the offline policy optimization does not need to be repeated to update $\mu$ and $\tilde J$ in Eq.~\eqref{eq:rollout_approximation} for the new model. By contrast, methods based on deep RL require extensive offline retraining to adjust the policy after model updates \cite{deep_rl_cyber_sec}.

\subsection{Summary of Our Method for Computing Security Policies}
In summary, our method for computing security policies is illustrated in Fig.~\ref{fig:framework} and involves three components:
\begin{enumerate}
\item \textit{Offline policy computation via Eq.~\eqref{eq:approximation_1}}.
  \begin{itemize}
  \item[] At the core of our method is the computation of a \textit{base security policy} through DP in an aggregated belief space. This computation offers theoretical guarantees and scales to large systems.
  \end{itemize}
\item \textit{Online belief estimation via Eq.~\eqref{eq:estimate_belief}}.
  \begin{itemize}
  \item[] Our method uses network logs and metrics to estimate a probabilistic belief about the system state through \textit{particle filtering}. This belief quantifies the likelihood of potential system compromises and serves as the basis for selecting appropriate security controls.
  \end{itemize}
\item \textit{Online policy adaptation via Eq.~\eqref{eq:rollout}.}
  \begin{itemize}
  \item[] During online operation, our method adapts the base policy to system changes through lookahead optimization and rollout based on a given model. This procedure ensures that the adapted policy improves the base policy under general conditions.
  \end{itemize}  
\end{enumerate}

\begin{figure}[H]
  \centering
\scalebox{0.76}{
    \input{tikz/pyramid.tex}
  }
  \caption{The three computational layers of our method.}
  \label{fig:pyramid}
\end{figure}
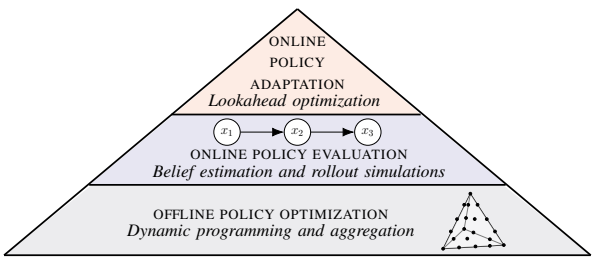

From an architectural point of view, our method extends current methods for computing security policies, which predominantly use offline (deep) RL [cf.~\S\ref{sec:related_work}], with online rollout and policy adaptation; see Fig.~\ref{fig:pyramid}.
\tikzexternaldisable
\tikzexternalenable
\section{Experimental Evaluation}\label{sec:evaluation}
In this section, we present an experimental evaluation of our method. We start by applying it to an instantiation of the example POMDP with a small state space, which allows us to illustrate Props.~\ref{prop:aggregation_bound}-\ref{prop:improvement}. We then assess the practical applicability of our method by applying it to an instantiation of the example POMDP with a practical system configuration in our testbed. Lastly, we evaluate the performance of our method on the CAGE-2 benchmark \cite{cage_challenge_2_announcement}, which enables a comparison with state-of-the-art methods for computing network security policies. The experimental setup is described in Appendix~\ref{appendix:hyperparameters}. Source code of our implementation is available at \cite{rollout_software}.

\subsection{Numerical Illustrations Based on the Running Example}\label{sec:example_eval}
The example POMDP models a networked system with $K$ service replicas; see \S\ref{sec:use_case}. A state $i=(i^1,\hdots,i^K)$ of this POMDP represents the replicas' compromise statuses, where $i^l=1$ if replica $l$ is compromised and $i^l=0$ otherwise. Similarly, a control $u=(u^1,\hdots,u^K)$ represents recovery actions, where $u^l=1$ means to recover replica $l$ and $u^l=0$ means no recovery. We define the cost function as
\begin{align}
g(i, u, j) = \sum_{l=1}^{K}\overbrace{w^{l}i^l(1-u^l)}^{\text{intrusion cost}} + \overbrace{u^l(1-i^l)}^{\text{recovery cost}},\label{example_pomdp_cost}
\end{align}
where $w^{l}$ is a weighting factor for the intrusion cost of replica $l$. Hence, costs are incurred for unmitigated intrusions ($i^l=1$) and unnecessary recovery actions ($u^l=1$ and $i^l=0$).

\begin{figure}[H]
  \centering
\scalebox{0.8}{
    \input{tikz/obs_dist.tex}
  }
  \caption{Observation distribution for each replica $l$ in the example POMDP.}
  \label{fig:obs_dist}
\end{figure}
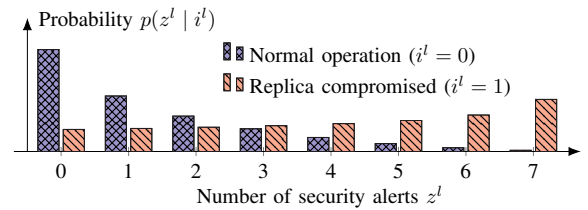

The observations of the POMDP correspond to the number of security alerts generated by an IDS. In the next section, we present an evaluation where these alerts are measured from an operational system. However, for the numerical illustrations presented in this section, we define the alert distribution as
\begin{align*}
p(z \mid i, u) &= \prod_{l=1}^Kp(z^l \mid i^l), && \text{for all }z\in Z,i\in X,u\in U,
\end{align*}  
where each $p(z^l \mid i^l)$ follows the Beta-binomial distribution shown in Fig.~\ref{fig:obs_dist}. This distribution reflects that alerts may occur during normal operation but are more likely during attacks; see Appendix~\ref{appendix:hyperparameters} for details about this distribution. 

\tikzexternaldisable
\tikzexternalenable

The transition probabilities $p_{ij}(u)$ are defined as follows. If replica $l$ is compromised ($i^l=1$), then it remains so until recovery is applied ($u^l=1$), at which point the state $i^l$ is set to $0$. Otherwise, the probability that it becomes compromised is $\min\{0.2(1+\mathcal{N}_l(i)), 1\}$, where $\mathcal{N}_l(i)$ is the number of compromised neighbors of replica $l$ in the network.

\vspace{2mm}

\noindent\textbf{\textit{Instantiation of our method}.} We instantiate our method for the example POMDP with the zone-based feature space defined in \S\ref{sec:belief_aggregation} and different discretization resolutions $\rho$; cf.~Eq.~\eqref{eq:aggregate_belief_space}. Specifically, we group the replicas into $V<K$ zones and define a feature state as $y=(y^1,\hdots,y^V)$, where $y^v=1$ if any replica in zone $v$ is compromised and $y^v=0$ otherwise. This yields a feature space $\mathcal{F}$ of size $2^{V}$, which can be substantially smaller than the number of states, which is $n=2^K$. We assume that the network follows the line topology shown in \S\ref{sec:belief_aggregation} and that neighboring replicas are grouped into zones of equal size. For example, if $K=4$ and $V=2$, then replicas $1$ and $2$ are in zone $1$, whereas replicas $3$ and $4$ are in zone $2$.

We define the aggregation probabilities as
\begin{align*}
  \phi_{iy} &= 
\begin{dcases}
  1 & \text{if } y^v=\max\{i^l \mid \text{replica l is in zone $v$}\} \text{ for all }v,\\
  0 & \text{otherwise},
\end{dcases}
\end{align*}
for all states $i \in X$ and feature states $y \in \mathcal{F}$. Similarly, we define the disaggregation probabilities as
\begin{align*}
  d_{yi} &= 
\begin{dcases}
  1, & \text{if } i^l=y^v \text{for all replicas $l$ in each zone $v$},\\
  0, & \text{otherwise},
\end{dcases}
\end{align*}
for all states $i \in X$ and feature states $y \in \mathcal{F}$. Finally, we define the weight in Eq.~\eqref{example_pomdp_cost} to be $w^{l}=2l$ for each replica $l$.

\vspace{2mm}

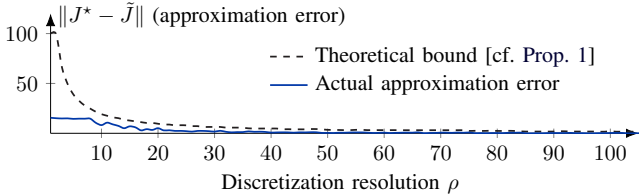
\begin{figure}[H]
  \centering
\scalebox{0.83}{
    \input{tikz/rep21.tex}
  }
  \caption{Comparison between the theoretical error bound in Prop.~\ref{prop:aggregation_bound} and the actual error of the approximation $\Tilde{J}$ [cf.~Eq.~\eqref{eq:approximation_1}] when applied to the example POMDP with $K=1$ and varying discretization resolutions $\rho$; cf.~Eq.~\eqref{eq:aggregate_belief_space}.}
  \label{fig:rep21}
\end{figure}

\begin{figure}[H]
  \centering
\scalebox{0.8}{
    \input{tikz/rep20.tex}
  }
  \caption{Number of representative feature beliefs [cf.~Eq.~\eqref{eq:aggregate_belief_space}] in function of the discretization resolution; curves relate to state spaces of different sizes.}\label{fig:rep22}
\end{figure}
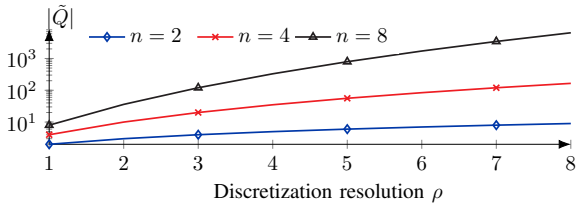

\noindent\textbf{\textit{Numerical illustrations.}}
We start by analyzing how close the bound in Prop.~\ref{prop:aggregation_bound} is to the actual approximation error, i.e., $\norm{\tilde{J}-J^{\star}}$. Although this quantity is generally intractable to evaluate due to its dependence on the optimal cost function $J^{\star}$, the case of a single replica ($K=1$) yields a sufficiently small POMDP to permit exact computation of $J^{\star}$ via the incremental pruning algorithm \cite{incremental_pruning_pomdp}. As shown in Fig.~\ref{fig:rep21}, the bound is not tight but becomes increasingly accurate when the resolution $\rho$ increases, as asserted in Prop.~\ref{thm:consistent_1}. However, increasing $\rho$ also causes the number of representative feature beliefs to grow, which is illustrated in Fig.~\ref{fig:rep22}. Hence, $\rho$ governs a trade-off between computational expedience and approximation error.

\tikzexternaldisable
\tikzexternalenable

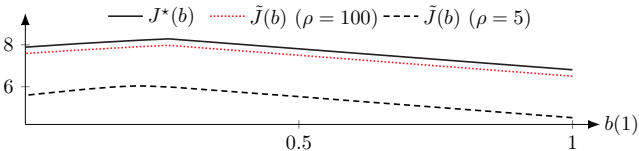
\begin{figure}[H]
  \centering
\scalebox{0.77}{
    \input{tikz/value_fun.tex}
  }
  \caption{Comparison between the optimal cost function $J^{\star}$ for the example POMDP and the approximation $\Tilde{J}$ [cf.~Eq.~\eqref{eq:approximation_1}]. The number of service replicas is $K=1$. Hence, $b(1)$ is the belief of system compromise.}
  \label{fig:value_fun}
\end{figure}

\tikzexternaldisable
\tikzexternalenable
Next, Fig.~\ref{fig:value_fun} shows the structure of the optimal cost function $J^{\star}$ and the cost function approximation $\Tilde{J}$; cf.~Eq.~\eqref{eq:approximation_1}. Interestingly, even when the difference between them is significant, they have a similar structure. We also note that $\Tilde{J} \leq J^{\star}$. Although not guaranteed in our setup, $\Tilde{J}$ can serve as a lower bound for $J^{\star}$ under certain conditions; see \cite[Prop. 6]{lihambert}.
\tikzexternaldisable
\tikzexternalenable

We now turn our attention to Prop.~\ref{prop:improvement}. Figure~\ref{fig:rollout_eval} shows the performance of the base policy $\mu$ [cf.~Eq.~\eqref{eq:approximation_1}] and the (adapted) rollout policy $\tilde{\mu}$ [cf.~Eq.~\eqref{eq:rollout}] for varying discretization resolutions $\rho$. We observe that $\tilde{\mu}$ incurs a lower cost than $\mu$, with the difference being dramatic in some cases. A theoretical explanation for the large cost reduction is provided by Bertsekas in \cite{bertsekas2022lessons}, where it is shown that the rollout computation performs one step of Newton's method for solving Bellman's equation. As a consequence, if the base policy is sufficiently close to optimal, the rollout policy can exhibit a superlinear rate of convergence to the optimal, just like Newton’s method in classical optimization.

\tikzexternaldisable
\begin{figure}[H]
  \centering
\scalebox{0.77}{
    \input{tikz/rollout_eval.tex}
  }
  \caption{Performance of rollout when applied to the example POMDP with $K=V=4$. The rollout and lookahead horizons are $m=10$ and $\ell=1$. The base policy is computed with varying discretization resolution $\rho$; cf.~Eq.~\eqref{eq:aggregate_belief_space}.}
  \label{fig:rollout_eval}
\end{figure}
\tikzexternalenable 

We now analyze the quality of the zone-based feature space defined above. We assess its quality in two ways: computational scalability and approximation error. Figure~\ref{fig:scalability} shows the offline computation time for the cost function approximation $\Tilde{J}$ [cf.~Eq.~\eqref{eq:approximation_1}] as a function of the number of replicas $K$ and zones $V$. When neither state nor belief aggregation is used, the computation becomes impractical for $K > 2$, which is expected due to the high complexity of solving POMDPs exactly. Similarly, when using only belief aggregation (i.e., $V=K$), the computation time grows exponentially with $K$ and becomes impractical for $K > 5$. By contrast, when using both state and belief aggregation (i.e., $V < K$), the computation time remains manageable even as $K$ increases. This demonstrates the scalability of feature-based aggregation.

However, the cost of this scalability is a higher approximation error, as shown in Fig.~\ref{fig:feature_cost}. We observe that the feature space obtained with $V=2$ zones yields the best trade-off between scalability and approximation error. In particular, it leads to less than 5\% approximation error compared to the optimal solution with $K=1$ replicas and less than 6\% error compared to the solution obtained with $K=V=4$ replicas.

\tikzexternaldisable
\begin{figure}[H]
  \centering
\scalebox{0.83}{
    \input{tikz/scalability.tex}
  }
  \caption{Offline compute time to obtain the cost function approximation $\tilde{J}$ through Eq.~\eqref{eq:approximation_1} for the example POMDP with varying number of zones $V$ and sizes of the feature space (i.e., varying $|\mathcal{F}|$). The discretization resolution is defined as $\rho=2$; cf.~Eq.~\eqref{eq:aggregate_belief_space}. (The computation time for $V=K$ and no aggregation becomes impractical for large values of $K$.)}\label{fig:scalability}
\end{figure}
\tikzexternalenable

\tikzexternaldisable
\begin{figure}[H]
  \centering
\scalebox{0.83}{
    \input{tikz/feature_cost.tex}
  }
  \caption{Performance of the base policy $\mu$ [cf.~Eq.~\eqref{eq:approximation_1}] when applied to the example POMDP with varying number of zones $V$ and sizes of the feature space (i.e., varying $|\mathcal{F}|$). The discretization resolution is defined as $\rho=2$; cf.~Eq.~\eqref{eq:aggregate_belief_space}. (Computing the optimal cost is impractical for $K > 1$.)}\label{fig:feature_cost}
\end{figure}
\tikzexternalenable
\vspace{2mm}

\noindent\textbf{\textit{Automatic feature extraction.}} The zone-based feature space described above is motivated by engineering intuition. However, such intuition may not always suffice to construct effective features. In these situations, effective features can be learned directly from data using deep learning techniques. For example, given a simulator of the POMDP and a policy $\pi$, one can generate a dataset of state–cost pairs and train a neural network to approximate a function $H_{\pi}: X \mapsto \Re$, where the value $H_{\pi}(i)$ represents the expected cost incurred under policy $\pi$ when the belief state is initialized as $b(i)=1$. The learned function can then be used to inform the design of feature states, as originally proposed by the last author \cite{bertsekas2018featurebasedaggregationdeepreinforcement}.

We apply this approach to the example POMDP as follows. We first generate state–cost pairs by simulating the POMDP under a threshold policy that recovers a replica whenever the belief of compromise exceeds $0.7$. We then train a neural network to approximate this policy's cost function. For the experiments reported in this paper, we use a neural network with three hidden layers of $64$ neurons each and ReLU activations. From the trained network, we extract the output of the final hidden layer to obtain a feature vector for each state $i \in X$. To construct the feature space, we cluster these vectors using K-means with $|\mathcal{F}|=4$ clusters \cite{MacQueen1967}, where each cluster defines a feature state $y$. Finally, we define the aggregation and disaggregation probabilities by assigning each state to the feature state that is nearest in Euclidean distance.

Figure~\ref{fig:feature_extraction} compares the cost of the base policy $\mu$ [cf.~Eq.~\eqref{eq:approximation_1}] when using feature states derived via deep learning versus zone-based feature states. We observe that for small numbers of replicas ($K$), both approaches yield comparable performance. However, as $K$ increases, the zone-based feature states lead to significantly lower cost.
\tikzexternaldisable
\begin{figure}[H]
  \centering
\scalebox{0.83}{
    \input{tikz/feature_extraction.tex}
  }
  \caption{Performance of the base policy $\mu$ [cf.~Eq.~\eqref{eq:approximation_1}] when applied to the example POMDP with zone-based features and features learned through a neural network. The discretization resolution is defined as $\rho=2$; cf.~Eq.~\eqref{eq:aggregate_belief_space}.}\label{fig:feature_extraction}
\end{figure}
\tikzexternalenable

\subsection{Testbed Evaluation}\label{sec:testbed_eval}
We now complement the analytical and numerical evaluations presented in the previous sections by evaluating our method on an operational system. To facilitate this evaluation, we deploy the networked system described in \S\ref{sec:use_case} on our testbed and subject it to a variety of cyberattacks. These attacks produce system measurements and logs, which we use to identify the parameters of the example POMDP. We then apply our method to the identified POMDP to compute a security policy. Finally, we deploy and execute the computed policy in our testbed and evaluate its performance against real cyberattacks, as well as its adaptability to system changes.

\vspace{2mm}

\noindent\textbf{\textit{Testbed setup.}} We deploy the networked system described in \S\ref{sec:use_case} on three physical servers in our testbed: two Supermicro 7049 and one Dell R740 2U. We run $K=8$ virtual service replicas on these servers. Server $1$ hosts replicas $1-4$; server $2$ hosts replicas $5-8$; and server $3$ emulates the cloud gateway. The configurations of the service replicas are listed in Table~\ref{tab:servers} and details of our testbed are available in Appendix~\ref{app:testbed}.

\begin{table}[H]
  \centering
\scalebox{0.81}{
  \begin{tabular}{llll}
\rowcolor{lightgray}
  {\textit{Replica}} & {\textit{Operating system}} & {\textit{Background services}} & {\textit{Vulnerabilities}}  \\ \midrule
  $1$  & Deb 9.2 & Apache2 & CWE-89\\
  $2$  & Deb Jessie & FTP & CVE-2015-3306\\
  $3$  & Ubuntu 20 & SSH, Spark & CWE-1391 \\
  $4$  & Deb Jessie & Phpmailer & CVE-2016-10033\\
  $5$  & Deb Wheezy & NGINX & CVE-2014-6271\\
  $6$  & Deb Jessie & SSH, GRPC & CWE-1391, CVE-2010-0426\\
  $7$  & Deb Jessie & SSH, Spring boot & CVE-2015-5602, CWE-1391\\
  $8$  & Deb Jessie & Postgresql, Samba & CVE-2017-7494\\
\end{tabular}
}
\caption{Replicas of the networked system described in \S\ref{sec:use_case}. Vulnerabilities are identified using CVE \cite{cve} and CWE \cite{cwe} identifiers.}\label{tab:servers}
\end{table}

\vspace{2mm}

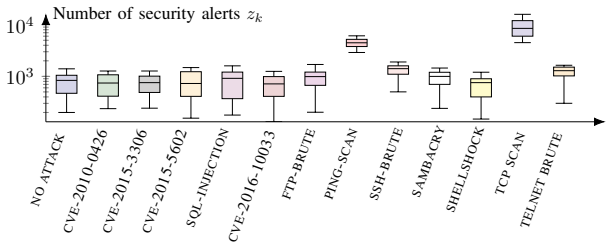
\begin{figure}
  \centering
\scalebox{0.75}{
    \input{tikz/security_alerts.tex}
  }
  \caption{Box plots of the empirical distributions (based on $24,386$ measurements) of security alerts in our testbed under different attacks (indicated on the x-axis); see Appendix~\ref{app:testbed} for details about the attacks and the data collection. Each box represents the interquartile range (IQR) of the distribution, with the median shown as a horizontal line; vertical lines extend to points within 1.5 times the IQR. The empirical measurements are available at \cite{rollout_software}.}
  \label{fig:security_alerts}
\end{figure}

\noindent\textbf{\textit{System identification.}} We identify the observation distribution of the example POMDP from measurement data obtained by running a sequence of emulated attacks and controls on our testbed. We define the length of a time step in our testbed to be $30$ seconds. During each step, we execute attacks against a subset of replicas; see Appendix~\ref{appendix:hyperparameters} for the list of attacks. We then measure the observation $z_k$ from our testbed by reading log files. We repeat this procedure for $24,386$ time steps and use the empirical distribution of security alerts over those time steps to define the observation distribution $p(z \mid i, u)$ in the POMDP. Figure~\ref{fig:security_alerts} shows the estimated distribution.

\vspace{2mm}

\noindent\textbf{\textit{Evaluation scenarios.}}
We consider two evaluation scenarios, both of which represent executions of the example POMDP.
\begin{enumerate}
\item \setword{\textsc{stationary system}}{testbed_scenario_1}: In this scenario, the system operates under the same conditions throughout.
\item \setword{\textsc{non-stationary system}}{testbed_scenario_2}: This scenario is divided into two time intervals. From time step $k=0$ to $k=200$, the system operates under the same conditions as in the first scenario. After time step $k=200$, background processes are started on each service replica, which alters the alert distribution [cf. Fig.~\ref{fig:security_alerts}] and requires policy adaptation.
\end{enumerate}

\tikzexternaldisable
\tikzexternalenable

\vspace{2mm}

\noindent\textbf{\textit{Methods for comparison}.} We compare our method with two baseline policies: (\textit{i}) a periodic policy that recovers replicas every fifth time step; and (\textit{ii}) a policy computed through PPO \cite[Alg. 1]{ppo}, which is a popular method in related work.

\vspace{2mm}

\noindent\textbf{\textit{Evaluation metrics}.} We compare the computed policies using three metrics: the cost in Eq.~\eqref{example_pomdp_cost}, the frequency with which the policies initiate recovery, and the average time-to-recovery, i.e., the average time from compromise to recovery initiation.

We compare the methods in terms of compute time and adaptation time, which refers to the time required to obtain a fully adapted policy. To evaluate the degree of adaptation of a policy $\mu$, we use the adaptation-completion metric
\begin{align}
A(\mu) = \frac{J_{0}(b_0)-J_{\mu}(b_0)}{J_0(b_0)-J_{1}(b_0)},   \label{eq:adaptation_metric}
\end{align}
where $J_0$ is the cost function at the start of adaptation and $J_1$ is the cost function of a fully adapted policy. Since the optimal cost is unknown, we define $J_1$ to be the cost of the best known policy. Here $b_0$ is the known initial belief, i.e., $J_{\mu}(b_0)$ is the expected cost of policy $\mu$ at the start of the evaluation.

\vspace{2mm}

\noindent\textbf{\textit{Instantiation of our method}.} We use identity aggregation (i.e., $X=\mathcal{F}$) with discretization resolution $\rho=2$, which leads to $32,896$ representative feature beliefs; cf.~Eq.~\eqref{eq:aggregate_belief_space}.

\vspace{2mm}

\noindent\textbf{\textit{Evaluation results.}}
The compute times and the cost values of each method are listed in Table~\ref{tab:testbed_results}. We observe that our method and PPO \cite{ppo} achieve the lowest cost for Scenario $1$, significantly outperforming the baselines. However, the results of PPO exhibit a higher variability than our method, which we attribute to PPO's tendency to converge to different local optima. Moreover, PPO requires four times more offline compute time than our method. We also note that the performance of our method improves when increasing the rollout and lookahead horizons ($m$ and $\ell$), at the expense of more online compute.

\begin{figure}
  \centering
\scalebox{0.75}{
    \input{tikz/adaption_eval.tex}
  }
  \caption{Policy adaptation time after a system change in our testbed (Scenario $2$). The metric $A(\mu_t)$ is calculated according to Eq.~\eqref{eq:adaptation_metric} with $J_1(b_0)=20.92$, where $\mu_t$ is the policy after $t$ seconds of adaptation.}
  \label{fig:adaption_eval}
\end{figure}
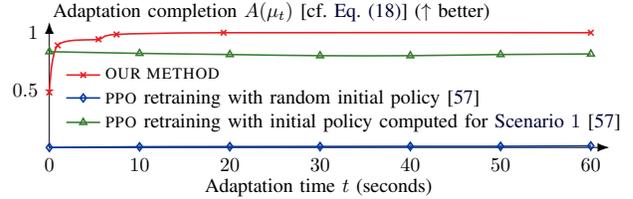

In the results from Scenario $2$, we observe that our method outperforms all other methods. We explain this improvement by our method's ability to adapt policies to changes. The adaptation time of our method and the baselines are shown in Fig.~\ref{fig:adaption_eval}. We find that effective policy adaptation takes $19$ seconds with our method and computing hardware, compared to more than $30$ minutes with PPO using the same hardware.

\begin{figure}
  \centering
\scalebox{0.75}{
    \input{tikz/recovery_frequency.tex}
  }
  \caption{Results from the testbed evaluation. The time-to-recovery and recovery frequency are averaged across $1000$ time steps.}
  \label{fig:recovery_frequency}
\end{figure}
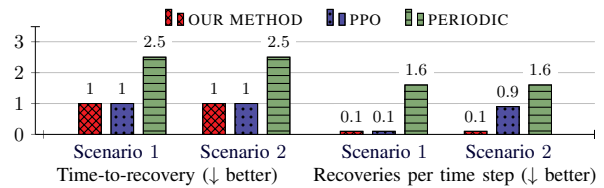

Lastly, Fig.~\ref{fig:recovery_frequency} shows the average time-to-recovery and the recovery frequency of the policies. We observe that the policies computed by our method and PPO achieve the lowest time-to-recovery, indicating that they initiate recovery more promptly after compromise. At the same time, they maintain a low recovery frequency when compared to the periodic policy, which also has a high time-to-recovery. However, the recovery frequency of PPO increases significantly in Scenario $2$.

\tikzexternaldisable
\tikzexternalenable

\begin{table}
  \centering
  \scalebox{0.85}{
\begin{tabular}{llll} \toprule
\rowcolor{lightgray}
  {\textit{Method}} & {\textit{Offline compute}} & {\textit{Online compute}} & {\textit{Cost ($\downarrow$ better)}} \\ \midrule
      \multicolumn{4}{c}{\textbf{\textit{Scenario 1 results (\textsc{stationary system})}.}} \\  
 \rowcolor{lightgreen}  
  Ours, $m=0,\ell=1$ & $17\text{ min}$ & $0.93\text{ sec}$  & $34.61\text{ }(0.32)$\\
 \rowcolor{lightgreen}  
  Ours, $m=0,\ell=2$ & $17\text{ min}$ & $12.87\text{ sec}$ & $27.76\text{ }(0.38)$\\
 \rowcolor{lightgreen}  
  Ours, $m=10,\ell=1$ & $17\text{ min}$ & $5.45\text{ sec}$ & $34.35\text{ }(0.32)$\\
 \rowcolor{lightgreen}  
  Ours, $m=10,\ell=2$ & $17\text{ min}$ & $16.41\text{ sec}$ & $27.67\text{ }(0.38)$\\
 \rowcolor{lightgreen}  
  Ours, $m=20,\ell=1$ & $17\text{ min}$ & $7.45\text{ sec}$  & $23.42\text{ }(0.51)$\\
 \rowcolor{lightgreen}  
  Ours, $m=20,\ell=2$ & $17\text{ min}$ & $19.31\text{ sec}$ & $\bm{20.12}\text{ }(0.45)$\\
  Base policy [Eq.~\eqref{eq:approximation_1}] & $17\text{ min}$ & $0.01\text{ sec}$ & $106.00\text{ }(0.32)$\\
  PPO \cite[Alg. 1]{ppo} & $80\text{ min}$ & $0.01\text{ sec}$ & $\bm{19.71}\text{ }(9.32)$\\
  Periodic & $0\text{ min}$ & $0.01\text{ sec}$ & $168.09\text{ }(0.22)$\\
  \midrule
\multicolumn{4}{c}{\textbf{\textit{Scenario 2 results (\textsc{non-stationary system})}.}} \\    
 \rowcolor{lightgreen}  
  Ours, $m=0,\ell=1$ & $17\text{ min}$ & $0.93\text{ sec}$ & $38.83\text{ }(1.12)$\\
 \rowcolor{lightgreen}  
  Ours, $m=0,\ell=2$ & $17\text{ min}$ & $12.87\text{ sec}$ & $29.76\text{ }(0.53)$\\
 \rowcolor{lightgreen}  
  Ours, $m=10,\ell=1$ & $17\text{ min}$ & $5.45\text{ sec}$ & $31.31\text{ }(0.51)$\\
 \rowcolor{lightgreen}  
  Ours, $m=10,\ell=2$ & $17\text{ min}$ & $16.41\text{ sec}$ & $26.67\text{ }(0.53)$\\
 \rowcolor{lightgreen}  
  Ours, $m=20,\ell=1$ & $17\text{ min}$ & $7.45\text{ sec}$ & $23.61\text{ }(0.69)$\\
 \rowcolor{lightgreen}  
  Ours, $m=20,\ell=2$ & $17\text{ min}$ & $19.31\text{ sec}$ & $\bm{20.92}\text{ }(0.48)$\\
  Base policy [Eq.~\eqref{eq:approximation_1}] & $17\text{ min}$ & $0.01\text{ sec}$ & $114.48\text{ }(0.32)$\\
  PPO \cite[Alg. 1]{ppo} & $80\text{ min}$ & $0.01\text{ sec}$ & $49.71\text{ }(13.67)$\\
  Periodic & $0\text{ min}$ & $0.01\text{ sec}$ & $168.09\text{ }(0.22)$\\  
  \bottomrule\\
\end{tabular}}
\caption{Testbed results. Numbers in the last column indicate the mean and (standard deviation) from $5$ evaluations. The best results are in bold.}\label{tab:testbed_results}
\end{table}

\tikzexternaldisable
\mybox{\textbf{Takeaway from the testbed evaluation.}}{Black!5}{Black!2}{
When applied to the use case in \S\ref{sec:use_case}, our method yields security policies that recover more effectively from attacks than periodic recovery and adapt faster to system changes than those learned with PPO.
}
\tikzexternalenable

\tikzexternaldisable
\tikzexternalenable
\subsection{CAGE-2 Evaluation}
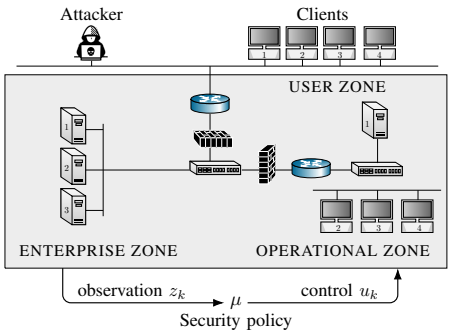
\begin{figure}[H]
  \centering
\scalebox{1.2}{
 \input{tikz/cage_network.tex}
  }
  \caption{The CAGE-2 benchmark problem \cite{cage_challenge_2_announcement}: compute a security policy $\mu$ to protect a system against an attacker while maintaining services for clients.}
  \label{fig:cage_2_topology}
\end{figure}

To compare our method with the state-of-the-art methods for computing security policies, we apply it to the CAGE-2 benchmark \cite{cage_challenge_2_announcement}. CAGE-2 involves a (simulated) networked system segmented into \textit{zones} with \textit{nodes} (servers and workstations) offering services to clients through a gateway, which is also accessible to an attacker; see Fig.~\ref{fig:cage_2_topology}. The system emits network statistics, which the security policy $\mu$ uses to prescribe security controls. These controls are applied to specific nodes of the system and can be grouped into four categories: intrusion analysis, decoy deployment, malware removal, or secure reset (which disrupts service). Each service disruption and node compromise incurs a predefined cost; the problem is to find a security policy that minimizes this cost. When formulated as a POMDP, CAGE-2 has $145$ controls, over $10^{47}$ states, and over $10^{25}$ observations \cite{hammar2024optimaldefenderstrategiescage2}.

\tikzexternaldisable
\tikzexternalenable

\vspace{2mm}

\noindent\textbf{\textit{Methods for comparison}.} Over $35$ methods have been evaluated against the CAGE-2 benchmark. We compare our method against the current state-of-the-art methods, namely: Cardiff \cite{vyas2023automated} and C-POMCP \cite[Alg. 1]{hammar2024optimaldefenderstrategiescage2}. We also compare it against four baseline methods: PPO \cite[Alg. 1]{ppo}, PPG \cite[Alg. 1]{ppg}, DQN \cite[Alg. 1]{mnih2015humanlevel}, and POMCP \cite[Alg. 1]{pomcp}.

\vspace{2mm}

\noindent\textbf{\textit{Evaluation scenarios}.} 
\begin{enumerate}
\item \setword{\textsc{stationary system}}{cage_scenario_1}: This is the standard CAGE-2 scenario where the system dynamics are stationary.
\item \setword{\textsc{non-stationary system}}{cage_scenario_2}: This scenario is divided into two time intervals. In the first interval ($[0, 20]$), the system behaves as in the stationary case. In the second interval, which starts at $k=20$, the decoys become ineffective (e.g., known to the attacker). As a consequence, the security policy must be adapted to remain effective.
\end{enumerate}

\vspace{2mm}

\noindent\textbf{\textit{Evaluation metrics}.} We compare the computed policies in terms of cost (calculated by the CAGE-2 simulator) and further compare the methods in terms of compute and adaptation times, with adaptation time calculated using Eq.~\eqref{eq:adaptation_metric}.

\vspace{2mm}

\noindent\textbf{\textit{Instantiation of our method}.} The size of the state space in CAGE-2 exceeds $10^{47}$, making it impractical to estimate beliefs over it. Therefore, we map each state into a \textit{feature state} with three components: \textsc{attacker-state}, \textsc{attacker-target}, and \textsc{decoy-state}. The first two components represent the attacker's location in the network and its target node. The last component is the configuration of the decoys. These feature states lead to a feature belief space $Q$ of dimension $427,500$, which we discretize with resolution $\rho=1$; cf.~Eq.~\eqref{eq:aggregate_belief_space}. We define the aggregation probability $\phi_{jy}$ to be $1$ if the feature state $y$ is consistent with the state $j$ and $0$ otherwise. Finally, we define the disaggregation probabilities $\{d_{yi}\mid i \in X\}$ to be uniform over states $i$ that are aggregated to the feature state $y$, i.e., states where $\phi_{iy}=1$.

\begin{table*}
  \centering
  \scalebox{0.93}{
\begin{tabular}{lcclllc}
\toprule
\rowcolor{lightgray}
\textit{Method} & \textit{Rollout $m$} & \textit{Offline/Online compute (min/s)} & \textit{State estimation} & \textit{Lookahead} $\ell$ & \textit{Base policy} $\mu$ & \textit{Cost ($\downarrow$ better)} \\
  \midrule
$\mu$ [Eq.~\eqref{eq:approximation_1}] & - & $8.5/0.01$ & Particle filter [Eq.~\eqref{eq:estimate_belief}] & - & - & $15.19\text{ }(0.82)$ \\
\midrule
PPO \cite[Alg. 1]{ppo} & - & $1000/0.01$ & Latest observation & - & - & $280\text{ }(114)$ \\
PPO \cite[Alg. 1]{ppo} & - & $1000/0.01$ & Particle filter [Eq.~\eqref{eq:estimate_belief}] & - & - & $119\text{ }(58)$ \\
\midrule
PPG \cite[Alg. 1]{ppg} & - & $1000/0.01$ & Latest observation & - & - & $338\text{ }(147)$ \\
PPG \cite[Alg. 1]{ppg} & - & $1000/0.01$ & Particle filter [Eq.~\eqref{eq:estimate_belief}] & - & - & $299\text{ }(108)$ \\
\midrule
DQN \cite[Alg. 1]{mnih2015humanlevel} & - & $1000/0.01$ & Latest observation & - & - & $479\text{ }(267)$ \\
  DQN \cite[Alg. 1]{mnih2015humanlevel} & - & $1000/0.01$ & Particle filter [Eq.~\eqref{eq:estimate_belief}] & - & - & $462\text{ }(244)$ \\
\midrule
\rowcolor{lightblue}
Cardiff \cite{vyas2023automated} & - & $300/0.01$ & Latest observation & - & - & $\bm{13.69}\text{ }(0.53)$ \\
Cardiff \cite{vyas2023automated} & - & $300/0.01$ & Particle filter [Eq.~\eqref{eq:estimate_belief}] & - & - & $\bm{13.31}\text{ }(0.87)$ \\
\midrule
POMCP \cite[Alg. 1]{pomcp} & - & $0/0.05$ & Particle filter [Eq.~\eqref{eq:estimate_belief}] & - & Random & $38.71\text{ }(2.0)$ \\
POMCP \cite[Alg. 1]{pomcp} & - & $0/0.1$ & Particle filter [Eq.~\eqref{eq:estimate_belief}] & - & Random & $38.02\text{ }(0.53)$ \\
POMCP \cite[Alg. 1]{pomcp} & - & $0/0.5$ & Particle filter [Eq.~\eqref{eq:estimate_belief}] & - & Random & $34.92\text{ }(0.96)$ \\
POMCP \cite[Alg. 1]{pomcp} & - & $0/1$ & Particle filter [Eq.~\eqref{eq:estimate_belief}] & - & Random & $34.50\text{ }(0.65)$ \\
POMCP \cite[Alg. 1]{pomcp} & - & $0/5$ & Particle filter [Eq.~\eqref{eq:estimate_belief}] & - & Random & $33.06\text{ }(0.21)$ \\
POMCP \cite[Alg. 1]{pomcp} & - & $0/15$ & Particle filter [Eq.~\eqref{eq:estimate_belief}] & - & Random & $30.88\text{ }(1.41)$ \\
POMCP \cite[Alg. 1]{pomcp} & - & $0/30$ & Particle filter [Eq.~\eqref{eq:estimate_belief}] & - & Random & $29.51\text{ }(2.00)$ \\
\midrule
  C-POMCP \cite[Alg. 1]{hammar2024optimaldefenderstrategiescage2} & - & $0/0.05$ & Particle filter [Eq.~\eqref{eq:estimate_belief}] & - & Random & $25.05\text{ }(3.02)$ \\
  C-POMCP \cite[Alg. 1]{hammar2024optimaldefenderstrategiescage2} & - & $0/0.1$ & Particle filter [Eq.~\eqref{eq:estimate_belief}] & - & Random & $21.28\text{ }(0.72)$ \\
  C-POMCP \cite[Alg. 1]{hammar2024optimaldefenderstrategiescage2} & - & $0/0.5$ & Particle filter [Eq.~\eqref{eq:estimate_belief}] & - & Random & $18.08\text{ }(1.32)$ \\
  C-POMCP \cite[Alg. 1]{hammar2024optimaldefenderstrategiescage2} & - & $0/1$ & Particle filter [Eq.~\eqref{eq:estimate_belief}] & - & Random & $17.42\text{ }(1.08)$ \\
  C-POMCP \cite[Alg. 1]{hammar2024optimaldefenderstrategiescage2} & - & $0/5$ & Particle filter [Eq.~\eqref{eq:estimate_belief}] & - & Random & $\bm{13.23}\text{ }(0.43)$ \\
  \rowcolor{lightblue}
  C-POMCP \cite[Alg. 1]{hammar2024optimaldefenderstrategiescage2} & - & $0/15$ & Particle filter [Eq.~\eqref{eq:estimate_belief}] & - & Random & $\bm{12.98}\text{ }(1.55)$ \\
  C-POMCP \cite[Alg. 1]{hammar2024optimaldefenderstrategiescage2} & - & $0/30$ & Particle filter [Eq.~\eqref{eq:estimate_belief}] & - & Random & $\bm{13.32}\text{ }(0.18)$ \\
  \midrule
 \rowcolor{lightgreen}
  Our method [Eq.~\eqref{eq:rollout}] & $0$ & $8.5/0.01$ & Particle filter [Eq.~\eqref{eq:estimate_belief}] & $1$ & $\mu$ [Eq.~\eqref{eq:approximation_1}] & $\bm{13.32}\text{ }(0.65)$ \\
\rowcolor{lightgreen}
  Our method [Eq.~\eqref{eq:rollout}] & $0$ & $8.5/0.95$ & Particle filter [Eq.~\eqref{eq:estimate_belief}] & $2$ & $\mu$ [Eq.~\eqref{eq:approximation_1}] & $\bm{13.24}\text{ }(0.57)$ \\
\rowcolor{lightgreen}
  Our method [Eq.~\eqref{eq:rollout}] & $10$ & $8.5/2.39$ & Particle filter [Eq.~\eqref{eq:estimate_belief}] & $1$ & $\mu$ [Eq.~\eqref{eq:approximation_1}] & $\bm{13.28}\text{ }(0.72)$ \\
\rowcolor{lightgreen}
  Our method [Eq.~\eqref{eq:rollout}] & $10$ & $8.5/8.29$ & Particle filter [Eq.~\eqref{eq:estimate_belief}] & $2$ & $\mu$ [Eq.~\eqref{eq:approximation_1}] & $\bm{13.23}\text{ }(0.62)$ \\
\rowcolor{lightgreen}
  Our method [Eq.~\eqref{eq:rollout}] & $20$ & $8.5/6.41$ & Particle filter [Eq.~\eqref{eq:estimate_belief}] & $1$ & $\mu$ [Eq.~\eqref{eq:approximation_1}] & $\bm{13.25}\text{ }(0.78)$ \\
\rowcolor{lightgreen}
  Our method [Eq.~\eqref{eq:rollout}] & $20$ & $8.5/14.80$ & Particle filter [Eq.~\eqref{eq:estimate_belief}] & $2$ & $\mu$ [Eq.~\eqref{eq:approximation_1}] & $\bm{13.23}\text{ }(0.57)$ \\
\bottomrule
\end{tabular}
}
\caption{Evaluation results on CAGE-2 (Scenario $1$). Rows relate to different methods; columns indicate performance metrics and configurations; green rows relate to our method (see Fig.~\ref{fig:framework}); blue rows relate to the previous state-of-the-art methods; results that are within the margin of statistical equivalence to the state-of-the-art are highlighted in bold ($\downarrow$ better); numbers in the last column indicate the mean and the (standard deviation) from $1000$ evaluations. The cost is calculated using CAGE-2's internal cost function (commit \textsc{9421c8e}) with the B-line attacker and $100$ time steps. (Since current deep RL methods do not use belief states, we report their performance both with and without the particle filter in Eq.~\eqref{eq:estimate_belief} to enable a fair comparison.)}\label{tab:cage2_results}
\end{table*}
\begin{table*}
  \centering
  \scalebox{0.93}{
\begin{tabular}{lcclllc}
\toprule
\rowcolor{lightgray}
\textit{Method} & \textit{Rollout $m$} & \textit{Offline/Online compute (min/s)} & \textit{State estimation} & \textit{Lookahead} $\ell$ & \textit{Base policy} $\mu$ & \textit{Cost ($\downarrow$ better)} \\
  \midrule
  $\mu$ [Eq.~\eqref{eq:approximation_1}] & - & $8.5/0.01$ & Particle filter [Eq.~\eqref{eq:estimate_belief}] & - & - & $61.72\text{ }(3.96)$ \\
  \midrule
PPO \cite[Alg. 1]{ppo} & - & $1000/0.01$ & Latest observation & - & - & $341\text{ }(133)$ \\
  PPO \cite[Alg. 1]{ppo} & - & $1000/0.01$ & Particle filter [Eq.~\eqref{eq:estimate_belief}] & - & - & $326\text{ }(116)$ \\
  \midrule
PPG \cite[Alg. 1]{ppg} & - & $1000/0.01$ & Latest observation & - & - & $328\text{ }(178)$ \\
PPG \cite[Alg. 1]{ppg} & - & $1000/0.01$ & Particle filter [Eq.~\eqref{eq:estimate_belief}] & - & - & $312\text{ }(163)$ \\
  \midrule
DQN \cite[Alg. 1]{mnih2015humanlevel} & - & $1000/0.01$ & Latest observation & - & - & $516\text{ }(291)$ \\
  DQN \cite[Alg. 1]{mnih2015humanlevel} & - & $1000/0.01$ & Particle filter [Eq.~\eqref{eq:estimate_belief}] & - & - & $492\text{ }(204)$ \\
  \midrule
\rowcolor{lightblue}
  Cardiff \cite{vyas2023automated} & - & $300/0.01$ & Latest observation & - & - & $57.45\text{ }(2.44)$ \\
Cardiff \cite{vyas2023automated} & - & $300/0.01$ & Particle filter [Eq.~\eqref{eq:estimate_belief}] & - & - & $56.45\text{ }(2.81)$ \\
\midrule
POMCP \cite[Alg. 1]{pomcp} & - & $0/0.05$ & Particle filter [Eq.~\eqref{eq:estimate_belief}] & - & Random & $66.80\text{ }(4.80)$ \\
POMCP \cite[Alg. 1]{pomcp} & - & $0/0.1$ & Particle filter [Eq.~\eqref{eq:estimate_belief}] & - & Random & $57.12\text{ }(4.62)$ \\
POMCP \cite[Alg. 1]{pomcp} & - & $0/0.5$ & Particle filter [Eq.~\eqref{eq:estimate_belief}] & - & Random & $55.43\text{ }(3.99)$ \\
POMCP \cite[Alg. 1]{pomcp} & - & $0/1$ & Particle filter [Eq.~\eqref{eq:estimate_belief}] & - & Random & $53.71\text{ }(3.84)$ \\
POMCP \cite[Alg. 1]{pomcp} & - & $0/5$ & Particle filter [Eq.~\eqref{eq:estimate_belief}] & - & Random & $52.23\text{ }(3.81)$ \\
POMCP \cite[Alg. 1]{pomcp} & - & $0/15$ & Particle filter [Eq.~\eqref{eq:estimate_belief}] & - & Random & $53.08\text{ }(3.78)$ \\
POMCP \cite[Alg. 1]{pomcp} & - & $0/30$ & Particle filter [Eq.~\eqref{eq:estimate_belief}] & - & Random & $53.18\text{ }(3.42)$ \\
\midrule
  C-POMCP \cite[Alg. 1]{hammar2024optimaldefenderstrategiescage2} & - & $0/0.05$ & Particle filter [Eq.~\eqref{eq:estimate_belief}] & - & Random & $61.05\text{ }(5.84)$ \\
  C-POMCP \cite[Alg. 1]{hammar2024optimaldefenderstrategiescage2} & - & $0/0.1$ & Particle filter [Eq.~\eqref{eq:estimate_belief}] & - & Random & $57.46\text{ }(4.53)$ \\
  C-POMCP \cite[Alg. 1]{hammar2024optimaldefenderstrategiescage2} & - & $0/0.5$ & Particle filter [Eq.~\eqref{eq:estimate_belief}] & - & Random & $51.18\text{ }(4.37)$ \\
  C-POMCP \cite[Alg. 1]{hammar2024optimaldefenderstrategiescage2} & - & $0/1$ & Particle filter [Eq.~\eqref{eq:estimate_belief}] & - & Random & $44.52\text{ }(3.75)$ \\
  C-POMCP \cite[Alg. 1]{hammar2024optimaldefenderstrategiescage2} & - & $0/5$ & Particle filter [Eq.~\eqref{eq:estimate_belief}] & - & Random & $41.61\text{ }(3.34)$ \\
  C-POMCP \cite[Alg. 1]{hammar2024optimaldefenderstrategiescage2} & - & $0/15$ & Particle filter [Eq.~\eqref{eq:estimate_belief}] & - & Random & $40.84\text{ }(2.92)$ \\
  \rowcolor{lightblue}
  C-POMCP \cite[Alg. 1]{hammar2024optimaldefenderstrategiescage2} & - & $0/30$ & Particle filter [Eq.~\eqref{eq:estimate_belief}] & - & Random & $\bm{38.71}\text{ }(2.38)$ \\
  \midrule
\rowcolor{lightgreen}
  Our method [Eq.~\eqref{eq:rollout}] & $0$ & $8.5/0.01$ & Particle filter [Eq.~\eqref{eq:estimate_belief}] & $1$ & $\mu$ [Eq.~\eqref{eq:approximation_1}] & $58.23\text{ }(1.67)$ \\
\rowcolor{lightgreen}
  Our method [Eq.~\eqref{eq:rollout}] & $0$ & $8.5/0.95$ & Particle filter [Eq.~\eqref{eq:estimate_belief}] & $2$ & $\mu$ [Eq.~\eqref{eq:approximation_1}] & $51.87\text{ }(1.42)$ \\
\rowcolor{lightgreen}
  Our method [Eq.~\eqref{eq:rollout}] & $10$ & $8.5/2.39$ & Particle filter [Eq.~\eqref{eq:estimate_belief}] & $1$ & $\mu$ [Eq.~\eqref{eq:approximation_1}] & $44.38\text{ }(1.76)$ \\
\rowcolor{lightgreen}
  Our method [Eq.~\eqref{eq:rollout}] & $10$ & $8.5/8.29$ & Particle filter [Eq.~\eqref{eq:estimate_belief}] & $2$ & $\mu$ [Eq.~\eqref{eq:approximation_1}] & $\bm{38.81}\text{ }(1.68)$ \\
\rowcolor{lightgreen}
  Our method [Eq.~\eqref{eq:rollout}] & $20$ & $8.5/6.41$ & Particle filter [Eq.~\eqref{eq:estimate_belief}] & $1$ & $\mu$ [Eq.~\eqref{eq:approximation_1}] & $\bm{39.05}\text{ }(2.04)$ \\
\rowcolor{lightgreen}
  Our method [Eq.~\eqref{eq:rollout}] & $20$ & $8.5/14.80$ & Particle filter [Eq.~\eqref{eq:estimate_belief}] & $2$ & $\mu$ [Eq.~\eqref{eq:approximation_1}] & $\bm{37.89}\text{ }(1.54)$ \\
\bottomrule
\end{tabular}
}
\caption{Evaluation results on CAGE-2 (Scenario 2). Rows relate to different methods; columns indicate performance metrics and configurations; green rows relate to our method (see Fig.~\ref{fig:framework}); blue rows relate to the previous state-of-the-art methods; the best results are highlighted in bold ($\downarrow$ better); numbers in the last column indicate the mean and the (standard deviation) from $1000$ evaluations. The cost is calculated using CAGE-2's internal cost function (commit \textsc{9421c8e}) with the B-line attacker and $100$ time steps.}\label{tab:cage2_results_2}
\end{table*}

\vspace{2mm}

\noindent\textbf{\textit{Scenario $1$ results (\textsc{stationary system})}.}
The results are presented in Table~\ref{tab:cage2_results}. We observe that our method achieves performance on par with the state-of-the-art methods in terms of cost. We also find that our method requires less offline compute time than methods based on deep RL (e.g., PPO, PPG, and DQN); see the third column of Table~\ref{tab:cage2_results}. Further, it requires less online compute time than POMCP \cite{pomcp} and C-POMCP \cite{hammar2024optimaldefenderstrategiescage2}. The performance of our method improves only slightly when increasing the lookahead horizon $\ell$ from $1$ to $2$ and when increasing the rollout horizon $m$ in Eq.~\eqref{eq:rollout} from $0$ to $20$; see the second and fifth columns in Table~\ref{tab:cage2_results}. This suggests that the performance achieved with $\ell=1$ and $m=0$ may be near optimal. Finally, we note that the results of PPO, DQN, and PPG exhibit high variability. This variability may be due to their tendency to converge to different local optima depending on the random seed.

\vspace{2mm}

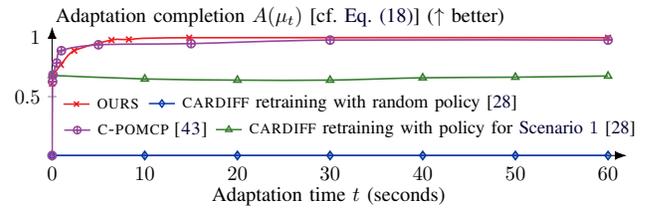
\begin{figure}
  \centering
\scalebox{0.77}{
    \input{tikz/adaption_eval_2.tex}
  }
  \caption{Policy adaptation time after a system change in CAGE-2 (Scenario $2$). The metric $A(\mu_t)$ is calculated according to Eq.~\eqref{eq:adaptation_metric} with $J_1(b_0)=37.89$, where $\mu_t$ is the policy after $t$ seconds of adaptation.}
  \label{fig:adaption_eval2}
\end{figure}

\noindent\textbf{\textit{Scenario $2$ results (\textsc{non-stationary system})}.} The results are presented in Table~\ref{tab:cage2_results_2}. While all methods incur an overall higher cost in this scenario due to the non-stationarity of the system, we find that our method adapts effectively to system changes and obtains the best results alongside C-POMCP \cite{hammar2024optimaldefenderstrategiescage2}. However, our method is more general than C-POMCP (which is customized for CAGE-2) and requires less online compute time. The policy adaptation of our method takes around $15$ seconds with our computing hardware, whereas the methods based on deep RL require minutes or hours of optimization to adapt; see Fig.~\ref{fig:adaption_eval2}. Unlike the first scenario, increasing the rollout and lookahead horizons ($m$ and $\ell$) in Eq.~\eqref{eq:rollout} substantially improves performance at the expense of increased computational effort; see the third column in Table~\ref{tab:cage2_results_2}.

\tikzexternaldisable
\myboxtwo{\textbf{Takeaway from the CAGE-2 evaluation.}}{Black!5}{Black!2}{
Our method achieves performance on par with the state-of-the-art on the CAGE-2 benchmark while being more computationally efficient and adaptive.
}
\tikzexternalenable

\tikzexternaldisable
\tikzexternalenable

\vspace{2cm}

\section{Discussion of the Evaluation Results}\label{sec:discussion}
The experimental evaluation highlights a key limitation of methods proposed in the research literature for computing security policies: they lack adaptability. In both the testbed and CAGE-2 evaluations, we find that most existing methods require lengthy retraining to adapt the policy to changes. This slow adaptation makes them unsuitable for modern networked systems, where configurations and workloads change frequently. Our method addresses this limitation by reducing the adaptation time, as shown in Fig.~\ref{fig:adaption_eval} and Fig.~\ref{fig:adaption_eval2}.

Another notable finding from our experiments is that while some of the current methods (e.g., deep RL methods) can yield effective security policies for stationary systems, they suffer from high variance and instability. For instance, in the testbed evaluation, we found the variance of PPO to be $10$ times higher than that of our method. This variability can be explained by the tendency of PPO to converge to different local optima depending on the random seed. Indeed, in our experiments, we often needed to restart PPO several times with different random seeds to discover an effective policy. Such sensitivity to the random seed poses a potential operational concern, as it may lead to inconsistent policy performance across deployments. By contrast, the offline computation of our method converges reliably in all cases and provides performance guarantees; see Props.~\ref{prop:aggregation_bound}-\ref{thm:consistent_1}.

The method that comes closest in performance to our method on the CAGE-2 benchmark is C-POMCP \cite[Alg. 1]{hammar2024optimaldefenderstrategiescage2}. However, this method is tailored for CAGE-2 and not generalizable. For example, it cannot be directly applied to our testbed. Furthermore, our method requires less online computation and offers stronger theoretical guarantees.

\vspace{2mm}

\noindent\textbf{\textit{Limitations}.} Our method has three main limitations. First, it requires a model or simulator of the target system to adapt the policy via rollout. As a result, its effectiveness depends on the accuracy with which this model captures the system dynamics. If the model is significantly misspecified, the quality of the adapted policy may deteriorate, as discussed in \S\ref{sec:rollout}. Second, the performance of our method depends on the design of the feature space used for aggregation. Although effective features can often be selected based on engineering intuition, poor feature design can lead to degraded policy performance. Third, our method assumes a static attacker model, and extending it to account for a dynamic attacker that adapts its strategy in response to the security policy remains an open challenge.
\section{Conclusion}\label{sec:conclusion}
Frequent adaptations of security policies are needed to keep pace with evolving security threats in networked systems. While RL is a promising approach to automate these adaptations, most methods proposed in the research literature lack performance guarantees and adapt slowly. Moreover, they have not been validated outside of simulation. This paper addresses these limitations by presenting and validating a scalable method for computing adaptive security policies with performance guarantees. It assumes a model or simulator of the target system and is based on three core ideas: (1) using particle filtering to estimate a belief about the system’s security state; (2) aggregating beliefs to enable scalable offline policy computation; and (3) using rollout for online policy adaptation.

We show both theoretically and experimentally that our method provides advantages over other methods proposed in the research literature. Unlike existing methods that lack performance guarantees, we derive a bound on the approximation error of our aggregation scheme; see Props.~\ref{prop:aggregation_bound} and \ref{thm:consistent_1}. Moreover, we establish conditions under which our rollout method efficiently adapts policies to system changes; see Prop.~\ref{prop:improvement}. Simulations show that our method obtains state-of-the-art performance on the CAGE-2 benchmark; see \tablesref{tab:cage2_results} and \ref{tab:cage2_results_2}. Additionally, testbed experiments demonstrate its practicality; see Table~\ref{tab:testbed_results}. While further testing remains to be done, these results indicate that our method provides a step towards reliable and automated adaptation of security policies in networked systems.

\vspace{2mm}

\noindent\textbf{\textit{Future work}.} Several directions for future work merit investigation. First, a formal and empirical analysis of our method's sensitivity to model misspecification would be valuable. Second, the identification procedure in \S\ref{sec:testbed_eval} could be extended into a more principled approach for estimating and updating model parameters from data. Third, the feature design in our method currently relies on domain knowledge; developing automated methods for feature extraction is a promising direction for future work. Fourth, our current POMDP formulation assumes a stochastic but non-adversarial system. Extending our approach to account for adversaries that adapt to the security policy is a natural direction for future work and would require a game-theoretic formulation.

\section{Acknowledgment}
This research is supported by the Swedish Research Council under contract 2024-06436.

\appendices

\section{Proof of Prop.~\ref{thm:consistent_1}}\label{app:proof_consistent}
It can be shown that the optimal cost function $J^{\star}:B \mapsto \Re$ is uniformly continuous; see e.g., \cite[Prop. 2.1]{10.5555/1195372}. Fix an arbitrary scalar $\gamma>0$. By uniform continuity, there exists a scalar $\delta > 0$ such that
\begin{align}\label{eq:uniform}
\norm{b-b'} < \delta \implies |J^{\star}(b)-J^{\star}(b')| < \gamma
\end{align}
for all $b,b^{\prime}\in B$, where $\norm{\cdot}$ denotes the maximum norm.

Since $X=\mathcal{F}$ by assumption, we have that $\Tilde{Q}$ is a finite subset of $B$. Therefore, the discretization in Eq.~\eqref{eq:aggregate_belief_space} partitions $B$ into grid cells $S_{\tilde q}$ with resolution $\rho \geq 1$; cf.~Eq.~\eqref{eq:partitioning}. Further, Eq.~\eqref{eq:belief_transition_1} implies that if $b \in S_{\tilde q}$, then 
\begin{align*}
\norm{b-\tilde{q}}=\min_{\tilde{q}^{\prime} \in \Tilde{Q}}\norm{b-\tilde{q}^{\prime}}.
\end{align*}
Because each belief coordinate $b(i)$ lies in $[0,1]$ and each representative feature belief coordinate $\tilde{q}(i)$ equals $\frac{\beta_i}{\rho}$ for some $\beta_i\in \{0,\dots,\rho\}$ [cf.~Eq.~\eqref{eq:aggregate_belief_space}], we have
\begin{align*}
\max_{b,b'\in S_{\tilde q}}\norm{b-b'} \leq \frac{2n}{\rho}, \qquad\text{for every } \tilde{q} \in \Tilde{Q}.
\end{align*}
Choose any $\rho$ such that $\frac{2n}{\rho} < \delta$. By Eq.~\eqref{eq:uniform}, we have
\begin{align*}
|J^{\star}(b)-J^{\star}(b^{\prime})| < \gamma,  \quad\text{for all } b,b'\in S_{\tilde q},\; \tilde q \in \Tilde{Q}.
\end{align*}  
Because $\gamma>0$ is arbitrary and there exists a large enough $\rho$ such that $\frac{n}{\rho} < \delta$ for any $\delta > 0$, we have
\begin{align*}
\lim_{\rho\to\infty}\max_{\tilde q \in \Tilde{Q}}\max_{b,b'\in S_{\tilde q}} |J^{\star}(b)-J^{\star}(b')| =0.
\end{align*}
Hence the constant $\epsilon$ in Prop.~\ref{prop:aggregation_bound} diminishes as $\rho \rightarrow \infty$. Invoking the error bound in Prop.~\ref{prop:aggregation_bound} completes the proof. \qed

\section{Experimental Setup}\label{appendix:hyperparameters}
All computations are performed on an M2 Ultra processor. The attacker actions in our testbed are listed in Table~\ref{tab:attacker_actions}. The hyperparameters are listed in Table~\ref{tab:hyperparams}. Notation is explained in Table~\ref{tab:notation}. We use the implementation of Cardiff described in \cite{vyas2023automated} and the implementation of C-POMCP described in \cite{hammar2024optimaldefenderstrategiescage2}. For PPO and DQN, we use the \textsc{stable-baselines} implementations \cite{stable-baselines3}. For PPG, we use the \textsc{clean-rl} implementation \cite{journals/jmlr/HuangDYBCMA22}. For POMCP, we use our implementation \cite{rollout_software}. We set the hyperparameters for these methods to be the same as those used in \cite{hammar2024optimaldefenderstrategiescage2}. Unless stated otherwise, we run PPO and PPG with a vector of the sample states of the particle filter as input. We identify the dynamics of the aggregate MDP in Fig.~\ref{fig:aggregation_5} through simulations of the original POMDP. We solve the aggregate MDP using value iteration (VI) with a convergence tolerance of $\epsilon = 10^{-6}$. Hyperparameters and feature selections are chosen based on domain knowledge and limited experimentation (1–3 trials), without extensive tuning. Reported results for all methods are averaged over multiple independent runs with different random seeds; unless otherwise specified, each method is run $5$ times.

\begin{table}
  \centering
    \vspace{0.24cm}  
  \scalebox{0.84}{
    \begin{tabular}{ll} \toprule
\rowcolor{lightgray}
    {\textit{Parameter(s)}} & {\textit{Values}} \\ \midrule
    Convergence threshold of VI & $0.1$.\\
    $L$ [Eq.~\eqref{eq:rollout_approximation}] & $20$. \\
      $M$, $\alpha$ & $50$, $0.99$ . \\
      Figure~\ref{fig:obs_dist} & $\mathrm{BetaBin}(7,1,0.7)$ when $i^l=1$.\\
    & $\mathrm{BetaBin}(7,0.7,3)$ when $i^l=0$.\\            
    \bottomrule\\
  \end{tabular}}
  \caption{Hyperparameters.}\label{tab:hyperparams}
\end{table}

\begin{table}
  \centering
  \scalebox{0.75}{
    \begin{tabular}{ll} \toprule
\rowcolor{lightgray}
      {\textit{Notation(s)}} & {\textit{Description}} \\ \midrule
      $X,U,Z,B, \mathcal{F}$ & State, control, observation, belief, and feature spaces; cf.~\S\ref{sec:preliminaries:model} and \S\ref{sec:belief_aggregation}. \\
      $Q, \Tilde{Q}$ & Feature belief and representative feature belief spaces; cf.~Eq.~\eqref{eq:aggregate_belief_space}. \\
      $\alpha, \rho$ & Discount factor and discretization resolution; cf.~Eq.~\eqref{eq:aggregate_belief_space}. \\
      $q, \tilde{q}$ & Feature belief and representative feature belief; cf.~Eq.~\eqref{eq:aggregate_belief_space}. \\ 
      $(i,j), (x,y), n$ & States, feature states, number of states; cf.~\S\ref{sec:preliminaries:model} and \S\ref{sec:belief_aggregation}. \\
      $b,u,z$ & Belief state, control, and observation; cf.~\S\ref{sec:preliminaries:model}.\\
      $b_k,u_k,z_k,i_k$ & Belief state, control, observation, and state at time $k$; cf.~\S\ref{sec:preliminaries:model}. \\      
      $F, g, p$ & Belief estimator, stage cost, and observation distribution; cf.~\S\ref{sec:preliminaries:model}. \\
      $p_{ij}(u)$ & Transition probabilities under control $u$; cf.~\S\ref{sec:preliminaries:model}.\\
      $\hat{b}$ & Belief state estimated through the particle filter; cf.~Eq.~\eqref{eq:estimate_belief}. \\
      $d_{yi}, \phi_{jy}$ & Disaggregation probabilities and aggregation probabilities; cf.~\S\ref{sec:belief_aggregation}. \\
      $\Phi$ & Aggregation mapping $\Phi: B \mapsto \Tilde{Q}$; cf.~Eq.~\eqref{eq:aggregation_mapping}. \\
      $\hat{g}, \hat{p}$ & Expected stage cost and observation probability given $b$; cf. \eqqsref{eq:hat_g}--\eqqqref{eq:hat_p}.\\      
      $J^{\star}, \mu^{\star}$ & Optimal cost function and optimal policy; cf.~Eq.~\eqref{eq:optimal_cost}. \\
      $J_{\mu}$ & Cost function of policy $\mu$; cf.~Eq.~\eqref{eq:optimal_cost}. \\      
      $\Tilde{J}$ & Cost function approximation; cf.~Eq.~\eqref{eq:approximation_1}. \\
      $r^{\star}, \pi^{\star}$ & Optimal cost function and policy in the aggregate MDP; cf.~Eq.~\eqref{eq:approximation_1}. \\
      $\mu, \tilde{\mu}$ & Base policy [cf.~Eq.~\eqref{eq:approximation_1}] and rollout policy [cf.~Eq.~\eqref{eq:rollout}]. \\
      $\ell,m,L$ & Lookahead and rollout horizons, and number of simulations; cf.~Eq.~\eqref{eq:rollout}. \\      
      $\norm{\cdot}, \Re, E\{\cdot\}$ & The maximum norm, the real numbers, and the expectation operator.\\
      $J_{\mu}, \Tilde{J}_{\mu}$ & Cost function of policy $\mu$ and estimated cost function of $\mu$; cf.~Eq.~\eqref{eq:rollout_approximation}.\\
      $S_{\tilde{q}}$ & Belief space partition of the representative feature belief $\tilde{q}$; cf.~Eq.~\eqref{eq:partitioning}. \\
      $\epsilon$ & Maximum variation of $J^{\star}$ within a partition $S_{\tilde{q}}$; cf.~Prop.~\ref{prop:aggregation_bound}.\\
      $K$ & Number of service replicas in the running example; cf.~\S\ref{sec:use_case}.\\
      $A(\mu)$ & The adaptation-completion metric of policy $\mu$; cf.~Eq.~\eqref{eq:adaptation_metric}.\\                  
    \bottomrule\\
  \end{tabular}}
  \caption{Notation.}\label{tab:notation}
\end{table}

\begin{table}
  \centering
  \scalebox{0.9}{  
\begin{tabular}{lll} \toprule
\rowcolor{lightgray}
  {\textit{Type}} & {\textit{Actions}} & {\textit{MITRE ATT\&CK} \textit{technique}} \\ \midrule
  Reconnaissance  & TCP SYN scan, UDP scan & T1046 service scanning\\
                  & TCP XMAS scan & T1046 service scanning \\
                  & Vulscan & T1595 active scanning \\
                  & ping-scan & T1018 system discovery\\\midrule
  Brute-force & Telnet, SSH & T1110 brute force\\
                  & FTP, Cassandra & T1110 brute force\\
                  &  IRC, MongoDB, MySQL & T1110 brute force\\
                  & SMTP, PostgreSQL & T1110 brute force\\\midrule
  Exploit & CVE-2017-7494 & T1210 service exploitation\\
                  &CVE-2015-3306 & T1210 service exploitation\\
                  & CVE-2010-0426 & T1068 privilege escalation\\
                  & CVE-2015-5602 & T1068 privilege escalation\\
                  & CVE-2015-1427 & T1210 service exploitation\\
                  & CVE-2014-6271 & T1210 service exploitation\\
                  & CVE-2016-10033 & T1210 service exploitation\\
                  & SQLJ injection (CWE-89) & T1210 service exploitation \\
  \bottomrule\\
\end{tabular}}
\caption{Attacker actions in our testbed; actions are identified by the corresponding CVEs \cite{cve} and CWEs \cite{cwe}; the actions are also linked to the corresponding attack techniques in \textsc{mitre att\&ck} \cite{strom2018mitre}.}\label{tab:attacker_actions}
\end{table}

\section{Testbed Setup}\label{app:testbed}
The network topology of the networked system that we run on our testbed is shown in Fig.~\ref{fig:service_chain} and the configuration is listed in Table~\ref{tab:servers}. Hosts and switches are emulated with Docker containers. Resource allocation to containers is enforced using cGroups. Network connectivity between containers is emulated with virtual links implemented by Linux bridges and network namespaces, which create logical copies of the physical host's network stack. Network conditions of virtual links are created using the NetEm module in the Linux kernel. We emulate connections between servers with full-duplex lossless connections of $1$ Gbit/s capacity in both directions. Similarly, we emulate connections between servers and clients with full-duplex connections of $100$ Mbit/s capacity and $0.1\%$ packet loss with random bursts of $1\%$ packet loss. These numbers are based on measurements on enterprise networks \cite{Paxson97end-to-endinternet}.

The client population is emulated through processes that access services on emulated hosts. Client arrivals are controlled by a Poisson process with exponentially distributed service times. The sequence of service invocations is selected uniformly at random. Similarly, the attacker is emulated by programs that select actions from the list in Table~\ref{tab:attacker_actions}. The source code of our emulation platform is available at \cite{rollout_software}.

\bibliographystyle{IEEEtran}
\bibliography{references, url}

\end{document}

%% file: config.tex
\usepackage[noadjust]{cite}
\usepackage[bookmarks=false]{hyperref}
\hypersetup{
    colorlinks,
    linkcolor={blue!20!black},
    citecolor={blue!20!black},
    urlcolor={blue!50!black}
}
\usepackage[font=footnotesize]{subcaption}
\usepackage[font=footnotesize]{caption}
\usepackage{caption}
\usepackage[utf8]{inputenc}
\usepackage{graphicx,url}

\usepackage{booktabs}
\usepackage{diagbox}
\graphicspath{ {./images/} }
\usepackage{multirow}
\usepackage{hhline}
\usepackage{xspace}
\usepackage{graphicx}
\usepackage{amssymb,amsmath,mathtools}
\usepackage{amsthm}
\newtheorem{proposition}{Proposition} %

\usepackage{bm}                 %
\usepackage{pifont}

\theoremstyle{plain}

\usepackage{color,soul}
\usepackage[dvipsnames]{xcolor}
\usepackage{bbm}
\usepackage{stmaryrd}
\usepackage{textcomp}
\usepackage{lipsum,lineno}
\usepackage{float}
\makeatletter
\newif\if@restonecol
\makeatother
\usepackage{amsmath}
\usepackage{kbordermatrix}
\usepackage{mathrsfs}
\usepackage[shortlabels]{enumitem}
\newcommand\numeq[1]%
{\stackrel{\scriptscriptstyle(\mkern-1.5mu#1\mkern-1.5mu)}{=}}
\newcommand\numeqq[1]%
{\stackrel{\scriptscriptstyle(\mkern-1.5mu#1\mkern-1.5mu)}{\triangleq}}
\newcommand\numleq[1]%
{\stackrel{\scriptscriptstyle(\mkern-1.5mu#1\mkern-1.5mu)}{\leq}}
\newcommand\numgeq[1]%
{\stackrel{\scriptscriptstyle(\mkern-1.5mu#1\mkern-1.5mu)}{\geq}}
\newcommand\numimp[1]%
{\stackrel{\scriptscriptstyle(\mkern-1.5mu#1\mkern-1.5mu)}{\implies}}
\newcommand\norm[1]{\lVert#1\rVert}

\usepackage[titlenumbered,ruled,linesnumbered]{algorithm2e}
\SetAlCapNameFnt{\footnotesize}
\SetAlCapFnt{\footnotesize}

\let\oldnl\nl
\newcommand{\nonl}{\renewcommand{\nl}{\let\nl\oldnl}}
\usepackage{algpseudocode}

\usepackage{fancyhdr}
\usepackage{tabularx}
\usepackage{dsfont}
\usepackage{listings}
\usepackage[most]{tcolorbox}
\lstdefinestyle{mystyle}{
    backgroundcolor=\color{backcolour},
    commentstyle=\color{codegreen},
    keywordstyle=\color{magenta},
    numberstyle=\tiny\color{codegray},
    stringstyle=\color{codepurple},
    basicstyle=\ttfamily\footnotesize,
    breakatwhitespace=false,
    breaklines=true,
    captionpos=b,
    keepspaces=true,
    showspaces=false,
    showstringspaces=false,
    showtabs=false,
    tabsize=2,
    xleftmargin=50pt,
    xrightmargin=50pt
  }
\usepackage{tikz,pgfplots}
\usepackage{pgfplots}
\usepackage{pgfplotstable}
\definecolor{gray2}{HTML}{ededed}
\definecolor{gray3}{HTML}{F5F5F5}
\definecolor{RoyalAzure}{rgb}{0.0, 0.22, 0.66}
\definecolor{lightgray}{gray}{0.9}
\definecolor{lightgray}{gray}{0.9}
\definecolor{lightgreen}{rgb}{0.88, 1, 0.88}
\definecolor{lightred}{rgb}{1, 0.88, 0.88}
\definecolor{lightblue}{rgb}{0.88, 0.94, 1}
\usepackage{colortbl}
\usetikzlibrary{pgfplots.groupplots}
\usepgfplotslibrary{colorbrewer}
\pgfplotsset{compat = 1.15, cycle list/Set1-8}
\usetikzlibrary{pgfplots.statistics, pgfplots.colorbrewer}
\usetikzlibrary{pgfplots.groupplots}
\usetikzlibrary{shapes.geometric,backgrounds,patterns, trees}
\usetikzlibrary{3d,decorations.text,shapes.arrows,positioning,fit,backgrounds}
\usetikzlibrary{positioning, decorations.pathmorphing, shapes}
\usetikzlibrary{decorations.pathreplacing}
\usetikzlibrary{shapes.geometric,backgrounds,patterns, trees}
\usetikzlibrary{spy}
\usetikzlibrary{arrows.meta,
                bending,
                intersections,
                quotes,
                shapes.geometric}
              \usetikzlibrary{automata, positioning}
              \usepgfplotslibrary{fillbetween}
\usetikzlibrary{shapes,arrows}
\usetikzlibrary{arrows.meta}
\usetikzlibrary{positioning}
\tikzset{set/.style={draw,circle,inner sep=0pt,align=center}}
\usetikzlibrary{automata, positioning}
  \usetikzlibrary{shapes,shadows}
  \tikzstyle{abstractbox} = [draw=black, fill=white, rectangle,
  inner sep=10pt, style=rounded corners, drop shadow={fill=black,
  opacity=1}]
\tikzstyle{abstracttitle} =[fill=white]
\usetikzlibrary{calc,positioning,shapes.geometric}
\usetikzlibrary{arrows.meta,arrows}

\DeclareMathOperator*{\argmin}{arg\,min}

\usetikzlibrary{matrix}
\tikzstyle{cblue}=[circle, draw, thin,fill=cyan!20, scale=0.8]
\tikzstyle{qgre}=[rectangle, draw, thin,fill=green!20, scale=0.8]
\tikzstyle{rpath}=[ultra thick, red, opacity=0.4]
\tikzstyle{legend_isps}=[rectangle, rounded corners, thin,
                       fill=gray!20, text=blue, draw]

\tikzstyle{legend_overlay}=[rectangle, rounded corners, thin,
                           top color= white,bottom color=green!25,
                           minimum width=2.5cm, minimum height=0.8cm,
                           pinegreen]
\tikzstyle{legend_phytop}=[rectangle, rounded corners, thin,
                          top color= white,bottom color=cyan!25,
                          minimum width=2.5cm, minimum height=0.8cm,
                          royalblue]
\tikzstyle{legend_general}=[rectangle, rounded corners, thin,
                          top color= white,bottom color=lavander!25,
                          minimum width=2.5cm, minimum height=0.8cm,
                          violet]
                          \usetikzlibrary{matrix}

\colorlet{myRed}{red!20}
\tikzset{
  rows/.style 2 args={/utils/temp/.style={row ##1/.append style={nodes={#2}}},
    /utils/temp/.list={#1}},
  columns/.style 2 args={/utils/temp/.style={column ##1/.append style={nodes={#2}}},
    /utils/temp/.list={#1}}}
\usetikzlibrary{backgrounds,calc,shadings,shapes.arrows,shapes.symbols,shadows}
\definecolor{switch}{HTML}{006996}

\makeatletter
\pgfkeys{/pgf/.cd,
  parallelepiped offset x/.initial=2mm,
  parallelepiped offset y/.initial=2mm
}
\pgfdeclareshape{parallelepiped}
{
  \inheritsavedanchors[from=rectangle] 
  \inheritanchorborder[from=rectangle]
  \inheritanchor[from=rectangle]{north}
  \inheritanchor[from=rectangle]{north west}
  \inheritanchor[from=rectangle]{north east}
  \inheritanchor[from=rectangle]{center}
  \inheritanchor[from=rectangle]{west}
  \inheritanchor[from=rectangle]{east}
  \inheritanchor[from=rectangle]{mid}
  \inheritanchor[from=rectangle]{mid west}
  \inheritanchor[from=rectangle]{mid east}
  \inheritanchor[from=rectangle]{base}
  \inheritanchor[from=rectangle]{base west}
  \inheritanchor[from=rectangle]{base east}
  \inheritanchor[from=rectangle]{south}
  \inheritanchor[from=rectangle]{south west}
  \inheritanchor[from=rectangle]{south east}
  \backgroundpath{
    \southwest \pgf@xa=\pgf@x \pgf@ya=\pgf@y
    \northeast \pgf@xb=\pgf@x \pgf@yb=\pgf@y
    \pgfmathsetlength\pgfutil@tempdima{\pgfkeysvalueof{/pgf/parallelepiped
      offset x}}
    \pgfmathsetlength\pgfutil@tempdimb{\pgfkeysvalueof{/pgf/parallelepiped
      offset y}}
    \def\ppd@offset{\pgfpoint{\pgfutil@tempdima}{\pgfutil@tempdimb}}
    \pgfpathmoveto{\pgfqpoint{\pgf@xa}{\pgf@ya}}
    \pgfpathlineto{\pgfqpoint{\pgf@xb}{\pgf@ya}}
    \pgfpathlineto{\pgfqpoint{\pgf@xb}{\pgf@yb}}
    \pgfpathlineto{\pgfqpoint{\pgf@xa}{\pgf@yb}}
    \pgfpathclose
    \pgfpathmoveto{\pgfqpoint{\pgf@xb}{\pgf@ya}}
    \pgfpathlineto{\pgfpointadd{\pgfpoint{\pgf@xb}{\pgf@ya}}{\ppd@offset}}
    \pgfpathlineto{\pgfpointadd{\pgfpoint{\pgf@xb}{\pgf@yb}}{\ppd@offset}}
    \pgfpathlineto{\pgfpointadd{\pgfpoint{\pgf@xa}{\pgf@yb}}{\ppd@offset}}
    \pgfpathlineto{\pgfqpoint{\pgf@xa}{\pgf@yb}}
    \pgfpathmoveto{\pgfqpoint{\pgf@xb}{\pgf@yb}}
    \pgfpathlineto{\pgfpointadd{\pgfpoint{\pgf@xb}{\pgf@yb}}{\ppd@offset}}
  }
}

\makeatletter
\tikzset{anchor/.append code=\let\tikz@auto@anchor\relax,
  add font/.code=%
    \expandafter\def\expandafter\tikz@textfont\expandafter{\tikz@textfont#1},
  left delimiter/.style 2 args={append after command={\tikz@delimiter{south east}
    {south west}{every delimiter,every left delimiter,#2}{south}{north}{#1}{.}{\pgf@y}}}}
\tikzstyle{sms} = [rectangle callout, draw,very thick, rounded corners, minimum height=20pt]
\makeatletter
\tikzset{anchor/.append code=\let\tikz@auto@anchor\relax,
  add font/.code=%
    \expandafter\def\expandafter\tikz@textfont\expandafter{\tikz@textfont#1},
  left delimiter/.style 2 args={append after command={\tikz@delimiter{south east}
    {south west}{every delimiter,every left delimiter,#2}{south}{north}{#1}{.}{\pgf@y}}}}
\tikzstyle{sms} = [rectangle callout, draw,very thick, rounded corners, minimum height=20pt]
\usetikzlibrary{positioning,calc}
\tikzstyle{block} = [rectangle, draw,
text width=10.5em, text centered, rounded corners, minimum height=4em]
\tikzstyle{line} = [draw, -latex]
\tikzset{
  mybackground51/.style={execute at end picture={
      \begin{scope}[on background layer]
        \draw[black, rounded corners=2ex, fill=gray2] (current bounding box.south west)
        rectangle (current bounding box.north east);
        \node[draw,fill=white,ellipse,anchor=west,inner sep=1pt,minimum width=1ex] at (current bounding box.north
        west){#1};
      \end{scope}
    }},
}
\tikzset{
  mybackground9/.style={execute at end picture={
        \begin{scope}[on background layer]
          \draw[black,fill=black!5,rounded corners=6ex] (current bounding box.south west)
                    rectangle (current bounding box.north east);
          \node[draw,fill=white,ellipse,anchor=west,inner sep=1pt,minimum width=4ex] at (current bounding box.north
                   west){#1};
        \end{scope}
    }},
}

\tikzset{
  mybackground13/.style={execute at end picture={
        \begin{scope}[on background layer]
          \draw[black, fill=gray2, rounded corners=4ex] (current bounding box.south west)
                    rectangle (current bounding box.north east);
          \node[draw,fill=white,ellipse,anchor=west,inner sep=1pt,minimum width=4ex] at (current bounding box.north
                   west){#1};
        \end{scope}
    }},
}
\tikzset{
  mybackground14/.style={execute at end picture={
        \begin{scope}[on background layer]
          \draw[black, rounded corners=2ex] (current bounding box.south west)
                    rectangle (current bounding box.north east);
          \node[draw,fill=white,ellipse,anchor=west,inner sep=1pt,minimum width=4ex] at (current bounding box.north
                   west){#1};
        \end{scope}
    }},
}

\tikzset{
  mybackground6/.style={execute at end picture={
        \begin{scope}[on background layer]
          \draw[black,rounded corners=1ex, line width=0.15mm] (current bounding box.south west)
                    rectangle (current bounding box.north east);
          \node[draw,fill=white,ellipse,anchor=west,inner sep=1pt,minimum width=4ex] at (current bounding box.north
                   west){#1};
        \end{scope}
c    }},
}

\tikzset{
  mybackground11/.style={execute at end picture={
        \begin{scope}[on background layer]
          \draw[black, fill=Black!80!Sepia!9, rounded corners=6ex] (current bounding box.south west)
                    rectangle (current bounding box.north east);
          \node[draw,fill=white,ellipse,anchor=west,inner sep=1pt,minimum width=4ex] at (current bounding box.north
                   west){#1};
        \end{scope}
    }},
}

\tikzset{
  mybackground15/.style={execute at end picture={
        \begin{scope}[on background layer]
          \draw[black, fill=Black!80!Sepia!9, rounded corners=3ex] (current bounding box.south west)
                    rectangle (current bounding box.north east);
          \node[draw,fill=white,ellipse,anchor=west,inner sep=1pt,minimum width=4ex] at (current bounding box.north
                   west){#1};
        \end{scope}
    }},
}

\tikzset{
  mybackground12/.style={execute at end picture={
        \begin{scope}[on background layer]
          \draw[black, fill=Black!40!Emerald!30, rounded corners=3ex, line width=0.3mm] (current bounding box.south west)
                    rectangle (current bounding box.north east);
        \end{scope}
    }},
}
\tikzset{
  mybackground18/.style={execute at end picture={
      \begin{scope}[on background layer]
        \draw[black, fill=gray3, rounded corners=3.5ex] (current bounding box.south west)
        rectangle (current bounding box.north east);
        \node[draw,fill=white,ellipse,anchor=west,inner sep=1pt,minimum width=4ex] at (current bounding box.north
        west){#1};
      \end{scope}
    }}
}
\tikzset{
  mybackground58/.style={execute at end picture={
        \begin{scope}[on background layer]
          \draw[black, fill=blue!40!black!5, rounded corners=1ex] (current bounding box.south west)
                    rectangle (current bounding box.north east);
          \node[draw,fill=white,ellipse,anchor=west,inner sep=1pt,minimum width=4ex, rounded corners=1ex] at (current bounding box.north
                   west){#1};
        \end{scope}
    }},
}
\tikzset{l3 switch/.style={
    parallelepiped,fill=switch, draw=white,
    minimum width=0.75cm,
    minimum height=0.75cm,
    parallelepiped offset x=1.75mm,
    parallelepiped offset y=1.25mm,
    path picture={
      \node[fill=white,
        circle,
        minimum size=6pt,
        inner sep=0pt,
        append after command={
          \pgfextra{
            \foreach \angle in {0,45,...,360}
            \draw[-latex,fill=white] (\tikzlastnode.\angle)--++(\angle:2.25mm);
          }
        }
      ]
       at ([xshift=-0.75mm,yshift=-0.5mm]path picture bounding box.center){};
    }
  },
  ports/.style={
    line width=0.3pt,
    top color=gray!20,
    bottom color=gray!80
  },
  rack switch/.style={
    parallelepiped,fill=white, draw,
    minimum width=1.25cm,
    minimum height=0.25cm,
    parallelepiped offset x=2mm,
    parallelepiped offset y=1.25mm,
    xscale=-1,
    path picture={
      \draw[top color=gray!5,bottom color=gray!40]
      (path picture bounding box.south west) rectangle
      (path picture bounding box.north east);
      \coordinate (A-west) at ([xshift=-0.2cm]path picture bounding box.west);
      \coordinate (A-center) at ($(path picture bounding box.center)!0!(path
        picture bounding box.south)$);
      \foreach \x in {0.275,0.525,0.775}{
        \draw[ports]([yshift=-0.05cm]$(A-west)!\x!(A-center)$)
          rectangle +(0.1,0.05);
        \draw[ports]([yshift=-0.125cm]$(A-west)!\x!(A-center)$)
          rectangle +(0.1,0.05);
       }
      \coordinate (A-east) at (path picture bounding box.east);
      \foreach \x in {0.085,0.21,0.335,0.455,0.635,0.755,0.875,1}{
        \draw[ports]([yshift=-0.1125cm]$(A-east)!\x!(A-center)$)
          rectangle +(0.05,0.1);
      }
    }
  },
  server/.style={
    parallelepiped,
    fill=white, draw,
    minimum width=0.35cm,
    minimum height=0.75cm,
    parallelepiped offset x=3mm,
    parallelepiped offset y=2mm,
    xscale=-1,
    path picture={
      \draw[top color=gray!5,bottom color=gray!40]
      (path picture bounding box.south west) rectangle
      (path picture bounding box.north east);
      \coordinate (A-center) at ($(path picture bounding box.center)!0!(path
        picture bounding box.south)$);
      \coordinate (A-west) at ([xshift=-0.575cm]path picture bounding box.west);
      \draw[ports]([yshift=0.1cm]$(A-west)!0!(A-center)$)
        rectangle +(0.2,0.065);
      \draw[ports]([yshift=0.01cm]$(A-west)!0.085!(A-center)$)
        rectangle +(0.15,0.05);
      \fill[black]([yshift=-0.35cm]$(A-west)!-0.1!(A-center)$)
        rectangle +(0.235,0.0175);
      \fill[black]([yshift=-0.385cm]$(A-west)!-0.1!(A-center)$)
        rectangle +(0.235,0.0175);
      \fill[black]([yshift=-0.42cm]$(A-west)!-0.1!(A-center)$)
        rectangle +(0.235,0.0175);
    }
  },
}

\usetikzlibrary{calc, shadings, shadows, shapes.arrows}
\tikzset{cross/.style={cross out, draw=black, minimum size=2*(#1-\pgflinewidth), inner sep=0pt, outer sep=0pt},
cross/.default={1pt}}
\tikzset{%
  interface/.style={draw, rectangle, rounded corners, font=\LARGE\sffamily},
  ethernet/.style={interface, fill=yellow!50},
  serial/.style={interface, fill=green!70},
  speed/.style={sloped, anchor=south, font=\large\sffamily},
  route/.style={draw, shape=single arrow, single arrow head extend=4mm,
    minimum height=1.7cm, minimum width=3mm, white, fill=switch!20,
    drop shadow={opacity=.8, fill=switch}, font=\tiny}
}
\newcommand*{\shift}{1.3cm}
\newcommand*{\router}[1]{
\begin{tikzpicture}
  \coordinate (ll) at (-3,0.5);
  \coordinate (lr) at (3,0.5);
  \coordinate (ul) at (-3,2);
  \coordinate (ur) at (3,2);
  \shade [shading angle=90, left color=switch, right color=white] (ll)
    arc (-180:-60:3cm and .75cm) -- +(0,1.5) arc (-60:-180:3cm and .75cm)
    -- cycle;
  \shade [shading angle=270, right color=switch, left color=white!50] (lr)
    arc (0:-60:3cm and .75cm) -- +(0,1.5) arc (-60:0:3cm and .75cm) -- cycle;
  \draw [thick] (ll) arc (-180:0:3cm and .75cm)
    -- (ur) arc (0:-180:3cm and .75cm) -- cycle;
  \draw [thick, shade, upper left=switch, lower left=switch,
    upper right=switch, lower right=white] (ul)
    arc (-180:180:3cm and .75cm);
  \node at (0,0.5){\color{blue!60!black}\Huge #1};
  \begin{scope}[yshift=2cm, yscale=0.28, transform shape]
    \node[route, rotate=45, xshift=\shift] {\strut};
    \node[route, rotate=-45, xshift=-\shift] {\strut};
    \node[route, rotate=-135, xshift=\shift] {\strut};
    \node[route, rotate=135, xshift=-\shift] {\strut};
  \end{scope}
\end{tikzpicture}}

\makeatletter
\pgfdeclareradialshading[tikz@ball]{cloud}{\pgfpoint{-0.275cm}{0.4cm}}{%
  color(0cm)=(tikz@ball!75!white);
  color(0.1cm)=(tikz@ball!85!white);
  color(0.2cm)=(tikz@ball!95!white);
  color(0.7cm)=(tikz@ball!89!black);
  color(1cm)=(tikz@ball!75!black)
}
\tikzoption{cloud color}{\pgfutil@colorlet{tikz@ball}{#1}%
  \def\tikz@shading{cloud}\tikz@addmode{\tikz@mode@shadetrue}}
\makeatother

\tikzset{my cloud/.style={
     cloud, draw, aspect=2,
     cloud color={gray!5!white}
  }
}

\usetikzlibrary{backgrounds}
\usetikzlibrary{patterns}
\pgfmathdeclarefunction{gauss}{3}{%
  \pgfmathparse{1/(#3*sqrt(2*pi))*exp(-((#1-#2)^2)/(2*#3^2))}%
}
\pgfmathdeclarefunction{cdf}{3}{%
  \pgfmathparse{1/(1+exp(-0.07056*((#1-#2)/#3)^3 - 1.5976*(#1-#2)/#3))}%
}
\pgfmathdeclarefunction{fq}{3}{%
  \pgfmathparse{1/(sqrt(2*pi*#1))*exp(-(sqrt(#1)-#2/#3)^2/2)}%
}
\pgfmathdeclarefunction{fq0}{1}{%
  \pgfmathparse{1/(sqrt(2*pi*#1))*exp(-#1/2)}%
}
\renewcommand\thetable{\arabic{table}}
\usepackage{etoolbox}

\usepackage{wrapfig}
\newcommand{\mybox}[4]{
    \begin{figure}[H]
        \centering
    \begin{tikzpicture}
      \node[anchor=text,text width=\columnwidth-1.2cm, draw, rounded corners, line width=1pt, fill=#3, inner sep=3.5mm] (big) {\\#4};
        \node[draw, rounded corners, line width=.5pt, fill=#2, anchor=west, xshift=5mm] (small) at (big.north west) {#1};
    \end{tikzpicture}
    \end{figure}

  }

\newcommand{\myboxtwo}[4]{
    \begin{figure}[H]
        \centering
    \begin{tikzpicture}
      \node[anchor=text,text width=\columnwidth-1.2cm, draw, rounded corners, line width=1pt, fill=#3, inner sep=3.5mm] (big) {\\#4};
        \node[draw, rounded corners, line width=.5pt, fill=#2, anchor=west, xshift=5mm] (small) at (big.north west) {#1};
    \end{tikzpicture}
    \end{figure}

  }  

\makeatletter
\newcommand{\setword}[2]{%
  \phantomsection
  #1\def\@currentlabel{\unexpanded{#1}}\label{#2}%
}
\makeatother

\newcommand{\figref}[1]{\hyperref[#1]{Fig.~\ref*{#1}}}
\newcommand{\figsref}[1]{\hyperref[#1]{Figs.~\ref*{#1}}}
\newcommand{\Figref}[1]{\hyperref[#1]{Figure~\ref*{#1}}}
\newcommand{\Figsref}[1]{\hyperref[#1]{Figures~\ref*{#1}}}
\newcommand{\tableref}[1]{\hyperref[#1]{Table~\ref*{#1}}}
\newcommand{\tablesref}[1]{\hyperref[#1]{Tables~\ref*{#1}}}
\newcommand{\appendixref}[1]{\hyperref[#1]{Appendix~\ref*{#1}}}
\newcommand{\theoremref}[1]{\hyperref[#1]{Thm.~\ref*{#1}}}
\newcommand{\Theoremref}[1]{\hyperref[#1]{Theorem~\ref*{#1}}}
\newcommand{\lemmaref}[1]{\hyperref[#1]{Lemma~\ref*{#1}}}
\newcommand{\propref}[1]{\hyperref[#1]{Prop.~\ref*{#1}}}
\newcommand{\propsref}[1]{\hyperref[#1]{Props.~\ref*{#1}}}
\newcommand{\Propsref}[1]{\hyperref[#1]{Propositions~\ref*{#1}}}
\newcommand{\Propref}[1]{\hyperref[#1]{Proposition~\ref*{#1}}}
\newcommand{\corref}[1]{\hyperref[#1]{Cor.~\ref*{#1}}}
\newcommand{\Corref}[1]{\hyperref[#1]{Corollary~\ref*{#1}}}
\newcommand{\scenarioref}[1]{\hyperref[#1]{Scenario~\ref*{#1}}}
\newcommand{\Scenarioref}[1]{\hyperref[#1]{\textsc{scenario}~\ref*{#1}}}
\newcommand{\probref}[1]{\hyperref[#1]{Prob.~\ref*{#1}}}
\newcommand{\Probref}[1]{\hyperref[#1]{Problem~\ref*{#1}}}
\newcommand{\gameref}[1]{\hyperref[#1]{Game~\ref*{#1}}}
\newcommand{\chapterref}[1]{\hyperref[#1]{Chapter~\ref*{#1}}}
\newcommand{\sectionref}[1]{\hyperref[#1]{\S\ref*{#1}}}
\newcommand{\Algref}[1]{\hyperref[#1]{Algorithm ~\ref*{#1}}}
\newcommand{\myalgref}[1]{\hyperref[#1]{Alg.~\ref*{#1}}}
\newcommand{\Myalgref}[1]{\hyperref[#1]{Algorithm~\ref*{#1}}}
\newcommand{\defref}[1]{\hyperref[#1]{Def.~\ref*{#1}}}
\newcommand{\Defref}[1]{\hyperref[#1]{Definition~\ref*{#1}}}
\newcommand{\assumptionref}[1]{\hyperref[#1]{Assumption~\ref*{#1}}}
\newcommand{\remarkref}[1]{\hyperref[#1]{Remark~\ref*{#1}}}
\newcommand{\exampleref}[1]{\hyperref[#1]{Ex.~\ref*{#1}}}
\newcommand{\eqqref}[1]{\hyperref[#1]{Eq.~(\ref*{#1})}}
\newcommand{\eqqqref}[1]{\hyperref[#1]{(\ref*{#1})}}
\newcommand{\eqqsref}[1]{\hyperref[#1]{Eqs.~(\ref*{#1})}}
\newcommand{\Eqqref}[1]{\hyperref[#1]{Equation~(\ref*{#1})}}

\usepackage{arydshln}

\newcommand{\cmark}{\textcolor{OliveGreen}{\ding{51}}}  
\newcommand{\xmark}{\textcolor{Red}{\ding{55}}}    
\usepackage{lettrine}

\usepackage{fontawesome5}
\usepackage{listofitems} 
\tikzstyle{mynode}=[thick,draw=blue,fill=blue!20,circle,minimum size=22]

\newtcolorbox{examplebox}{
  colback=black!5!white,
  colframe=black!30!black!70,
  title=Example POMDP: Intrusion recovery,
  fonttitle=\bfseries,
  sharp corners
}

\newtcolorbox{examplefeatures}{
  float,
  colback=black!5!white,
  colframe=black!30!black!70,
  title=Example feature space for aggregation,
  fonttitle=\bfseries,
  sharp corners
}

\newtcolorbox{exampleboxtwo}{
  colback=black!5!white,
  colframe=black!30!black!70,
  title=Example feature space,
  fonttitle=\bfseries,
  sharp corners
}

\newtcolorbox{frameworkbox}{
  colback=black!5!white,
  colframe=black!30!black!70,
  title=Framework summary,
  fonttitle=\bfseries,
  sharp corners
}

\usepackage{mdframed}

\definecolor{grad1}{RGB}{0,150,150}
\definecolor{grad2}{RGB}{0,100,200}
\definecolor{grad3}{RGB}{80,0,150}

\usepackage{tikz-3dplot}
\tdplotsetmaincoords{70}{110}

\newcommand{\legendSymbol}[1]{%
    \tikz[baseline=-0.5ex]{
      \draw[Red,thick, mark=none, solid] (0,0) -- (0.5cm,0);
      \path[Red,thick, mark=x, mark repeat=1] plot coordinates {(0.25cm,0)};      
    }%
  }

\newcommand{\legendSymboltwo}[1]{%
    \tikz[baseline=-0.5ex]{
      \draw[RoyalAzure,thick, mark=none, solid] (0,0) -- (0.5cm,0);
      \path[RoyalAzure,thick, mark=diamond, mark repeat=1] plot coordinates {(0.25cm,0)};      
    }%
  }

  \newcommand{\legendSymbolthree}[1]{%
    \tikz[baseline=-0.5ex]{
      \draw[Purple,thick, mark=none, solid] (0,0) -- (0.5cm,0);
      \path[Purple,thick, mark=oplus, mark repeat=1] plot coordinates {(0.25cm,0)};      
    }%
  }
  \newcommand{\legendSymbolfour}[1]{%
    \tikz[baseline=-0.5ex]{
      \draw[OliveGreen,thick, mark=none, solid] (0,0) -- (0.5cm,0);
      \path[OliveGreen,thick, mark=triangle, mark repeat=1] plot coordinates {(0.25cm,0)};      
    }%
  }

%% file: latex-commands.tex
\newcommand{\acro}[1]{\textsc{#1}\xspace}

\newcommand{\ppo}{\acro{ppo}}

%% file: tikz/rollout_framework11.tex
\begin{tikzpicture}

\node[scale=1] (node1) at (0,0)
{
  \begin{tikzpicture}
\node[scale=1] (node1) at (0,0)
{
  \begin{tikzpicture}
\node [scale=1] (node1) at (2,0) {
  \begin{tikzpicture}[fill=white, >=stealth,mybackground51={\small \textsc{networked system}},
        node distance=3cm,
    database/.style={
      cylinder,
      cylinder uses custom fill,
      shape border rotate=90,
      aspect=0.25,
      draw}]
    \tikzset{
node distance = 9em and 4em,
sloped,
   box/.style = {%
    shape=rectangle,
    rounded corners,
    draw=blue!40,
    fill=blue!15,
    align=center,
    font=\fontsize{12}{12}\selectfont},
 arrow/.style = {%
    line width=0.1mm,
    shorten >=1mm, shorten <=1mm,
    font=\fontsize{8}{8}\selectfont},
}

\draw[draw=none] (3,0.25) rectangle node (m1){} (7.3,2);
\end{tikzpicture}
};

\end{tikzpicture}
};
  \node[scale=0.8](test) at (0.05,-0.15) {
    \begin{tikzpicture}
\node[server, scale=0.6](s1) at (0.6,-0.1) {};
\node[server, scale=0.6](s2) at (2,0.3) {};
\node[server, scale=0.6](s3) at (3.6,0.3) {};
\node[server, scale=0.6](s4) at (5,-0.5) {};
\node[server, scale=0.6](s5) at (3.6,-0.9) {};
\node[server, scale=0.6](s6) at (2,-0.9) {};

\node[inner sep=0pt,align=center] (hacker2) at (0.95,-0.8)
  {\scalebox{0.1}{
     \includegraphics{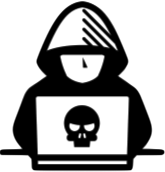}
   }
 };
\draw[-, color=black] (s1) to (s2);
\draw[-, color=black] (s1) to (s3);
\draw[-, color=black] (s1) to (s4);
\draw[-, color=black] (s1) to (s5);
\draw[-, color=black] (s1) to (s6);
\draw[-, color=black] (s2) to (s3);
\draw[-, color=black] (s2) to (s4);
\draw[-, color=black] (s2) to (s5);
\draw[-, color=black] (s2) to (s6);
\draw[-, color=black] (s3) to (s4);
\draw[-, color=black] (s3) to (s5);
\draw[-, color=black] (s3) to (s6);
\draw[-, color=black] (s4) to (s5);
\draw[-, color=black] (s4) to (s6);
\draw[-, color=black] (s5) to (s6);
\end{tikzpicture}
};

\draw[-{Latex[width=1.4mm]}, color=black, line width=0.2mm, rounded corners] (2.15,-0.15) to (3.8,-0.15) to (3.8,-1.4);

\node[inner sep=0pt,align=center, scale=0.8, color=black] (hacker) at (4.4,-0.79) {
  System\\
  metrics\\
  $z_k$
};

\node[inner sep=0pt,align=center, scale=0.8, color=black] (hacker) at (1.53,0.375) {
State $i_k$
};

\node[inner sep=0pt,align=center, scale=0.8, color=black] (hacker) at (-1.9,-1.2) {
Adapted security policy
};

\node[inner sep=0pt,align=center, scale=0.8, color=black] (hacker) at (-3.75,-1.2) {
$\tilde{\mu}(b_k)$
};

\node[inner sep=0pt,align=center, scale=0.8, color=black] (hacker) at (-3,0.2) {
Control $u_k$
};

\draw[-{Latex[width=1.4mm]}, color=black, line width=0.2mm, rounded corners] (2.53,-2.45) to (1,-2.45);
\node[inner sep=0pt,align=center, scale=0.8, color=black] (hacker) at (1.78,-2.75) {
  Belief $b_k$
};

\node[inner sep=0pt,align=center, scale=0.8, color=black] (hacker) at (-1.8,-3.6) {
  Base policy $\mu$
};
\node[inner sep=0pt,align=center, scale=0.8, color=black] (hacker) at (1.12,-3.6) {
Cost function approximation $\Tilde{J}$
};

\draw[-{Latex[width=1.2mm]}, color=black, line width=0.2mm, rounded corners] (-0.9,-3.85) to (-0.9,-3.45);
\draw[-{Latex[width=1.4mm]}, color=black, line width=0.2mm, rounded corners] (-3,-2.5) to (-3.75,-2.5) to (-3.75,-1.4);
\draw[-{Latex[width=1.4mm]}, color=black, line width=0.2mm, rounded corners] (-3.75,-0.95) to (-3.75,0) to (-2.2, 0);

\node [scale=1] (node1) at (-1,-2.4) {
  \begin{tikzpicture}[fill=white, >=stealth,mybackground51={\small \textsc{rollout}},
        node distance=3cm,
    database/.style={
      cylinder,
      cylinder uses custom fill,
      shape border rotate=90,
      aspect=0.25,
      draw}]
    \tikzset{
node distance = 9em and 4em,
sloped,
   box/.style = {%
    shape=rectangle,
    rounded corners,
    draw=blue!40,
    fill=blue!15,
    align=center,
    font=\fontsize{12}{12}\selectfont},
 arrow/.style = {%
    line width=0.1mm,
    shorten >=1mm, shorten <=1mm,
    font=\fontsize{8}{8}\selectfont},
}

\draw[draw=none] (0,0) rectangle node (m1){} (4,1.8);
\end{tikzpicture}
};

\node [scale=1] (node1) at (4.1,-2.4) {
  \begin{tikzpicture}[fill=white, >=stealth,mybackground51={\small \textsc{particle filter}},
        node distance=3cm,
    database/.style={
      cylinder,
      cylinder uses custom fill,
      shape border rotate=90,
      aspect=0.25,
      draw}]
    \tikzset{
node distance = 9em and 4em,
sloped,
   box/.style = {%
    shape=rectangle,
    rounded corners,
    draw=blue!40,
    fill=blue!15,
    align=center,
    font=\fontsize{12}{12}\selectfont},
 arrow/.style = {%
    line width=0.1mm,
    shorten >=1mm, shorten <=1mm,
    font=\fontsize{8}{8}\selectfont},
}

\draw[draw=none] (-0.2,0.3) rectangle node (m1){} (2.95,2);
\end{tikzpicture}
};


  \node[scale=0.85, rotate=0](test) at (0.05,-1.95) {
    \begin{tikzpicture}
\node[scale=0.5, rotate=10](c1) at (2.13,0.72) {
    \begin{tikzpicture}
\draw[->, x=0.15cm,y=0.15cm, line width=0.7mm, Black!40!Red!90, dashed]
        (3,0) sin (5,1) cos (7,0) sin (9,-1) cos (11,0)
        sin (13,1) cos (15,0) sin (17,-1)
        ;
    \end{tikzpicture}
};

\node[scale=0.5, rotate=2](c1) at (2.17,0.5) {
    \begin{tikzpicture}
\draw[->, x=0.15cm,y=0.15cm, line width=0.7mm, Black!40!Red!90, dashed]
        (3,0) sin (5,1) cos (7,0) sin (9,-1) cos (11,0)
        sin (13,1) cos (15,0) sin (17,-1)
        ;
    \end{tikzpicture}
  };

\node[scale=0.5, rotate=-10](c1) at (2.17,0.3) {
    \begin{tikzpicture}
\draw[->, x=0.15cm,y=0.15cm, line width=0.7mm, Black!40!Red!90, dashed]
        (3,0) sin (5,1) cos (7,0) sin (9,-1) cos (11,0)
        sin (13,1) cos (15,0) sin (17,-1)
        ;
    \end{tikzpicture}
  };
  \node[draw,circle, minimum width=1mm, scale=0.4, fill=black](start) at (1.6,0.55) {};
\node[scale=0.6](c1) at (2.62,0.5) {
  \begin{tikzpicture}
\draw (3,0) ellipse (0.25cm and 0.85cm);
    \end{tikzpicture}
  };

\draw[draw=none] (0,0) rectangle node (m1){} (3.5,1.5);
    \end{tikzpicture}
};

\node[inner sep=0pt,align=center, scale=0.7, color=black] (hacker) at (-1.6,-2.12) {
  (1) Evaluation of the base\\
  policy through rollouts
};

\node[inner sep=0pt,align=center, scale=0.7, color=black] (hacker) at (0,-2.32) {
$b_k$
};

\node[inner sep=0pt,align=center, scale=0.7, color=black] (hacker) at (0.8,-2.73) {
$\Tilde{J}$
};

\node[inner sep=0pt,align=center, scale=0.7, color=black] (hacker) at (-1.4,-2.9) {
  (2) Policy adaptation through\\
  lookahead optimization
};
  \end{tikzpicture}
};

\node[scale=1.02] (kth_cr) at (3.2,-1.18)
{
\begin{tikzpicture}

  \def\B{11};
  \def\Bs{3.0};
  \def\xmax{\B+3.2*\Bs};
  \def\ymin{{-0.1*gauss(\B,\B,\Bs)}};
  \def\h{0.08*gauss(\B,\B,\Bs)};
  \def\N{50}

  \begin{axis}[every axis plot post/.append style={
      mark=none,domain=0:20,
      samples=\N,smooth},
               xmin=0, xmax=20,
               ymin=0, ymax={1.1*gauss(\B,\B,\Bs)},
               axis lines=middle,
               axis line style=thick,
               enlargelimits=upper, 
               ticks=none,
               every axis x label/.style={at={(current axis.right of origin)},anchor=north},
               width=4cm,
               height=2.4cm,
               clip=false
              ]

    \addplot[Blue,thick,name path=B] {gauss(x,\B,\Bs)};
    \path[name path=xaxis](0,0) -- (20,0);
    \addplot[Blue!25] fill between[of=xaxis and B];
  \end{axis}
\node[draw,circle, minimum width=0.1cm, scale=0.3, fill=black!30](p1) at (0.1,0) {};
\node[draw,circle, minimum width=0.1cm, scale=0.3, fill=black!30](p2) at (0.35,0) {};
\node[draw,circle, minimum width=0.1cm, scale=0.3, fill=black!30](p3) at (0.6,0) {};
\node[draw,circle, minimum width=0.1cm, scale=0.3, fill=black!30](p4) at (0.8,0) {};
\node[draw,circle, minimum width=0.1cm, scale=0.3, fill=black!30](p5) at (0.92,0) {};
\node[draw,circle, minimum width=0.1cm, scale=0.3, fill=black!30](p6) at (1.05,0) {};
\node[draw,circle, minimum width=0.1cm, scale=0.3, fill=black!30](p7) at (1.17,0) {};
\node[draw,circle, minimum width=0.1cm, scale=0.3, fill=black!30](p8) at (1.3,0) {};
\node[draw,circle, minimum width=0.1cm, scale=0.3, fill=black!30](p9) at (1.43,0) {};
\node[draw,circle, minimum width=0.1cm, scale=0.3, fill=black!30](p10) at (1.6,0) {};
\node[draw,circle, minimum width=0.1cm, scale=0.3, fill=black!30](p11) at (1.8,0) {};
\node[draw,circle, minimum width=0.1cm, scale=0.3, fill=black!30](p12) at (2,0) {};
\node[draw,circle, minimum width=0.1cm, scale=0.3, fill=black!30](p13) at (2.2,0) {};

\draw[-, thick] (p3) to (0.6, 0.13);
\draw[-, thick] (p4) to (0.8, 0.325);
\draw[-, thick] (p5) to (0.92, 0.455);
\draw[-, thick] (p6) to (1.05, 0.6);
\draw[-, thick] (p7) to (1.17, 0.662);
\draw[-, thick] (p8) to (1.3, 0.662);
\draw[-, thick] (p9) to (1.43, 0.555);
\draw[-, thick] (p10) to (1.6, 0.35);
\draw[-, thick] (p11) to (1.8, 0.13);

\draw[-, thick, color=Blue] (1.5, 0.8) to (1.8, 0.8);
\node[draw,circle, minimum width=0.1cm, scale=0.3, fill=black!30](p12) at (1.66,0.62) {};
\node[inner sep=0pt,align=center, scale=0.62, color=black] (hacker) at (2.25,0.62) {
   Particle
 };
\node[inner sep=0pt,align=center, scale=0.62, color=black] (hacker) at (2.37,0.8) {
   Probability
};

\end{tikzpicture}
};

\node[scale=0.75] (node1) at (-0.54,-3.85)
{
\begin{tikzpicture}
\node[scale=1] (tab1) at (0,-0)
{
  \begin{tikzpicture}[scale=0.45,every node/.style={minimum size=1cm},on grid]
    \begin{scope}[
    	yshift=0,every node/.append style={
    	    yslant=0.5,xslant=-1},yslant=0.5,xslant=-1
          ]
        \fill[white,fill opacity=.9] (0,0) rectangle (5,5);
        \draw[black,very thick] (0,0) rectangle (5,5);
        \draw[step=1mm, red!50,thin] (3,1) grid (4,2);
      \end{scope}
  \end{tikzpicture}
};
\draw[-{Latex[width=1.4mm]}, color=black, line width=0.2mm, rounded corners] (0.95,0) to (0.95,0.55);
\node[scale=1] (tab1) at (0,1.2)
{
  \begin{tikzpicture}[scale=0.45,every node/.style={minimum size=1cm},on grid]
    \begin{scope}[
    	yshift=0,every node/.append style={
    	    yslant=0.5,xslant=-1},yslant=0.5,xslant=-1
          ]
          \fill[white,fill opacity=.9] (0,0) rectangle (5,5);
          \draw[black,very thick, inner color=white,outer color=red!80] (0,0) rectangle (5,5);
        \draw[black,very thick] (0,0) rectangle (5,5);
        \draw[step=10mm, black] (0,0) grid (5,5);
      \end{scope}
  \end{tikzpicture}
};

\node[inner sep=0pt,align=center, scale=0.95, color=black] (hacker) at (2.4,-0.75) {
  Belief space
};
\node[inner sep=0pt,align=center, scale=0.95, color=black] (hacker) at (3.8,0.8) {
  Aggregate belief space
};

\draw[->, color=black, bend right=30, dashed] (1.85, 1.5) to (0.5,2.2);

\node[inner sep=0pt,align=center, scale=0.95, color=black] (hacker) at (3.45,1.9) {
  \textit{Dynamic programming}
};

  \end{tikzpicture}
};

\draw[-, color=black, dashed, line width=0.35mm] (-4.6, -2.45) to (4.5,-2.45);

\node[inner sep=0pt,align=center, scale=0.95, color=black] (hacker) at (3.5,-2.7) {
  \textsc{offline}
};
\node[inner sep=0pt,align=center, scale=0.95, color=black] (hacker) at (3.5,-2.2) {
  \textsc{online}
};

\end{tikzpicture}

%% file: tikz/stats.tex
\begin{tikzpicture}

\pgfplotstableread{
0 1
10 14
20 28
30 27
40 15
50 10
60 5
}\datatablee

\pgfplotstableread{
2017 14376
2018 15589
2019 17414
2020 19113
2021 21339
2022 23762
2023 26511
2024 29643
2025 33353
2026 37600
2027 41838
2028 46077
}\datatableee

\node[scale=1] (box) at (0,5.4) {
\begin{tikzpicture}
\node[scale=0.8] (box) at (0.65,0.35) {
\begin{tikzpicture}
\draw[fill=Red] (0,-2.35) rectangle node (m1){} (0.35,-2.65);
\node[inner sep=0pt,align=center, scale=0.9, rotate=0, opacity=1, color=black] (obs) at (1,-2.5)
{
  Daily
};

\draw[fill=Red!50] (2,-2.35) rectangle node (m1){} (2.35,-2.65);
\node[inner sep=0pt,align=center, scale=0.9, rotate=0, opacity=1, color=black] (obs) at (3.1,-2.5)
{
  Weekly
};

\draw[fill=Black!20] (4,-2.35) rectangle node (m1){} (4.35,-2.65);
\node[inner sep=0pt,align=center, scale=0.9, rotate=0, opacity=1, color=black] (obs) at (5.1,-2.5)
{
  Monthly
};

\draw[fill=OliveGreen!20] (6,-2.35) rectangle node (m1){} (6.35,-2.65);
\node[inner sep=0pt,align=center, scale=0.9, rotate=0, opacity=1, color=black] (obs) at (7.2,-2.5)
{
  Quarterly
};

\draw[fill=OliveGreen!50] (8,-2.35) rectangle node (m1){} (8.35,-2.65);
\node[inner sep=0pt,align=center, scale=0.9, rotate=0, opacity=1, color=black] (obs) at (9.2,-2.5)
{
  Annually
};
\end{tikzpicture}
};

\node[scale=0.8] (box) at (0.05,0) {
\begin{tikzpicture}
\node[inner sep=0pt,align=center, scale=1, color=black] (hacker) at (-0.7,-2.5) {
Priorities
};
\draw[fill=Red] (0,-2.35) rectangle node (m1){} (2.4,-2.65);
\draw[fill=Red!50] (2.4,-2.35) rectangle node (m1){} (5.4,-2.65);
\draw[fill=Black!20] (5.4,-2.35) rectangle node (m1){} (7.6,-2.65);
\draw[fill=OliveGreen!20] (7.6,-2.35) rectangle node (m1){} (9.5,-2.65);
\draw[fill=OliveGreen!50] (9.5,-2.35) rectangle node (m1){} (10.5,-2.65);
\node[inner sep=0pt,align=center, scale=0.9, rotate=0, opacity=1, color=white] (obs) at (1.1,-2.5)
{
  27\%
};
\node[inner sep=0pt,align=center, scale=0.9, rotate=0, opacity=1, color=white] (obs) at (3.65,-2.5)
{
  34\%
};
\node[inner sep=0pt,align=center, scale=0.9, rotate=0, opacity=1, color=black] (obs) at (6.7,-2.5)
{
  20\%
};
\node[inner sep=0pt,align=center, scale=0.9, rotate=0, opacity=1, color=black] (obs) at (8.75,-2.5)
{
  15\%
};
\node[inner sep=0pt,align=center, scale=0.9, rotate=0, opacity=1, color=black] (obs) at (10.1,-2.5)
{
  4\%
};
\end{tikzpicture}
};

\node[scale=0.8] (box) at (0.27,-0.4) {
\begin{tikzpicture}
\node[inner sep=0pt,align=center, scale=1, color=black] (hacker) at (-0.4,-2.5) {
Tools
};
\draw[fill=Red] (0,-2.35) rectangle node (m1){} (2.15,-2.65);
\draw[fill=Red!50] (2.15,-2.35) rectangle node (m1){} (4.3,-2.65);
\draw[fill=Black!20] (4.3,-2.35) rectangle node (m1){} (6.45,-2.65);
\draw[fill=OliveGreen!20] (6.45,-2.35) rectangle node (m1){} (8.4,-2.65);
\draw[fill=OliveGreen!50] (8.4,-2.35) rectangle node (m1){} (10.5,-2.65);
\node[inner sep=0pt,align=center, scale=0.9, rotate=0, opacity=1, color=white] (obs) at (1.2,-2.5)
{
  22\%
};
\node[inner sep=0pt,align=center, scale=0.9, rotate=0, opacity=1, color=white] (obs) at (3.6,-2.5)
{
  21\%
};
\node[inner sep=0pt,align=center, scale=0.9, rotate=0, opacity=1, color=black] (obs) at (5.75,-2.5)
{
  19\%
};
\node[inner sep=0pt,align=center, scale=0.9, rotate=0, opacity=1, color=black] (obs) at (7.7,-2.5)
{
  18\%
};
\node[inner sep=0pt,align=center, scale=0.9, rotate=0, opacity=1, color=black] (obs) at (9.7,-2.5)
{
  20\%
};
\end{tikzpicture}
};
\node[scale=0.8] (box) at (-0.08,-0.8) {
\begin{tikzpicture}
\node[inner sep=0pt,align=center, scale=1, color=black] (hacker) at (-0.85,-2.5) {
Technology
};
\draw[fill=Red] (0,-2.35) rectangle node (m1){} (1.8,-2.65);
\draw[fill=Red!50] (1.8,-2.35) rectangle node (m1){} (4.6,-2.65);
\draw[fill=Black!20] (4.6,-2.35) rectangle node (m1){} (6.9,-2.65);
\draw[fill=OliveGreen!20] (6.9,-2.35) rectangle node (m1){} (9.2,-2.65);
\draw[fill=OliveGreen!50] (9.2,-2.35) rectangle node (m1){} (10.5,-2.65);
\node[inner sep=0pt,align=center, scale=0.9, rotate=0, opacity=1, color=white] (obs) at (1,-2.5)
{
  16\%
};
\node[inner sep=0pt,align=center, scale=0.9, rotate=0, opacity=1, color=white] (obs) at (3.4,-2.5)
{
  30\%
};
\node[inner sep=0pt,align=center, scale=0.9, rotate=0, opacity=1, color=black] (obs) at (5.85,-2.5)
{
  20\%
};
\node[inner sep=0pt,align=center, scale=0.9, rotate=0, opacity=1, color=black] (obs) at (8.1,-2.5)
{
  20\%
};
\node[inner sep=0pt,align=center, scale=0.9, rotate=0, opacity=1, color=black] (obs) at (10,-2.5)
{
  14\%
};
\end{tikzpicture}
};
\end{tikzpicture}
};

\end{tikzpicture}

%% file: tikz/outages2.tex
\begin{tikzpicture}
  \node[scale=1] (box) at (0,5.4) {
  \begin{tikzpicture}
    \definecolor{grad1}{RGB}{0,150,150}
    \definecolor{grad2}{RGB}{0,100,200}
    \definecolor{grad3}{RGB}{80,0,150}
    \definecolor{grad4}{RGB}{255,100,100}

    \node[scale=0.8] (box) at (0,0) {
    \begin{tikzpicture}

    \node[inner sep=0pt,align=center,scale=1,color=black] at (-1.30,  0.00) {\textit{Network failure}};
    \draw[fill=grad2!60!black!80,draw=black,line width=0.23mm] (0,-0.15) rectangle (7.5, 0.15);
    \node[inner sep=0pt,align=center,scale=0.9,color=white]  at (4.30,  0.00) {18.7\%};

    \draw[fill=grad2!60!black!80,draw=black,line width=0.23mm] (0,-0.25) rectangle (6.3,-0.55);
    \node[inner sep=0pt,align=center,scale=0.9,color=white]  at (3.90,-0.40) {15.5\%};
    \node[inner sep=0pt,align=center,scale=1,color=black]     at (-1.73,-0.40) {\textit{Cloud provider failure}};

    \draw[fill=Red,draw=black,line width=0.23mm]            (0,-0.65) rectangle (6.1,-0.95);
    \node[inner sep=0pt,align=center,scale=0.9,color=white] at (3.70,-0.80) {14.9\%};
    \node[inner sep=0pt,align=center,scale=1,color=black]   at (-1.70,-0.80) {\textcolor{Red}{\textit{Configuration change}}};

    \draw[fill=Red,draw=black,line width=0.23mm]            (0,-1.05) rectangle (5.7,-1.35);
    \node[inner sep=0pt,align=center,scale=0.9,color=white] at (3.55,-1.20) {14.2\%};
    \node[inner sep=0pt,align=center,scale=1,color=black]   at (-1.37,-1.20) {\textcolor{Red}{\textit{Software change}}};

    \draw[fill=grad2!60!black!80,draw=black,line width=0.23mm] (0,-1.45) rectangle (5.2,-1.75);
    \node[inner sep=0pt,align=center,scale=0.9,color=white]   at (3.30,-1.60) {12.9\%};
    \node[inner sep=0pt,align=center,scale=1,color=black]     at (-1.37,-1.60) {\textit{Hardware failure}};

    \draw[fill=grad2!60!black!80,draw=black,line width=0.23mm] (0,-1.85) rectangle (4.9,-2.15);
    \node[inner sep=0pt,align=center,scale=0.9,color=white]   at (3.10,-2.00) {12\%};
    \node[inner sep=0pt,align=center,scale=1,color=black]     at (-1.10,-2.00) {\textit{Power failure}};

    \draw[fill=grad2!60!black!80,draw=black,line width=0.23mm] (0,-2.25) rectangle (4.7,-2.55);
    \node[inner sep=0pt,align=center,scale=0.9,color=white]   at (2.90,-2.40) {11.8\%};
    \node[inner sep=0pt,align=center,scale=1,color=black]     at (-1.10,-2.40) {\textit{Cyberattack}};

    \end{tikzpicture}
    };
  \end{tikzpicture}
  };
\end{tikzpicture}

%% file: tikz/service_chain.tex
      \begin{tikzpicture}[fill=white, >=stealth,
    node distance=3cm,
    database/.style={
      cylinder,
      cylinder uses custom fill,
      shape border rotate=90,
      aspect=0.25,
      draw}]

    \tikzset{
node distance = 9em and 4em,
sloped,
   box/.style = {%
    shape=rectangle,
    rounded corners,
    draw=blue!40,
    fill=blue!15,
    align=center,
    font=\fontsize{12}{12}\selectfont},
 arrow/.style = {%
    line width=0.1mm,
    -{Triangle[length=5mm,width=2mm]},
    shorten >=1mm, shorten <=1mm,
    font=\fontsize{8}{8}\selectfont},
}
\node[draw,circle, minimum width=10mm, scale=0.6, fill=gray2](n2) at (1.5,0) {};
\node[draw,circle, minimum width=10mm, scale=0.6, fill=gray2](n3) at (3,0) {};
\node[draw,circle, minimum width=10mm, scale=0.6, fill=gray2](n4) at (4.5,0) {};
\node[draw,circle, minimum width=10mm, scale=0.6, fill=gray2](n5) at (6,1) {};
\node[draw,circle, minimum width=10mm, scale=0.6, fill=gray2](n6) at (6,0) {};
\node[draw,circle, minimum width=10mm, scale=0.6, fill=gray2](n7) at (6,-1) {};
\node[draw,circle, minimum width=10mm, scale=0.6, fill=gray2](n8) at (7.55,1) {};
\node[draw,circle, minimum width=10mm, scale=0.6, fill=gray2](n9) at (7.55,0) {};
\node[draw,circle, minimum width=10mm, scale=0.6, fill=gray2, dashed](n10) at (7.55,-1) {};

\node[draw,circle, minimum width=10mm, scale=0.6, fill=gray2](n13) at (0,0.65) {};
\node[draw,circle, minimum width=10mm, scale=0.6, fill=gray2](n14) at (0,-0.65) {};
\node[draw,circle, minimum width=10mm, scale=0.6, fill=gray2](n15) at (0.8,-0.65) {};
\node[scale=0.8] (square) at (6.8,0)
{
  \begin{tikzpicture}
    \draw[-, color=black] (0,0) to (3,0) to (3, 3.35) to (0,3.35) to (0,0);
    \draw[-, color=black, dashed] (0,1.1) to (3,1.1);
    \draw[-, color=black, dashed] (0,2.35) to (3,2.35);
    \end{tikzpicture}
  };
\draw[->, color=black, line width=0.25mm, rounded corners] (n13) to (0.5, 0) to (n2);
\draw[->, color=black, line width=0.25mm, rounded corners] (n14) to (0.5, 0) to (n2);    
\draw[-, color=black, line width=0.25mm, rounded corners] (n15) to (0.79, 0);    
\draw[->, color=black, line width=0.25mm] (n2) to (n3);
\draw[->, color=black, line width=0.25mm] (n3) to (n4);
\draw[->, color=black, line width=0.25mm] (n4) to (5.6,0);
\draw[-, color=black, line width=0.2mm] (n5) to (n8);
\draw[-, color=black, line width=0.2mm] (n6) to (n9);
\draw[-, color=black, line width=0.2mm] (n7) to (n10);
\draw[-, color=black, line width=0.2mm] (n10) to (n9);
\draw[-, color=black, line width=0.2mm] (n8) to (n9);
\draw[-, color=black, line width=0.2mm] (n6) to (n7);
\draw[-, color=black, line width=0.2mm] (n6) to (n5);
\node[server, scale=0.45](s1) at (6.06,-0.05) {};
\node[server, scale=0.45](s1) at (6.06,-1.05) {};
\node[server, scale=0.45](s1) at (6.06,0.95) {};
\node[server, scale=0.45](s1) at (7.61,0.95) {};
\node[server, scale=0.45, color=Red](s19) at (7.61,-1.05) {};
\node[server, scale=0.45](s1) at (7.61,-0.05) {};
\node[inner sep=0pt,align=center] (gpu1) at (3.06,0)
  {\scalebox{0.45}{
     \includegraphics{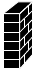}
   }
 };

\node[inner sep=0pt,align=center] (idps1) at (4.56,0)
  {\scalebox{0.05}{
     \includegraphics{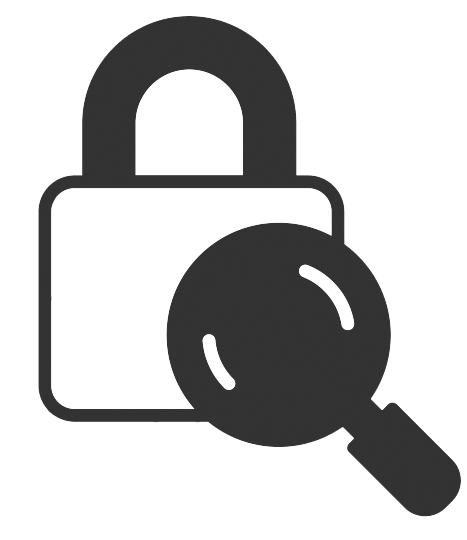}
   }
 };
\node[inner sep=0pt,align=center, scale=0.8, color=black] (hacker) at (1.53,0.5) {
Gateway
};
\node[inner sep=0pt,align=center, scale=0.8, color=black] (hacker) at (3.03,0.5) {
Firewall
};
\node[inner sep=0pt,align=center, scale=0.8, color=black] (hacker) at (4.53,0.52) {
IDS
};
\node[inner sep=0pt,align=center, scale=0.8, color=black] (hacker) at (5.08,-0.85) {
Service\\
replicas
};
\node[inner sep=0pt,align=center, scale=0.6, color=black] (hacker) at (6.8,-0.65) {
\textit{Compromised}
};

\node[inner sep=0pt,align=center, scale=0.6, color=black] (hacker) at (6.8,-1.2) {
\textit{Zone 3}
};
\node[inner sep=0pt,align=center, scale=0.6, color=black] (hacker) at (6.8,0.2) {
\textit{Zone 2}
};
\node[inner sep=0pt,align=center, scale=0.6, color=black] (hacker) at (6.8,1.2) {
\textit{Zone 1}
};

\draw[->, color=black, line width=0.2mm] (6.8, -0.73) to (7.25,-0.85);    

%


\node[inner sep=0pt,align=center] (idps1) at (0.07,0.65)
  {\scalebox{0.04}{
     \includegraphics{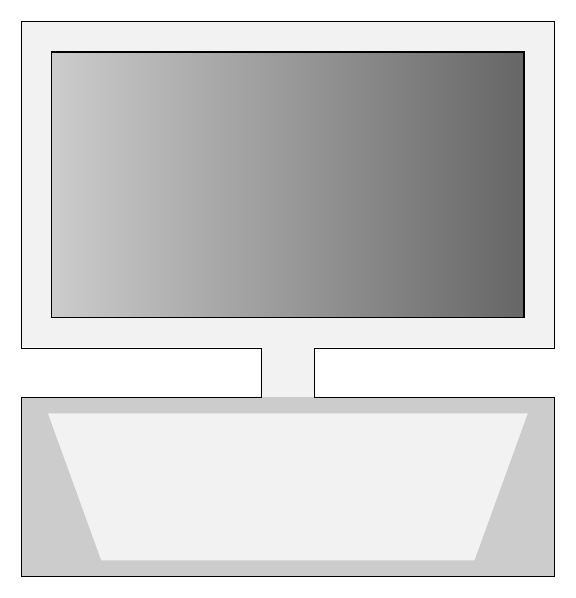}
   }
 }; 
\node[inner sep=0pt,align=center] (idps1) at (0.07,-0.65)
  {\scalebox{0.04}{
     \includegraphics{laptop3.pdf}
   }
 };
\node[inner sep=0pt,align=center] (idps1) at (0.85,-0.65)
  {\scalebox{0.08}{
     \includegraphics{hacker.png}
   }
 }; 
\node[inner sep=0pt,align=center, scale=1.2, color=black] (hacker) at (0.07,0.12) {
$\vdots$
};  
\node[inner sep=0pt,align=center, scale=0.8, color=black] (hacker) at (0.07,1.15) {
Clients
};
\node[inner sep=0pt,align=center, scale=0.8, color=black] (hacker) at (0.85,-1.12) {
Attacker
};
\node[scale=0.08](n1r) at (1.5,0) {\router{}};
\end{tikzpicture}

%% file: tikz/example.tex
      \begin{tikzpicture}[fill=white, >=stealth,
    node distance=3cm,
    database/.style={
      cylinder,
      cylinder uses custom fill,
      shape border rotate=90,
      aspect=0.25,
      draw}]

    \tikzset{
node distance = 9em and 4em,
sloped,
   box/.style = {%
    shape=rectangle,
    rounded corners,
    draw=blue!40,
    fill=blue!15,
    align=center,
    font=\fontsize{12}{12}\selectfont},
 arrow/.style = {%
    line width=0.1mm,
    -{Triangle[length=5mm,width=2mm]},
    shorten >=1mm, shorten <=1mm,
    font=\fontsize{8}{8}\selectfont},
}

\draw[rounded corners=0.5ex, fill=OliveGreen!50!black!2] (-0.6,1.3) rectangle node (m1){} (5.15,1.8);
\node[server, scale=1, color=Red](n0) at (0,0) {};
\node[server, scale=1](n1) at (2,0) {};
\node[server, scale=1](n2) at (5,0) {};
\draw[->, black, thick, line width=0.27mm] (-0.43,0.6) to (-0.43,1.3);
\draw[<-, black, thick, line width=0.27mm] (-0.2,0.6) to (-0.2,1.3);

\draw[->, black, thick, line width=0.27mm] (1.57,0.6) to (1.57,1.3);
\draw[<-, black, thick, line width=0.27mm] (1.8,0.6) to (1.8,1.3);

\draw[->, black, thick, line width=0.27mm] (4.57,0.6) to (4.57,1.3);
\draw[<-, black, thick, line width=0.27mm] (4.8,0.6) to (4.8,1.3);

\node[inner sep=0pt,align=center, scale=2.2, color=black] (hacker) at (3.5,0) {
$\hdots$
};
\node[inner sep=0pt,align=center] (hacker2) at (-0.8,0)
  {\scalebox{0.11}{
     \includegraphics{hacker.png}
   }
 };

 \draw[-, color=black] (0.2,0) to (1.5,0);

 \draw[-, color=black] (2.2,0) to (2.85,0);
 \draw[-, color=black] (3.83,0) to (4.5,0);

\node[inner sep=0pt,align=center, scale=1, color=black] (hacker) at (-0.65,0.85) {
$z^{1}$
};
\node[inner sep=0pt,align=center, scale=1, color=black] (hacker) at (1.35,0.85) {
$z^{2}$
};
\node[inner sep=0pt,align=center, scale=1, color=black] (hacker) at (4.35,0.85) {
$z^{K}$
};

\node[inner sep=0pt,align=center, scale=1, color=black] (hacker) at (0.17,0.85) {
$u^{1}$
};
\node[inner sep=0pt,align=center, scale=1, color=black] (hacker) at (2.17,0.85) {
$u^{2}$
};
\node[inner sep=0pt,align=center, scale=1, color=black] (hacker) at (5.17,0.85) {
$u^{K}$
};

\node[inner sep=0pt,align=center, scale=1.1, color=black] (hacker) at (2.4,1.5) {
$\mu$
};

\node[inner sep=0pt,align=center, scale=1, color=black] (hacker) at (0.8,-0.2) {
$i^{1}=1$
};
\node[inner sep=0pt,align=center, scale=1, color=black] (hacker) at (2.8,-0.2) {
$i^{2}=0$
};
\node[inner sep=0pt,align=center, scale=1, color=black] (hacker) at (5.9,-0.2) {
$i^{K}=0$
};

\end{tikzpicture}

%% file: tikz/example2.tex
      \begin{tikzpicture}[fill=white, >=stealth,
    node distance=3cm,
    database/.style={
      cylinder,
      cylinder uses custom fill,
      shape border rotate=90,
      aspect=0.25,
      draw}]

    \tikzset{
node distance = 9em and 4em,
sloped,
   box/.style = {%
    shape=rectangle,
    rounded corners,
    draw=blue!40,
    fill=blue!15,
    align=center,
    font=\fontsize{12}{12}\selectfont},
 arrow/.style = {%
    line width=0.1mm,
    -{Triangle[length=5mm,width=2mm]},
    shorten >=1mm, shorten <=1mm,
    font=\fontsize{8}{8}\selectfont},
}

\draw[rounded corners=0.5ex, fill=RoyalAzure!10!white] (-1.5,0.76) rectangle node (m1){} (2.4,-0.5);
\draw[rounded corners=0.5ex, fill=RoyalAzure!10!white] (4.25,0.76) rectangle node (m1){} (7.5,-0.5);

\draw[rounded corners=0.5ex, fill=OliveGreen!50!black!2] (-0.6,1.3) rectangle node (m1){} (7,1.8);
\node[server, scale=1, color=Red](n0) at (0,0) {};
\node[server, scale=1](n1) at (2,0) {};
\node[server, scale=1](n2) at (5,0) {};
\node[server, scale=1](n3) at (7,0) {};
\draw[->, black, thick, line width=0.27mm] (-0.43,0.6) to (-0.43,1.3);
\draw[<-, black, thick, line width=0.27mm] (-0.2,0.6) to (-0.2,1.3);

\draw[->, black, thick, line width=0.27mm] (1.57,0.6) to (1.57,1.3);
\draw[<-, black, thick, line width=0.27mm] (1.8,0.6) to (1.8,1.3);

\draw[->, black, thick, line width=0.27mm] (4.57,0.6) to (4.57,1.3);
\draw[<-, black, thick, line width=0.27mm] (4.8,0.6) to (4.8,1.3);

\draw[->, black, thick, line width=0.27mm] (6.57,0.6) to (6.57,1.3);
\draw[<-, black, thick, line width=0.27mm] (6.8,0.6) to (6.8,1.3);

\node[inner sep=0pt,align=center, scale=2.2, color=black] (hacker) at (3.5,0) {
$\hdots$
};
\node[inner sep=0pt,align=center] (hacker2) at (-0.8,0)
  {\scalebox{0.11}{
     \includegraphics{hacker.png}
   }
 };

 \draw[-, color=black] (0.2,0) to (1.5,0);

 \draw[-, color=black] (2.2,0) to (2.85,0);
 \draw[-, color=black] (3.83,0) to (4.5,0);
 \draw[-, color=black] (5.2,0) to (6.5,0);

\node[inner sep=0pt,align=center, scale=1, color=black] (hacker) at (-0.65,1) {
$z^{1}$
};
\node[inner sep=0pt,align=center, scale=1, color=black] (hacker) at (1.35,1) {
$z^{2}$
};
\node[inner sep=0pt,align=center, scale=1, color=black] (hacker) at (4.15,1) {
$z^{K-1}$
};
\node[inner sep=0pt,align=center, scale=1, color=black] (hacker) at (6.35,1) {
$z^{K}$
};

\node[inner sep=0pt,align=center, scale=1, color=black] (hacker) at (0.17,1) {
$u^{1}$
};
\node[inner sep=0pt,align=center, scale=1, color=black] (hacker) at (2.17,1) {
$u^{2}$
};
\node[inner sep=0pt,align=center, scale=1, color=black] (hacker) at (5.42,1) {
$u^{K-1}$
};
\node[inner sep=0pt,align=center, scale=1, color=black] (hacker) at (7.17,1) {
$u^{K}$
};

\node[inner sep=0pt,align=center, scale=1.1, color=black] (hacker) at (3.5,1.5) {
$\mu$
};

\node[inner sep=0pt,align=center, scale=1, color=black] (hacker) at (0.9,0.3) {
\textit{Zone $1$}
};
\node[inner sep=0pt,align=center, scale=1, color=black] (hacker) at (5.9,0.3) {
\textit{Zone $V$}
};

\node[inner sep=0pt,align=center, scale=1, color=black] (hacker) at (0.8,-0.75) {
$y^{1}=1$
};
\node[inner sep=0pt,align=center, scale=1, color=black] (hacker) at (5.9,-0.75) {
$y^{V}=0$
};

\end{tikzpicture}

%% file: tikz/aggregation_3.tex
\begin{tikzpicture}

\node[scale=0.8] (kth_cr) at (-6.1,0)
{
  \begin{tikzpicture}
\draw[-, black, thick, line width=0.6mm] (11,1) to (17,1) to (17,4) to (11,4) to (11,1);
  \end{tikzpicture}
};

\node[scale=0.5] (kth_cr) at (-7.1,-0.51)
{
  \begin{tikzpicture}
\filldraw[fill=Red!20, thick] 
  (0,0) .. controls (1,1.5) and (2,1.2) .. (2.5,0.5) 
        .. controls (3,-0.5) and (1,-1.5) .. (0,-1)
        .. controls (-1,-0.5) and (-0.5,0.5) .. (0,0);
  \end{tikzpicture}
};

\node[scale=0.3] (kth_cr) at (-4.9,-0.38)
{
  \begin{tikzpicture}
\filldraw[fill=gray!50, thick] 
  (0,0) .. controls (1,2.2) and (3,1.8) .. (2.5,0)
        .. controls (2.7,-1.5) and (1,-1) .. (0,-1)
        .. controls (-0.5,-0.5) and (-0.5,0.5) .. (0,0);
  \end{tikzpicture}
};

\node[scale=0.3] (kth_cr) at (-4.5,0.7)
{
  \begin{tikzpicture}
\filldraw[fill=Blue!20, thick] 
  (0,0) .. controls (0.8,1.2) and (1.5,1.2) .. (2,0.6)
        .. controls (2.2,0.2) and (1.5,-0.2) .. (1,-0.4)
        .. controls (0.5,-0.6) and (1.5,-1.6) .. (0,-1.2)
        .. controls (-0.7,-0.5) and (-0.4,0.5) .. (0,0);
  \end{tikzpicture}
};

\node[scale=0.3] (kth_cr) at (-7.9,0.5)
{
  \begin{tikzpicture}
\filldraw[fill=OliveGreen!20, thick] 
  (0,0) -- (1.5,2) .. controls (2.5,1.5) and (2.8,0.5) .. (2,0)
        .. controls (1.8,-0.2) and (1.8,-0.8) .. (2.5,-1.5)
        .. controls (1.5,-1.2) and (0,-0.5) .. (0,0);
  \end{tikzpicture}
};

\node[scale=0.8] (kth_cr) at (0,0)
{
  \begin{tikzpicture}
\draw[-, black, thick, line width=0.6mm] (11,1) to (15,1) to (15,4) to (11,4) to (11,1);
\draw[-, black, thick, line width=0.6mm, fill=Red!20] (11,4) to (13,4) to (13,2.5) to (11,2.5);
\draw[-, black, thick, line width=0.6mm] (11,2.5) to (15,2.5);
\draw[-, black, thick, line width=0.6mm] (13,4) to (13,1);
\draw[-, black, thick, line width=0.6mm, fill=Blue!20] (13,4) to (15,4) to (15,2.5) to (13,2.5);
\draw[-, black, thick, line width=0.6mm, fill=OliveGreen!20] (13,2.5) to (15,2.5) to (15,1) to (13,1);
\draw[-, black, thick, line width=0.6mm, fill=Black!20] (11,2.5) to (13,2.5) to (13,1) to (11,1);
  \end{tikzpicture}
};

\node[scale=0.94] (kth_cr) at (4.8,0.05)
{
  \begin{tikzpicture}
\node[scale=1] (kth_cr) at (-0.8,4)
{
\begin{tikzpicture}[tdplot_main_coords, scale=3]
  \coordinate (A) at (0,0,0);
  \coordinate (B) at (1,0,0);
  \coordinate (C) at (0.5,{sqrt(3)/2},0);
  \coordinate (D) at (0.5,{sqrt(3)/6},{sqrt(6)/3});

  \draw[line width=0.37mm] (A) -- (B) -- (C) -- cycle;
  \draw[line width=0.37mm] (A) -- (D);
  \draw[line width=0.37mm] (B) -- (D);
  \draw[line width=0.37mm] (C) -- (D);

  \fill[white, opacity=0.3] (A) -- (B) -- (C) -- cycle;
  \fill[white, opacity=0.3] (A) -- (B) -- (D) -- cycle;
  \fill[white, opacity=0.3] (A) -- (C) -- (D) -- cycle;
  \fill[white, opacity=0.3] (B) -- (C) -- (D) -- cycle;

  \newcommand{\BaryPoint}[4]{%
    \coordinate (temp) at
      ($#1*(A) + #2*(B) + #3*(C) + #4*(D)$);
    \node[draw,circle,fill=black,scale=0.4] at (temp) {};
  }

  \node[draw,circle, fill=black, scale=0.4] at (A) {};
  \node[draw,circle, fill=black, scale=0.4] at (B) {};
  \node[draw,circle, fill=black, scale=0.4] at (C) {};
  \node[draw,circle, fill=black, scale=0.4] at (D) {};

  \node[draw,circle, fill=black, scale=0.4] at (0.5, 0) {};
  \node[draw,circle, fill=black, scale=0.4] at (-0.5, 0) {};
  \node[draw,circle, fill=black, scale=0.4] at (-0.5, -0.15) {};
  \node[draw,circle, fill=black, scale=0.4] at (-1.1, -0.33) {};

\end{tikzpicture}
};
\node[draw,circle, fill=black, scale=0.4] at (-1.1, 3.15) {};
  \node[draw,circle, fill=black, scale=0.4] at (-1.5, 2.72) {};
  \node[draw,circle, fill=black, scale=0.4] at (-0.7, 2.77) {};
  \node[draw,circle, fill=black, scale=0.4] at (0, 2.8) {};
  \node[draw,circle, fill=black, scale=0.4] at (0.25,3.5) {};
  \node[draw,circle, fill=black, scale=0.4] at (-0.15,4.1) {};
  \node[draw,circle, fill=black, scale=0.4] at (-0.6,4.8) {};
  \node[draw,circle, fill=black, scale=0.4] at (-1.25,4.8) {};
  \node[draw,circle, fill=black, scale=0.4] at (-1.6,4.1) {};
  \node[draw,circle, fill=black, scale=0.4] at (-1.9,3.5) {};
  \node[draw,circle, fill=black, scale=0.4] at (-0.8,3.45) {};
  \node[draw,circle, fill=black, scale=0.4] at (0,3.12) {};

\draw[-, black, line width=0.25mm, opacity=0.4] (0.5, 4.3) to (-1.1, 3.15);
\draw[-, black, line width=0.25mm, opacity=0.4] (0.5, 4.3) to (-1.5, 2.72);
\draw[-, black, line width=0.25mm, opacity=0.4] (0.5, 4.3) to (-0.7, 2.77);
\draw[-, black, line width=0.25mm, opacity=0.4] (0.5, 4.3) to (0, 2.8);
\draw[-, black, line width=0.25mm, opacity=0.4] (0.5, 4.3) to (0.25,3.5);
\draw[-, black, line width=0.25mm, opacity=0.4] (0.5, 4.3) to (-0.15,4.1);
\draw[-, black, line width=0.25mm, opacity=0.4] (0.5, 4.3) to (-0.6,4.8);
\draw[-, black, line width=0.25mm, opacity=0.4] (0.5, 4.3) to (-1.25,4.8);
\draw[-, black, line width=0.25mm, opacity=0.4] (0.5, 4.3) to (-1.6,4.1);
\draw[-, black, line width=0.25mm, opacity=0.4] (0.5, 4.3) to (-1.9,3.5);
\draw[-, black, line width=0.25mm, opacity=0.4] (0.5, 4.3) to (-0.8,3.45);
\draw[-, black, line width=0.25mm, opacity=0.4] (0.5, 4.3) to (0,3.12);
  \end{tikzpicture}
};

\node[inner sep=0pt,align=center, scale=1.2, rotate=0, opacity=1] (obs) at (6,-1.6)
{
};
\node[inner sep=0pt,align=center, scale=1.2, rotate=0, opacity=1] (obs) at (4.8,-1.6)
{
Feature belief space $Q$
};
\node[inner sep=0pt,align=center, scale=1.2, rotate=0, opacity=1] (obs) at (0.1,-1.6)
{
Feature space $\mathcal{F}$
};

\node[inner sep=0pt,align=left, scale=1.2, rotate=0, opacity=1] (obs) at (7.35,0.45)
{
Representative\\
feature \\
beliefs $\tilde{q}\in\Tilde{Q}$
};

\node[inner sep=0pt,align=center, scale=1.2, rotate=0, opacity=1] (obs) at (-6,-1.6)
{
State space $X$
};

\draw[-{Latex[length=3mm]}, black, bend left=20, line width=0.5mm] (-7.1, -0.35) to (-1.1,0.6);
\draw[-{Latex[length=3mm]}, black, bend left=20, line width=0.5mm] (-0.87, 0.6) to (4.73,0.3);


\node[inner sep=0pt,align=center, scale=1.2, rotate=0, opacity=1] (obs) at (-0.8,0.3)
{
$y$
};

\node[inner sep=0pt,align=center, scale=1.2, rotate=0, opacity=1] (obs) at (-7.1,-0.52)
{
$i$
};
\node[inner sep=0pt,align=center, scale=1.2, rotate=0, opacity=1] (obs) at (-6.9,0.28)
{
$I_y$
};

\end{tikzpicture}

%% file: tikz/aggregation_9.tex
\begin{tikzpicture}

\node[scale=0.9](states) at (4,-0.045) {
\begin{tikzpicture}
\node[draw,circle, scale=1.8](b1) at (0,0.3) {};
\node[draw,circle, scale=1.8](b2) at (4,0.3) {};
\node[draw,circle, scale=1.8](b3) at (4,-0.6) {};
\draw[fill=Black!10, scale=0.5] (0,-3) ellipse (1cm and 0.5cm) node (bb1) {};
\draw[fill=Black!10, scale=0.5] (8,-3) ellipse (1cm and 0.5cm) node (bb2) {};

\draw[-{Latex[length=1.85mm]}, black, line width=0.25mm] (b1) to (b2);
\draw[-{Latex[length=1.85mm]}, black, line width=0.25mm, dashed] (0.5,-1.5) to (3.5,-1.5);
\draw[-{Latex[length=1.85mm]}, black, line width=0.25mm] (0, -1.25) to (b1);
\draw[-{Latex[length=1.85mm]}, black, line width=0.25mm] (b2) to (b3);
\draw[-{Latex[length=1.85mm]}, black, line width=0.25mm] (b3) to (4, -1.25);

\node[inner sep=0pt,align=center, scale=0.9, rotate=0, opacity=1] (obs) at (0.05,0.3)
{
$b$
};
\node[inner sep=0pt,align=center, scale=0.9, rotate=0, opacity=1] (obs) at (4.05,-0.6)
{
$q$
};
\node[inner sep=0pt,align=center, scale=0.72, rotate=0, opacity=1] (obs) at (4.525,-0.6)
{
$Q$
};
\node[inner sep=0pt,align=center, scale=0.72, rotate=0, opacity=1] (obs) at (2.93,-0.6)
{
Feature belief
};
\node[inner sep=0pt,align=center, scale=0.9, rotate=0, opacity=1] (obs) at (4.05,0.3)
{
$b^{\prime}$
};

\node[inner sep=0pt,align=center, scale=0.9, rotate=0, opacity=1] (obs) at (0.03,-1.5)
{
$\tilde{q}$
};
\node[inner sep=0pt,align=center, scale=0.9, rotate=0, opacity=1] (obs) at (4.03,-1.5)
{
$\tilde{q}^{\prime}$
};

\node[inner sep=0pt,align=center, scale=0.72, rotate=0, opacity=1] (obs) at (2.06,-1.8)
{
Representative feature beliefs
};
\node[inner sep=0pt,align=center, scale=0.72, rotate=0, opacity=1] (obs) at (0,-1.95)
{
$\tilde{Q}$
};
\node[inner sep=0pt,align=center, scale=0.72, rotate=0, opacity=1] (obs) at (4,-1.95)
{
$\tilde{Q}$
};
\node[inner sep=0pt,align=center, scale=0.72, rotate=0, opacity=1] (obs) at (0,0.75)
{
$B$
};
\node[inner sep=0pt,align=center, scale=0.72, rotate=0, opacity=1] (obs) at (4,0.75)
{
$B$
};
\node[inner sep=0pt,align=center, scale=0.72, rotate=0, opacity=1] (obs) at (2,0.45)
{
Original beliefs
};

\node[inner sep=0pt,align=center, scale=0.72, rotate=0, opacity=1] (obs) at (2.05,0.08)
{
$b^{\prime}=F(b,u,z),\hat{g}(b,u)$
};

\node[inner sep=0pt,align=center, scale=0.75, rotate=0, opacity=1] (obs) at (-0.45,-0.68)
{
\eqqref{eq:belief_transition_2}
};

\node[inner sep=0pt,align=center, scale=0.75, rotate=0, opacity=1] (obs) at (4.6,-0.15)
{
\eqqref{eq:belief_transition_1}
};
\node[inner sep=0pt,align=center, scale=0.75, rotate=0, opacity=1] (obs) at (4.6,-1.08)
{
\eqqref{eq:belief_transition_3}
};
\end{tikzpicture}
};

%
\end{tikzpicture}

%% file: tikz/rep28.tex
\begin{tikzpicture}

\node[scale=1] (kth_cr) at (9.7,0)
{
  \begin{tikzpicture}
    \draw[-{Latex[length=1.5mm]}, black, thick, line width=0.3mm] (0,0) to (9,0);
    \draw[-{Latex[length=1.5mm]}, black, thick, line width=0.3mm] (0,0) to (0,2.0);

    \node[inner sep=0pt,align=center, scale=0.9] at (9.5,1.35)
    {\textcolor{Red}{$J^{\star}(b)$}};
    
    \node[inner sep=0pt,align=center, scale=0.9] at (9.5,0.5)
    {\textcolor{Blue}{$\Tilde{J}(b)$}};

    \node[inner sep=0pt,align=center, scale=0.9] at (9.7,0)
    {Beliefs $b$};
    
    \node[inner sep=0pt,align=center, scale=0.9] at (0.45,1.93)
    {Cost};

    \draw[Red, thick]
      plot coordinates {
        (0,0.45) (1.5,1.5) (2.3,1.65) (3.5,1.85) (5.2, 2.0) (7, 1.8) (9, 1.35)};

    \draw[Blue, thick]
      plot coordinates {
        (0,0.08) (1.5,0.8) (2.5,0.95) (3.5,1.05) (5.6, 1.15) (7, 1.0) (9, 0.5)};

    \draw[
      decorate,
      decoration={brace, mirror, amplitude=4pt},
      thick
    ] 
    (0, -0.15) -- (2.7, -0.15)
    node[midway, below=5pt]{$S_{\tilde{q}}$};

    \draw[
      decorate,
      decoration={brace, mirror, amplitude=4pt},
      thick
    ] 
    (2.7, -0.15) -- (5.9, -0.15)
    node[midway, below=5pt]{$S_{\tilde{q}^{\prime}}$};

    \draw[
      decorate,
      decoration={brace, mirror, amplitude=4pt},
      thick
    ] 
    (5.9, -0.15) -- (9, -0.15)
    node[midway, below=5pt]{$S_{\tilde{q}^{\prime\prime}}$};  

    \draw[-, black, thick, line width=0.3mm, dashed] (0,0.45) to (2.7, 0.45);
    \draw[{Latex[length=1.5mm]}-{Latex[length=1.5mm]}, black, thick, line width=0.3mm] (2.7,0.45) to (2.7, 1.7);

    \node[inner sep=0pt,align=center, scale=0.9] at (2.55,1.2)
    {$\epsilon_1$};

    \node[inner sep=0pt,align=center, scale=0.9] at (5.48,1.8)
    {$\epsilon_2$};

    \draw[-, black, thick, line width=0.3mm, dashed] (2.75,1.65) to (5.2, 1.65);
    \draw[{Latex[length=1.5mm]}-{Latex[length=1.5mm]}, black, thick, line width=0.3mm] (5.2,1.6) to (5.2, 2.0);

    \draw[-, black, thick, line width=0.3mm, dashed] (5.9,1.35) to (9, 1.35);
    \draw[{Latex[length=1.5mm]}-{Latex[length=1.5mm]}, black, thick, line width=0.3mm] (5.9,1.35) to (5.9, 1.9);

    \node[inner sep=0pt,align=center, scale=1] at (6.25,1.6)
    {$\epsilon_3$};

    \node[inner sep=0pt,align=center, scale=1] at (4.7,0.5)
    {$\epsilon = \max \{\epsilon_1, \epsilon_2, \epsilon_3\}$};

  \end{tikzpicture}
};

\end{tikzpicture}

%% file: tikz/pyramid.tex
      \begin{tikzpicture}[fill=white, >=stealth,
    node distance=3cm,
    database/.style={
      cylinder,
      cylinder uses custom fill,
      shape border rotate=90,
      aspect=0.25,
      draw}]

    \tikzset{
node distance = 9em and 4em,
sloped,
   box/.style = {%
    shape=rectangle,
    rounded corners,
    draw=blue!40,
    fill=blue!15,
    align=center,
    font=\fontsize{12}{12}\selectfont},
 arrow/.style = {%
    line width=0.1mm,
    -{Triangle[length=5mm,width=2mm]},
    shorten >=1mm, shorten <=1mm,
    font=\fontsize{8}{8}\selectfont},
}

\fill[Black!8]
(0,0) -- (10,0) -- (8.5,1.2) -- (1.5,1.2) -- cycle;

\fill[Blue!8]
(1.45,1.2) -- (8.55,1.2) -- (7.1,2.4) -- (2.9,2.4) -- cycle;

\fill[Red!8]
(2.9,2.4) -- (7.1,2.4) -- (5,4.2) -- cycle;


\draw[-, line width=0.3mm] (0,0) -- (10,0);
\draw[-, line width=0.3mm] (0,0) -- (5,4.2);
\draw[-, line width=0.3mm] (5,4.2) -- (10,0);
\draw[-, line width=0.3mm] (1.45,1.2) -- (8.55,1.2);
\draw[-, line width=0.3mm] (2.85,2.4)  -- (7.1,2.4);

\node[inner sep=0pt,align=center, scale=0.85, color=black] (hacker) at (4.6,0.7) {
  \textsc{offline policy optimization}
};
\node[inner sep=0pt,align=center, scale=0.85, color=black] (hacker) at (4.6,0.4) {
  \textit{Dynamic programming and aggregation}
};

\node[inner sep=0pt,align=center, scale=0.85, color=black] (hacker) at (5.08,1.72) {
  \textsc{online policy evaluation}
};

\node[inner sep=0pt,align=center, scale=0.85, color=black] (hacker) at (5.08,1.4) {
  \textit{Belief estimation and rollout simulations}
};

\node[inner sep=0pt,align=center, scale=0.85, color=black] (hacker) at (5,3.28) {
  \textsc{online}\\
\textsc{policy}\\
  \textsc{adaptation}
};
\node[inner sep=0pt,align=center, scale=0.85, color=black] (hacker) at (5,2.62) {
  \textit{Lookahead optimization}
};

\node[draw,circle, minimum width=10mm, scale=0.45, fill=white](s0) at (3.8,2.1) {\Large$x_{1}$};
\node[draw,circle, minimum width=10mm, scale=0.45, fill=white](s1) at (5,2.1) {\Large$x_{2}$};
\node[draw,circle, minimum width=10mm, scale=0.45, fill=white](s2) at (6.2,2.1) {\Large$x_{3}$};

\draw[-{Latex[length=2mm]}] (s0) to (s1);
\draw[-{Latex[length=2mm]}] (s1) to (s2);

\node[scale=0.35] (kth_cr) at (8,0.58)
{
\begin{tikzpicture}
\node[scale=1] (kth_cr) at (-0.8,4)
{
\begin{tikzpicture}[tdplot_main_coords, scale=3]
  \coordinate (A) at (0,0,0);
  \coordinate (B) at (1,0,0);
  \coordinate (C) at (0.5,{sqrt(3)/2},0);
  \coordinate (D) at (0.5,{sqrt(3)/6},{sqrt(6)/3});

  \draw[line width=0.37mm] (A) -- (B) -- (C) -- cycle;
  \draw[line width=0.37mm] (A) -- (D);
  \draw[line width=0.37mm] (B) -- (D);
  \draw[line width=0.37mm] (C) -- (D);

  \fill[white, opacity=0.3] (A) -- (B) -- (C) -- cycle;
  \fill[white, opacity=0.3] (A) -- (B) -- (D) -- cycle;
  \fill[white, opacity=0.3] (A) -- (C) -- (D) -- cycle;
  \fill[white, opacity=0.3] (B) -- (C) -- (D) -- cycle;

  \newcommand{\BaryPoint}[4]{%
    \coordinate (temp) at
      ($#1*(A) + #2*(B) + #3*(C) + #4*(D)$);
    \node[draw,circle,fill=black,scale=0.4] at (temp) {};
  }

  \node[draw,circle, fill=black, scale=0.4] at (A) {};
  \node[draw,circle, fill=black, scale=0.4] at (B) {};
  \node[draw,circle, fill=black, scale=0.4] at (C) {};
  \node[draw,circle, fill=black, scale=0.4] at (D) {};

  \node[draw,circle, fill=black, scale=0.4] at (0.5, 0) {};
  \node[draw,circle, fill=black, scale=0.4] at (-0.5, 0) {};
  \node[draw,circle, fill=black, scale=0.4] at (-0.5, -0.15) {};
  \node[draw,circle, fill=black, scale=0.4] at (-1.1, -0.33) {};

\end{tikzpicture}
};
\node[draw,circle, fill=black, scale=0.4] at (-1.1, 3.15) {};
  \node[draw,circle, fill=black, scale=0.4] at (-1.5, 2.72) {};
  \node[draw,circle, fill=black, scale=0.4] at (-0.7, 2.77) {};
  \node[draw,circle, fill=black, scale=0.4] at (0, 2.8) {};
  \node[draw,circle, fill=black, scale=0.4] at (0.25,3.5) {};
  \node[draw,circle, fill=black, scale=0.4] at (-0.15,4.1) {};
  \node[draw,circle, fill=black, scale=0.4] at (-0.6,4.8) {};
  \node[draw,circle, fill=black, scale=0.4] at (-1.25,4.8) {};
  \node[draw,circle, fill=black, scale=0.4] at (-1.6,4.1) {};
  \node[draw,circle, fill=black, scale=0.4] at (-1.9,3.5) {};
  \node[draw,circle, fill=black, scale=0.4] at (-0.8,3.45) {};
  \node[draw,circle, fill=black, scale=0.4] at (0,3.12) {};
\end{tikzpicture}
 };

\end{tikzpicture}

%% file: tikz/obs_dist.tex
\begin{tikzpicture}
\begin{axis}[
    ybar,
    bar width=10pt,
    ymin=0,
    ymax=0.55,
    xtick={0,1,...,7},
    xticklabels={0,1,2,3,4,5,6,7},    
    xtick=data,
    axis y line=center,
    axis x line=bottom,
    ytick=\empty,
    ylabel={},
    axis line style={-{Latex[length=2mm]}},
    width=11cm,
    height=3.8cm,
    enlarge x limits=0.1,
    legend style={/tikz/every node/.style={anchor=west}, at={(0.9, 0.87)}, legend columns=1, draw=none, /tikz/column 2/.style={
                column sep=10pt,
              }},
]
\addplot+[Black,fill=Blue!40, postaction={
        pattern=crosshatch
      }] coordinates {
(0.5, 0.4204381193405085)
(1.5, 0.22890519830761025)
(2.5, 0.14592706392110147)
(3.5, 0.09381025537785098)
(4.5, 0.05784965748300809)
(5.5, 0.03262720682041655)
(6.5, 0.015497923239697894)
(7.5, 0.0049445755098083705)
  };

\addplot+[Black,fill=Red!40, postaction={
        pattern=north west lines
      }] coordinates {
(0.5, 0.0909090909090911)
(1.5, 0.09497964721845345)
(2.5, 0.09997857601942456)
(3.5, 0.10636018725470703)
(4.5, 0.1149839862213048)
(5.5, 0.12775998469033872)
(6.5, 0.15030586434157509)
(7.5, 0.21472266334510712)
  };  
  \legend{Normal operation ($i^l=0$), Replica compromised ($i^l=1$)}
\end{axis}
\node[inner sep=0pt,align=center, scale=1, rotate=0, opacity=1] (obs) at (5.15,-0.75)
{
  Number of security alerts $z^l$
};
\node[inner sep=0pt,align=center, scale=1, rotate=0, opacity=1] (obs) at (1.95,2.2)
{
  Probability $p(z^l \mid i^l)$
};
\end{tikzpicture}

%% file: tikz/rep21.tex
\begin{tikzpicture}

\pgfplotstableread{
1 99.99999999999991 15.59032754707749
2 99.2 15.131978257743441
3 66.40000000000009 14.878149381086327
4 49.599999999999824 14.987060306964262
5 39.20000000000009 14.498233460516527
6 32.80000000000027 14.70076842778196
7 28.00000000000009 14.657026545439903
8 24.799999999999912 14.73899051984282
9 21.60000000000009 10.533810113630164
10 19.200000000000355 8.219827741472114
11 17.600000000000357 10.728272365688564
12 16.000000000000178 8.964688431215862
13 15.200000000000088 7.446058298316169
14 13.600000000000444 5.094839370515018
15 12.800000000000178 7.013823736191245
16 12.000000000000266 5.99527585134838
17 11.200000000000355 2.9475991171371874
18 10.400000000000444 4.181468576298
19 10.399999999999912 3.394966258060059
20 9.600000000000355 5.170459281044618
21 8.800000000000267 2.9488590156075887
22 8.800000000000267 3.0741377100116267
23 8.000000000000533 2.421367248432425
24 8.139183462754566 2.462223405937806
25 8.0 2.1548714186169207
26 7.200000000000444 1.7048844577456457
27 7.200000000000267 2.4017490821188527
28 6.400000000000356 2.4132089673194272
29 6.400000000000533 2.1457036606504314
30 6.400000000000356 1.7069564146969807
31 6.399999999999823 2.5481752474610175
32 5.717158898083413 1.5573232372447592
33 5.717158898083413 0.9668791245237678
34 5.717158898083413 1.223935063034876
35 5.6000000000000885 1.2393117396258653
36 4.994343713374678 1.9506808463542278
37 4.855548279298456 1.5212401454694877
38 4.800000000000355 1.6287327962923346
39 4.839553308995543 1.3855034887966635
40 4.8000000000001775 0.7453204724509295
41 4.8 0.9102934836873473
42 4.127786596361457 0.9889761791353227
43 4.127786596361457 0.9397184986737717
44 4.000000000000444 0.6485949986162787
45 4.0000000000002665 0.9692215802081883
46 4.0000000000002665 0.9095333920205562
47 4.280942581266341 1.1073576481560838
48 4.280942581266341 1.0641370079217936
49 4.0000000000002665 0.41350580908560985
50 3.484061714578243 0.4660114845968799
51 3.484061714578243 0.3144516625920488
52 3.3294013165507477 0.623688767071032
53 3.200000000000533 0.606890682372029
54 3.200000000000533 0.8482398156819251
55 3.509690318970901 0.9071086615244255
56 3.509690318970901 0.9139511735246106
57 3.200000000000533 0.705991934594012
58 3.2000000000003554 0.35590197605812435
59 3.2000000000003554 0.4779053747679676
60 3.200000000000178 0.6681266096253271
61 3.2 0.31412348327215867
62 3.2 0.4661298086756638
63 3.1155644298966196 0.6123980843925771
64 3.1155644298966196 0.7263208339706022
65 3.1155644298966196 0.6590344785725275
66 3.1155644298966196 1.0117981086298133
67 3.1155644298966196 0.14417136183732993
68 3.1155644298966196 0.3690584782405484
69 3.1155644298966196 0.392340501786947
70 3.1155644298966196 0.6812810745040636
71 3.1155644298966196 0.6841710556976395
72 3.1155644298966196 0.4850777707318148
73 3.1155644298966196 0.8152449280632368
74 3.1155644298966196 0.579229965223437
75 2.4000000000002664 0.8486623387421925
76 2.535418626203343 0.20017186801266362
77 2.4000000000002664 0.26360307342160283
78 2.4575387997581166 0.17540709903825302
79 2.6510512128378827 0.3050806695172401
80 2.4000000000000887 0.3103802853457722
81 2.4000000000000887 0.2299169756295285
82 2.5273270197672537 0.5549180546179482
83 2.4000000000000887 0.5384055754339307
84 2.053480616338808 0.7288207133024702
85 1.7894405940529252 0.06930921474868512
86 1.7894405940529252 0.06563405852317139
87 1.7894405940529252 0.1129987120571947
88 1.7829112807307352 0.07716354326968933
89 1.7829112807307352 0.2822739242762733
90 1.7947751770032367 0.2897277817406305
91 2.053480616338808 0.2953684139441606
92 2.053480616338808 0.2963595635153453
93 1.829954359268447 0.4521614906817355
94 1.7947751770032367 0.1122237639881849
95 1.7947751770032367 0.2118842889059689
96 1.7947751770032367 0.053151949838918355
97 1.7947751770032367 0.026845471426657497
98 1.6543828974629788 0.20457212165601923
99 1.6702477966033398 0.1681886720437511
100 2.053480616338808 0.3625992106552349
101 1.6543828974629788 0.37232372087489907
102 1.6797359597024537 0.40742121476015747
103 1.6797359597024537 0.04148427482165218
104 1.6797359597024537 0.041503413104981846
105 1.829954359268447 0.10623353532139745
106 1.829954359268447 0.12423021161826675
107 1.6541073553097958 0.29527722219572894
108 2.053480616338808 0.2396156241929699
109 1.6849659687199632 0.29691358652099886
110 1.7894405940529252 0.2119243055630804
111 1.7894405940529252 0.14996645406711728
112 1.7894405940529252 0.08289990848889595
}\datatable


\pgfplotsset{/dummy/workaround/.style={/pgfplots/axis on top}}

\node[scale=1] (kth_cr) at (5.7,-0.19)
{
\begin{tikzpicture}
  \begin{axis}
[
        xmin=1,
        xmax=105,
        ymax=120,
        width=11cm,
        height =3.5cm,
        axis y line=center,
        axis x line=bottom,
        scaled y ticks=false,
        yticklabel style={
        /pgf/number format/fixed,
        /pgf/number format/precision=5
      },
        xlabel style={below right},
        ylabel style={above left},
        axis line style={-{Latex[length=2mm]}},
        smooth,
        legend style={at={(0.95,0.8)}},
        legend columns=1,
        legend style={
          /tikz/every node/.style={anchor=west},
          draw=none,
            /tikz/column 2/.style={
                column sep=5pt,
              }
              }
              ]
            \addplot[Black,name path=l1, thick, dashed] table [x index=0, y index=1, domain=0:1] {\datatable};              
            \addplot[RoyalAzure,name path=l1, thick] table [x index=0, y index=2, domain=0:1] {\datatable};
\legend{Theoretical bound [cf. \propref{prop:aggregation_bound}], Actual approximation error}
\end{axis}
\node[inner sep=0pt,align=center, scale=1, rotate=0, opacity=1] (obs) at (2.48,1.9)
{
  $\norm{J^{\star}-\Tilde{J}}$ (approximation error)
};
\node[inner sep=0pt,align=center, scale=1, rotate=0, opacity=1] (obs) at (4.7,-0.8)
{
  Discretization resolution $\rho$
};
\end{tikzpicture}
};
\end{tikzpicture}

%% file: tikz/rep20.tex
\begin{tikzpicture}

\pgfplotstableread{
0 1
1 2
2 3
3 4
4 5
5 6
6 7
7 8
8 9
}\statetwo

\pgfplotstableread{
0 1
1 4
2 10
3 20
4 35
5 56
6 84
7 120
8 165
}\statefour

\pgfplotstableread{
0 1
1 8
2 36
3 120
4 330
5 792
6 1716
7 3432
8 6435
}\stateeight

\pgfplotstableread{
1 6.12
2 6.12
3 4.12
4 4.62
5 4.01
6 3.93
7 3.45
8 3.33
9 3.04
10 2.85
11 2.73
12 2.5
13 2.44
14 2.28
15 2.25
16 2.14
17 2.03
18 1.96
19 1.87
20 1.83
21 1.53
22 1.25
23 1.26
24 0.96
25 0.9
26 0.73
27 0.73
28 0.62
29 0.62
30 0.57
31 0.53
32 0.54
33 0.46
34 0.48
35 0.4
36 0.41
37 0.42
38 0.4
39 0.41
40 0.35
41 0.36
42 0.35
43 0.35
44 0.35
45 0.35
46 0.37
47 0.36
48 0.35
49 0.33
50 0.33
51 0.3
52 0.29
53 0.3
54 0.3
55 0.31
56 0.29
57 0.28
58 0.27
59 0.27
60 0.27
61 0.28
62 0.27
63 0.28
64 0.28
65 0.28
66 0.28
67 0.26
68 0.27
69 0.27
70 0.28
71 0.29
72 0.29
73 0.29
74 0.28
75 0.28
76 0.28
77 0.28
78 0.28
79 0.28
80 0.27
81 0.27
82 0.28
83 0.27
84 0.28
85 0.27
86 0.26
87 0.26
88 0.26
89 0.27
90 0.26
91 0.26
92 0.26
93 0.26
94 0.26
95 0.25
96 0.25
97 0.25
98 0.25
99 0.25
100 0.26
101 0.26
102 0.26
103 0.26
104 0.25
105 0.25
106 0.25
107 0.26
108 0.26
109 0.26
110 0.26
111 0.26
112 0.26
113 0.25
114 0.24
115 0.24
116 0.24
117 0.25
118 0.25
119 0.25
120 0.24
}\solvetwo

\pgfplotstableread{
1 10.69
2 6.77
3 4.77
4 3.92
5 3.55
6 3.09
7 2.78
8 2.59
9 2.31
10 2.19
11 2.01
12 1.85
13 1.76
14 1.68
15 1.62
16 1.56
17 1.46
18 1.43
19 1.38
20 1.33
21 0.81
22 0.69
23 0.68
24 0.64
25 0.55
26 0.54
27 0.51
28 0.46
29 0.48
30 0.44
31 0.45
32 0.47
33 0.46
34 0.46
35 0.44
36 0.42
37 0.44
38 0.42
39 0.4
40 0.4
41 0.41
42 0.42
43 0.4
44 0.39
45 0.4
46 0.4
47 0.39
48 0.42
49 0.41
50 0.41
51 0.41
52 0.41
53 0.4
54 0.39
55 0.39
56 0.4
57 0.4
58 0.4
59 0.41
60 0.41
61 0.42
62 0.42
63 0.41
64 0.41
65 0.41
66 0.4
67 0.4
68 0.39
69 0.39
70 0.39
71 0.38
72 0.39
73 0.39
74 0.38
75 0.39
76 0.39
77 0.39
78 0.39
79 0.38
80 0.38
81 0.37
82 0.37
83 0.38
84 0.38
85 0.38
86 0.39
87 0.38
88 0.39
89 0.39
90 0.39
91 0.39
92 0.38
93 0.38
94 0.38
95 0.38
96 0.38
97 0.38
98 0.38
99 0.38
100 0.38
101 0.38
102 0.39
103 0.38
104 0.38
105 0.38
106 0.38
107 0.38
108 0.38
109 0.37
110 0.37
111 0.37
112 0.38
113 0.38
114 0.38
115 0.38
116 0.38
117 0.38
118 0.38
119 0.38
120 0.38
}\solvefour

\pgfplotstableread{
1 8.03
2 5.93
3 4.53
4 3.61
5 3.06
6 2.6
7 2.26
8 2.08
9 1.94
10 1.77
11 1.68
12 1.55
13 1.43
14 1.36
15 1.32
16 1.26
17 1.22
18 1.18
19 1.12
20 1.09
21 0.71
22 0.54
23 0.48
24 0.47
25 0.44
26 0.45
27 0.47
28 0.44
29 0.43
30 0.43
31 0.41
32 0.42
33 0.45
34 0.44
35 0.42
36 0.42
37 0.41
38 0.41
39 0.42
40 0.42
41 0.41
42 0.41
43 0.4
44 0.4
45 0.4
46 0.4
47 0.4
48 0.4
49 0.4
50 0.4
51 0.41
52 0.41
53 0.41
54 0.41
55 0.4
56 0.41
57 0.4
58 0.4
59 0.4
60 0.4
61 0.4
62 0.4
63 0.4
64 0.4
65 0.4
66 0.4
67 0.39
68 0.39
69 0.39
70 0.39
71 0.39
72 0.39
73 0.38
74 0.39
75 0.39
76 0.38
77 0.39
78 0.39
79 0.39
80 0.39
81 0.39
82 0.39
83 0.39
84 0.39
85 0.39
86 0.39
87 0.39
88 0.39
89 0.39
90 0.39
91 0.39
92 0.39
93 0.39
94 0.39
95 0.39
96 0.39
97 0.39
98 0.39
99 0.39
100 0.39
101 0.39
102 0.38
103 0.38
104 0.38
105 0.38
106 0.38
107 0.38
108 0.38
109 0.38
110 0.38
111 0.38
112 0.38
113 0.38
114 0.38
115 0.38
116 0.38
117 0.37
118 0.37
119 0.37
120 0.37
}\solveeight

\pgfplotstableread{
1 0.002092123031616211
2 0.003713846206665039
3 0.0038819313049316406
4 0.00777125358581543
5 0.011507034301757812
6 0.016611099243164062
7 0.02157282829284668
8 0.029363155364990234
9 0.03541398048400879
10 0.041932106018066406
}\vione

\pgfplotstableread{
1 0.1882920265197754
2 1.1791269779205322
3 4.848140001296997
4 16.423473834991455
5 47.63405084609985
6 126.22407865524292
7 289.82488918304443
8 645.9581429958344
9 1358.296599149704
}\vitwo

\pgfplotstableread{
1 0.8174553979430499
2 0.5277080386539204
3 0.4854515188566065
4 0.3271052118603257
5 0.21198936265286528
6 0.25357901350799983
7 0.22667558802944976
8 0.08996632869685334
9 0.20500989335070316
10 0.23030856291524912
11 0.21220307801542299
12 0.09035471292412017
13 0.16106149077505214
14 0.09563208319555039
15 0.07674626787741051
16 0.012792905270250843
17 0.07727795155661105
18 0.05024301265141917
19 0.026530299756261616
20 0.029875320741394262
21 0.02975385508367634
22 0.01108851154605639
23 0.000229497966178549
24 0.00935816225852154
25 0.008618224201830671
26 0.019104638089423698
27 0.046337821616816266
28 0.04541117345723611
29 0.015075318910745678
30 0.003344378758808886
31 0.010422806476308583
32 0.012655280641803185
33 0.01377959272979934
34 0.03074515584070675
35 0.010896367663508294
36 0.011058869433781537
37 0.016834969799636917
38 0.0521375981462969
39 0.05982065321407434
40 0.047564151673513494
41 0.07590018473852442
42 0.09230291773725319
43 0.05995467466802837
44 0.08983198287772234
45 0.08342883936188303
46 0.07944728812603918
47 0.07039239540798903
48 0.09259640119572113
49 0.05975773867719164
50 0.06595955715967924
}\policyone

\pgfplotstableread{
1 1.2583676643087585
2 0.7532216025677129
3 0.638458127809979
4 0.533530357052362
5 0.1507599166845921
6 0.06256073653859406
7 0.14183776087532485
8 0.131100091
9 0.1678192015342
10 0.16125102110
}\policytwo

\pgfplotstableread{
1 3.2019309997558594
2 3.339353084564209
3 3.1998019218444824
4 3.3643980026245117
5 3.187494993209839
6 3.3815529346466064
7 3.2713332176208496
8 3.269848108291626
9 3.3910391330718994
10 3.4765219688415527
}\policyevalone

\pgfplotstableread{
1 3.709481954574585
2 3.9555530548095703
3 3.9309158325195312
4 4.086724758148193
5 4.755553960800171
6 4.532117128372192
7 4.580238103866577
8 5.3591601848602295
9 5.988803148269653
10 6.23109192851254
}\policyevaltwo

\pgfplotstableread{
1 0.729
2 0.486
3 0.585
4 0.585
5 0.486
6 0.486
7 0.484
8 0.5
9 0.486
10 0.486
11 0.483
12 0.484
13 0.5
14 0.486
15 0.483
16 0.484
17 0.484
18 0.486
19 0.483
20 0.484
21 0.484
22 0.5
23 0.484
24 0.483
25 0.484
26 0.484
27 0.484
28 0.483
29 0.484
30 0.484
31 0.484
32 0.483
33 0.483
34 0.484
35 0.484
36 0.483
37 0.483
38 0.484
39 0.484
40 0.483
41 0.483
42 0.483
43 0.484
44 0.484
45 0.483
46 0.483
47 0.484
48 0.484
49 0.483
50 0.483
}\basecost

\pgfplotstableread{
1 0.484
2 0.484
3 0.483
4 0.483
5 0.484
6 0.484
7 0.482
8 0.483
9 0.484
10 0.484
11 0.482
12 0.482
13 0.483
14 0.484
15 0.482
16 0.482
17 0.482
18 0.484
19 0.482
20 0.482
21 0.482
22 0.483
23 0.483
24 0.482
25 0.482
26 0.482
27 0.483
28 0.482
29 0.482
30 0.482
31 0.482
32 0.482
33 0.482
34 0.482
35 0.482
36 0.482
37 0.482
38 0.482
39 0.482
40 0.482
41 0.482
42 0.482
43 0.482
44 0.483
45 0.482
46 0.482
47 0.482
48 0.483
49 0.482
50 0.482
}\rolloutcost

\pgfplotstableread{
1 0.484
2 0.482
3 0.483
4 0.483
5 0.482
6 0.482
7 0.482
8 0.482
9 0.482
10 0.482
11 0.482
12 0.482
13 0.482
14 0.482
15 0.482
16 0.482
17 0.482
18 0.482
19 0.482
20 0.482
21 0.482
22 0.482
23 0.482
24 0.482
25 0.482
26 0.482
27 0.482
28 0.482
29 0.482
30 0.482
31 0.482
32 0.482
33 0.482
34 0.482
35 0.482
36 0.482
37 0.482
38 0.482
39 0.482
40 0.482
41 0.482
42 0.482
43 0.482
44 0.482
45 0.482
46 0.482
47 0.482
48 0.482
49 0.482
50 0.482
}\rollouttwocost

\pgfplotstableread{
1 0.484
2 0.484
3 0.483
4 0.483
5 0.484
6 0.484
7 0.482
8 0.483
9 0.484
10 0.484
11 0.482
12 0.482
13 0.483
14 0.484
15 0.482
}\truncatedrolloutcost

\pgfplotstableread{
1 0.485
2 0.482
3 0.483
4 0.483
5 0.482
6 0.482
7 0.482
8 0.482
9 0.482
10 0.482
11 0.482
12 0.482
13 0.482
14 0.482
15 0.482
16 0.482
17 0.482
18 0.482
19 0.482
20 0.482
}\truncatedrolloutcosttwo

\pgfplotstableread{
1 0.584
2 0.482
3 0.483
4 0.483
5 0.482
6 0.482
7 0.482
8 0.483
9 0.482
10 0.482
}\cerolloutcost

\pgfplotstableread{
1 0.485
2 0.482
3 0.483
4 0.483
5 0.482
6 0.482
7 0.482
8 0.483
9 0.482
10 0.482
}\cerolloutcosttwo

\pgfplotstableread{
1 0.486
2 0.483
3 0.483
4 0.483
5 0.484
6 0.483
7 0.484
8 0.483
9 0.483
10 0.483
11 0.483
12 0.483
13 0.483
14 0.483
15 0.483
}\mcrolloutcosttwo

\pgfplotstableread{
1 0.486
2 0.484
3 0.483
4 0.483
5 0.484
6 0.484
7 0.484
8 0.483
9 0.484
10 0.484
11 0.485
12 0.483
13 0.484
14 0.484
15 0.485
}\mcrolloutcost


\pgfplotsset{/dummy/workaround/.style={/pgfplots/axis on top}}

\node[scale=1] (kth_cr) at (0,0.065)
{
\begin{tikzpicture}
  \begin{axis}
[
        xmin=1,
        xmax=8,
        ymax=8500,
        ymode=log,
        width=10.25cm,
        height =3.5cm,
        axis y line=center,
        axis x line=bottom,
        scaled y ticks=false,
        yticklabel style={
        /pgf/number format/fixed,
        /pgf/number format/precision=5
      },
        xlabel style={below right},
        ylabel style={above left},
        axis line style={-{Latex[length=2mm]}},
        legend style={at={(0.67,1.08)}},
        legend columns=3,
        legend style={
          draw=none,
            /tikz/column 2/.style={
                column sep=5pt,
              }
              }
              ]
              \addplot[RoyalAzure,name path=l1, thick, mark=diamond, mark repeat=2, samples=100] table [x index=0, y index=1, domain=0:1] {\statetwo};
              \addplot[Red,name path=l1, thick, mark=x, mark repeat=2, samples=100] table [x index=0, y index=1, domain=0:1] {\statefour};
              \addplot[Black,name path=l1, thick, mark=triangle, mark repeat=2, samples=100] table [x index=0, y index=1, domain=0:1] {\stateeight};
              \legend{$n=2$, $n=4$, $n=8$}
            \end{axis}
\node[inner sep=0pt,align=center, scale=1, rotate=0, opacity=1] (obs) at (0.25,2.12)
{
  $|\Tilde{Q}|$
};
\node[inner sep=0pt,align=center, scale=1, rotate=0, opacity=1] (obs) at (4.7,-0.8)
{
  Discretization resolution $\rho$
};
\end{tikzpicture}
};
\end{tikzpicture}

%% file: tikz/value_fun.tex
\begin{tikzpicture}
\pgfplotstableread{
0.0 7.88818
0.010101010101010102 7.9060062626262635
0.020202020202020204 7.923796969696969
0.030303030303030304 7.941425454545454
0.04040404040404041 7.958910202020202
0.05050505050505051 7.9763452525252525
0.06060606060606061 7.993780303030304
0.07070707070707072 8.011213636363637
0.08080808080808081 8.028632222222223
0.09090909090909091 8.04605
0.10101010101010102 8.061807373737373
0.11111111111111112 8.076841111111111
0.12121212121212122 8.091872121212122
0.13131313131313133 8.106901212121212
0.14141414141414144 8.121911818181818
0.15151515151515152 8.136719090909091
0.16161616161616163 8.151526363636364
0.17171717171717174 8.166306565656566
0.18181818181818182 8.180906363636364
0.19191919191919193 8.195500303030304
0.20202020202020204 8.210082424242424
0.21212121212121213 8.224664545454544
0.22222222222222224 8.237346666666667
0.23232323232323235 8.249315151515152
0.24242424242424243 8.261283636363636
0.25252525252525254 8.273251717171716
0.26262626262626265 8.284047474747474
0.27272727272727276 8.263845454545455
0.2828282828282829 8.243643434343435
0.29292929292929293 8.223441414141414
0.30303030303030304 8.203239393939395
0.31313131313131315 8.183037373737374
0.32323232323232326 8.162835353535353
0.33333333333333337 8.142633333333334
0.3434343434343435 8.122431313131315
0.3535353535353536 8.102229292929293
0.36363636363636365 8.082027272727274
0.37373737373737376 8.061825252525253
0.38383838383838387 8.041623232323232
0.393939393939394 8.021421212121211
0.4040404040404041 8.001219191919192
0.4141414141414142 7.981017171717172
0.42424242424242425 7.960815151515151
0.43434343434343436 7.940613131313132
0.4444444444444445 7.920411111111112
0.4545454545454546 7.900209090909092
0.4646464646464647 7.8800070707070695
0.4747474747474748 7.8598050505050505
0.48484848484848486 7.839603030303031
0.494949494949495 7.81940101010101
0.5050505050505051 7.79919898989899
0.5151515151515152 7.77899696969697
0.5252525252525253 7.75879494949495
0.5353535353535354 7.738592929292929
0.5454545454545455 7.718390909090909
0.5555555555555556 7.69818888888889
0.5656565656565657 7.677986868686869
0.5757575757575758 7.657784848484848
0.5858585858585859 7.637582828282829
0.595959595959596 7.617380808080808
0.6060606060606061 7.5971787878787875
0.6161616161616162 7.576976767676768
0.6262626262626263 7.556774747474748
0.6363636363636365 7.536572727272727
0.6464646464646465 7.516370707070708
0.6565656565656566 7.496168686868687
0.6666666666666667 7.475966666666666
0.6767676767676768 7.4557646464646465
0.686868686868687 7.435562626262627
0.696969696969697 7.415360606060607
0.7070707070707072 7.395158585858585
0.7171717171717172 7.374956565656566
0.7272727272727273 7.354754545454545
0.7373737373737375 7.334552525252525
0.7474747474747475 7.3143505050505055
0.7575757575757577 7.294148484848485
0.7676767676767677 7.273946464646464
0.7777777777777778 7.253744444444445
0.787878787878788 7.233542424242424
0.797979797979798 7.213340404040404
0.8080808080808082 7.193138383838384
0.8181818181818182 7.172936363636364
0.8282828282828284 7.152734343434344
0.8383838383838385 7.132532323232324
0.8484848484848485 7.112330303030303
0.8585858585858587 7.092128282828283
0.8686868686868687 7.0719262626262624
0.8787878787878789 7.051724242424243
0.888888888888889 7.031522222222223
0.8989898989898991 7.011320202020203
0.9090909090909092 6.991118181818182
0.9191919191919192 6.970916161616161
0.9292929292929294 6.950714141414141
0.9393939393939394 6.9305121212121215
0.9494949494949496 6.910310101010102
0.9595959595959597 6.890108080808082
0.9696969696969697 6.869906060606061
0.9797979797979799 6.84970404040404
0.98989898989899 6.82950202020202
1.0 6.8093
}\tenobs

\pgfplotstableread{
1.0 6.501701261079517
0.99 6.521701261079516
0.98 6.541701261079518
0.97 6.561701261079517
0.96 6.581701261079517
0.95 6.6017012610795165
0.94 6.621701261079517
0.93 6.6417012610795165
0.92 6.661701261079517
0.91 6.681701261079518
0.9 6.701701261079517
0.89 6.721701261079517
0.88 6.741701261079517
0.87 6.761701261079518
0.86 6.781701261079517
0.85 6.801701261079517
0.84 6.821701261079517
0.83 6.841701261079516
0.82 6.861701261079517
0.81 6.8817012610795185
0.8 6.901701261079518
0.79 6.9217012610795186
0.78 6.941701261079517
0.77 6.961701261079518
0.76 6.981701261079518
0.75 7.0017012610795195
0.74 7.021701261079517
0.73 7.041701261079518
0.72 7.061701261079518
0.71 7.081701261079518
0.7 7.101701261079517
0.69 7.121701261079516
0.68 7.141701261079519
0.67 7.161701261079518
0.66 7.1817012610795175
0.65 7.20170126107952
0.64 7.2217012610795175
0.63 7.241701261079518
0.62 7.2617012610795175
0.61 7.281701261079518
0.6 7.301701261079518
0.59 7.321701261079518
0.58 7.3417012610795185
0.57 7.361701261079517
0.56 7.3817012610795185
0.55 7.401701261079519
0.54 7.42170126107952
0.53 7.44170126107952
0.52 7.461701261079519
0.51 7.481701261079519
0.5 7.501701261079519
0.49 7.521701261079518
0.48 7.541701261079519
0.47 7.561701261079519
0.46 7.5817012610795205
0.45 7.601701261079518
0.44 7.621701261079519
0.43 7.6417012610795165
0.42 7.661701261079519
0.41 7.681701261079517
0.4 7.701701261079521
0.39 7.72170126107952
0.38 7.741701261079517
0.37 7.761701261079518
0.36 7.78170126107952
0.35 7.801701261079519
0.34 7.82170126107952
0.33 7.841701261079519
0.32 7.861701261079522
0.31 7.881701261079519
0.3 7.901701261079518
0.29 7.921701261079519
0.28 7.941701261079521
0.27 7.9617012610795195
0.26 7.9740428198831586
0.25 7.962553594933103
0.24 7.9505843018639
0.23 7.937111694532879
0.22 7.926623740631492
0.21 7.914603135679306
0.2 7.89744165963759
0.19 7.882176958042314
0.18 7.869137402718916
0.17 7.852418019911841
0.16 7.843116545242116
0.15 7.82565682094244
0.14 7.809539358796806
0.13 7.796778847676673
0.12 7.783607071883878
0.11 7.771208805205669
0.1 7.751420004252342
0.09 7.737389779002392
0.08 7.722039629414349
0.07 7.706250659576993
0.06 7.684616684666645
0.05 7.671649147517737
0.04 7.653703813498719
0.03 7.639553195312077
0.02 7.61750687779808
0.01 7.601861147589515
0.0 7.5872679262457865
}\bnone

\pgfplotstableread{
1.0 4.521918213654312
0.8 4.921918213654313
0.6 5.321918213654314
0.4 5.721918213654316
0.2 6.034358495202882
0.0 5.587627327736727
}\bnfive

\pgfplotsset{/dummy/workaround/.style={/pgfplots/axis on top}}

\node[scale=1] (kth_cr) at (0,2.15)
{
  \begin{tikzpicture}
    \begin{axis}[
      xmin=0,
      xmax=1.05,
      ymin=4.2,
      ymax=9.5,
      xtick={0.5, 1},
        axis lines=center,
        xlabel style={below right},
        ylabel style={above left},
        axis line style={-{Latex[length=2mm]}},
        smooth,
        legend style={at={(0.9,1.15)}},
        legend columns=3,
        legend style={
            /tikz/column 2/.style={
                column sep=5pt,
              },
              draw=none
            },
        width=11.5cm,
        height=3.5cm,
            ]
\addplot[Black,name path=l1, thick] table [x index=0, y index=1, domain=0:1] {\tenobs};
\addplot[Red,densely dotted, name path=l1, thick, densely dotted] table [x index=0, y index=1, domain=0:1] {\bnone};
\addplot[densely dashed,name path=l1, thick] table [x index=0, y index=1, domain=0:1] {\bnfive};
\legend{$J^{\star}(b)$, $\Tilde{J}(b) \text{ }(\rho=100)$, $\Tilde{J}(b) \text{ }(\rho=5)$}
\end{axis}
\node[inner sep=0pt,align=center, scale=1, rotate=0, opacity=1] (obs) at (10.36,0)
{
  $b(1)$
};
\end{tikzpicture}
};
\end{tikzpicture}

%% file: tikz/rollout_eval.tex
\begin{tikzpicture}

\pgfplotstableread{
1 0.0007
2 0.0031
3 0.0160
4 0.0991
5 0.5995
6 3.5346
7 21.170
8 122.590
9 744.018
10 4912.01
}\rolloutellone

\pgfplotstableread{
1 0.0057
2 0.1928
3 6.8623
4 252.75
5 10584.39
}\rolloutelltwo

\pgfplotstableread{
1 0.15
2 31.85
3 6592.10
}\rolloutellthree

\pgfplotstableread{
1 0.0044
5 0.0111
10 0.0191
15 0.0309
20 0.0356
25 0.0448
30 0.0506
35 0.0596
40 0.0663
45 0.0758
50 0.0868
55 0.0918
60 0.0995
65 0.1089
70 0.1145
75 0.1245
80 0.1311
85 0.1415
90 0.1465
95 0.1540
100 0.1663
}\rollouthorizonone

\pgfplotstableread{
1 0.0473
5 0.1257
10 0.2342
15 0.3278
20 0.4254
25 0.5320
30 0.6355
35 0.7199
40 0.8270
45 0.9189
50 1.0423
55 1.1190
60 1.2079
65 1.3179
70 1.4119
75 1.4983
80 1.6082
85 1.7008
90 1.7835
95 1.9150
100 2.0147
}\rollouthorizontwo

\pgfplotstableread{
1 1.3107
5 3.7309
10 6.8179
15 9.7937
20 12.8156
25 15.6759
30 18.7750
35 21.7000
40 24.6037
45 29
50 32
55 35
60 38
65 41
70 44
75 49
80 52
85 55
90 58
95 61
100 64
}\rollouthorizonthree

\pgfplotstableread{
1 41.647 4.901
2 20.985 4.901
3 18.847 4.901
4 16.338 4.901
5 14.745 4.901
6 10.244 4.901
7 10.244 4.901
8 10.244 4.901
9 7.085 4.901
10 7.085 4.901
11 7.085 4.901
12 7.085 4.901
13 7.085 4.901
14 7.085 4.901
15 7.085 4.901
16 6.824 4.901
17 5.71 3.813
18 5.71 3.813
19 4.314 2.725
20 4.314 2.725
21 4.314 2.725
22 4.314 2.725
23 4.314 2.725
24 4.314 2.725
25 4.314 2.725
26 2.62 2.62
27 2.62 2.62
28 2.62 2.62
29 2.62 2.62
30 2.62 2.62
}\rolloutaggone

\pgfplotstableread{
1 0.0017
2. 0.09
3 2.59
4 34.56
5 10000
}\rolloutcontrol

\pgfplotstableread{
1 0.006
2 0.01
3 0.05
4 0.44
5 36.32
}\multiagentcontrol

\pgfplotsset{/dummy/workaround/.style={/pgfplots/axis on top}}

\node[scale=1] (kth_cr) at (0,-7.5)
{
\begin{tikzpicture}
  \begin{axis}
[
        xmin=1,
        xmax=32,
        ymax=70,
        ymode=log,
        width=11.5cm,
        height=3.5cm,
        axis y line=center,
        axis x line=bottom,
        scaled y ticks=false,
        yticklabel style={
        /pgf/number format/fixed,
        /pgf/number format/precision=5
      },
        xlabel style={below right},
        ylabel style={above left},
        axis line style={-{Latex[length=2mm]}},
        legend style={at={(1,0.95)}},
        legend columns=2,
        legend style={
          draw=none,
            /tikz/column 2/.style={
                column sep=5pt,
              }
              }
              ]
              \addplot[RoyalAzure,name path=l1, thick, mark=diamond, mark repeat=1, samples=100] table [x index=0, y index=1, domain=0:1] {\rolloutaggone};
              \addplot[Red,name path=l1, thick, mark=x, mark repeat=1, samples=100] table [x index=0, y index=2, domain=0:1] {\rolloutaggone};
              \legend{Base policy $\mu$ [\eqqref{eq:approximation_1}], Rollout policy $\tilde{\mu}$ [\eqqref{eq:rollout}]}
            \end{axis}
\node[inner sep=0pt,align=center, scale=1, rotate=0, opacity=1] (obs) at (1.2,1.9)
{
  Cost ($\downarrow$ better)
};
\node[inner sep=0pt,align=center, scale=1, rotate=0, opacity=1] (obs) at (5.1,-0.7)
{
  Discretization resolution $\rho$ 
};
\end{tikzpicture}
};

\draw[-{Latex[length=2mm]}, line width=0.22mm, color=black] (-4.2,-6.78) to (-4.2,-7.68);
\node[inner sep=0pt,align=center, scale=0.8, rotate=0, opacity=1] (obs) at (-3.57,-7.37)
{
  8.5x\\
  reduction
};

\end{tikzpicture}

%% file: tikz/scalability.tex
\begin{tikzpicture}
\pgfplotstableread{
1 0.0011
2 0.0028
4 0.0151
6 0.0779
8 0.3838
10 1.7408
}\scalesimVone

\pgfplotstableread{
1 0.0011
2 0.0103
4 0.0575
6 0.2945
8 1.4602
10 6.7797
}\scalesimVtwo

\pgfplotstableread{
1 0.0012
2 0.0102
4 1.2510
6 600.000000
8 600.000000
10 600.000000
}\scalesimVK

\pgfplotstableread{
1 0.790000
2 600.000000
3 600.000000
4 600.000000
5 600.000000
6 600.000000
7 600.000000
8 600.000000
9 600.000000
10 600.000000
}\scalepomdpsolve

\pgfplotsset{/dummy/workaround/.style={/pgfplots/axis on top}}

\node[scale=1] (kth_cr) at (0,0.065)
{
\begin{tikzpicture}
  \begin{axis}
[
        ymax=50,
        ymode=log,
        width=10.25cm,
        height =3.5cm,
        axis y line=center,
        axis x line=bottom,
        scaled y ticks=false,
        yticklabel style={
        /pgf/number format/fixed,
        /pgf/number format/precision=5
      },
        xlabel style={below right},
        ylabel style={above left},
        axis line style={-{Latex[length=2mm]}},
        legend style={at={(0.85,-0.55)}, cells={anchor=west}},
        legend columns=2,
        legend style={
          draw=none,
            /tikz/column 2/.style={
                column sep=5pt,
              }
              }
              ]
            \addplot[RoyalAzure,name path=l1, thick, mark=diamond] table [x index=0, y index=1, domain=0:1] {\scalesimVtwo};
            \addplot[Red,name path=l1, thick, mark=x] table [x index=0, y index=1, domain=0:1] {\scalesimVone};
            \addplot[OliveGreen,name path=l1, thick, mark=triangle] table [x index=0, y index=1, domain=0:1] {\scalesimVK};
            \addplot[Black,name path=l1, thick, mark=o] table [x index=0, y index=1, domain=0:1] {\scalepomdpsolve};

              \legend{{$V=1,|\mathcal{F}|=2$}, {$V=2, |\mathcal{F}|=4$}, {$V=K,|\mathcal{F}|=2^K$}, No aggregation}
            \end{axis}
\node[inner sep=0pt,align=center, scale=1, rotate=0, opacity=1] (obs) at (1.8,2.15)
{
  Offline compute time (s)
};
\node[inner sep=0pt,align=center, scale=1, rotate=0, opacity=1] (obs) at (4.7,-0.8)
{
  Number of replicas $K$
};
\end{tikzpicture}
};
\end{tikzpicture}

%% file: tikz/feature_cost.tex
\begin{tikzpicture}

\node[scale=1] (kth_cr) at (0,-3)
{
\begin{tikzpicture}
\begin{axis}[
   ybar,
    title style={align=center},
    ticks=both,
    ymin=0,
    axis x line = bottom,
    axis y line = left,
    axis line style={-|},
    enlarge y limits={lower, value=0.1},
    enlarge y limits={upper, value=0.22},
    xtick=\empty,
    ymajorgrids,
    xticklabels={},
    legend style={nodes={scale=0.9, transform shape}, at={(-0.08, 1.23)}, align=left, anchor=west, legend columns=4, draw=none},
   x tick label style={align=center, yshift=-0.1cm},
    enlarge x limits=2.6,
    width=11cm,
    bar width=0.7cm,
    height=4cm,
    ]

\addplot+[
  draw=black, color=black,fill=OliveGreen!70,
  nodes near coords,
  every node near coord/.append style={
      anchor=south,
      shift={(axis direction cs:0,1.2)}, 
      font=\small\bfseries, fill=white,
scale=0.8,
  },
  error bars/.cd, y dir=both, y explicit,
] coordinates {
  (0,4.96)
};

\addplot+[
  draw=black, color=black,fill=Blue!50,
postaction={pattern=dots},
  nodes near coords,
  every node near coord/.append style={
      anchor=south,
      shift={(axis direction cs:0,2)}, 
      font=\small\bfseries, fill=white,
scale=0.85,
  },
  error bars/.cd, y dir=both, y explicit,
] coordinates {
  (1,4.96)
};

\addplot+[
  draw=black, color=black,fill=Red!90,
postaction={pattern=north west lines},
  nodes near coords,
  every node near coord/.append style={
      anchor=south,
      shift={(axis direction cs:0,2)}, 
      font=\small\bfseries, fill=white,
scale=0.85,
  },
  error bars/.cd, y dir=both, y explicit,
] coordinates {
  (2,4.96)
};

\addplot+[
  draw=black, color=black,fill=Black!30,
postaction={pattern=north east lines},
  nodes near coords,
  every node near coord/.append style={
      anchor=south,
      shift={(axis direction cs:0,2)}, 
      font=\small\bfseries, fill=white,
scale=0.85,
  },
  error bars/.cd, y dir=both, y explicit,
] coordinates {
  (3,4.76)
};

\addplot+[
  draw=black, color=black,fill=OliveGreen!70,
  nodes near coords,
  every node near coord/.append style={
      anchor=south,
      shift={(axis direction cs:0,1.8)}, 
      font=\small\bfseries, fill=white,
scale=0.8,
  },
  error bars/.cd, y dir=both, y explicit,
] coordinates {
  (20,13.23)
};
\addplot+[
  draw=black, color=black,fill=Blue!50,
postaction={pattern=dots},
  nodes near coords,
  every node near coord/.append style={
      anchor=south,
      shift={(axis direction cs:0,2.4)}, 
      font=\small\bfseries, fill=white,
scale=0.85,
  },
  error bars/.cd, y dir=both, y explicit,
] coordinates {
  (21,13.23)
};

\addplot+[
  draw=black, color=black,fill=Red!90,
postaction={pattern=north west lines},
  nodes near coords,
  every node near coord/.append style={
      anchor=south,
      shift={(axis direction cs:0,2.4)}, 
      font=\small\bfseries, fill=white,
scale=0.85,
  },
  error bars/.cd, y dir=both, y explicit,
] coordinates {
  (22,18.84)
};

\addplot+[
  draw=black, color=black,fill=OliveGreen!70,
  nodes near coords,
  every node near coord/.append style={
      anchor=south,
      shift={(axis direction cs:0,1.2)}, 
      font=\small\bfseries, fill=white,
scale=0.8,
  },
  error bars/.cd, y dir=both, y explicit,
] coordinates {
  (40,39.33)
};
\addplot+[
  draw=black, color=black,fill=Blue!50,
postaction={pattern=dots},
  nodes near coords,
  every node near coord/.append style={
      anchor=south,
      shift={(axis direction cs:0,1.4)}, 
      font=\small\bfseries, fill=white,
scale=0.85,
  },
  error bars/.cd, y dir=both, y explicit,
] coordinates {
  (41,41.57)
};

\addplot+[
  draw=black, color=black,fill=Red!90,
postaction={pattern=north west lines},
  nodes near coords,
  every node near coord/.append style={
      anchor=south,
      shift={(axis direction cs:0,1.4)}, 
      font=\small\bfseries, fill=white,
scale=0.85,
  },
  error bars/.cd, y dir=both, y explicit,
] coordinates {
  (42,57.21)
};

\legend{{$V=K,|\mathcal{F}|=2^K$$\quad$}, {$V=2,|\mathcal{F}|=4$$\quad$}, {$V=1,|\mathcal{F}|=2$$\quad$}, Optimal}
\end{axis}
\end{tikzpicture}
};

\node[inner sep=0pt,align=center, scale=0.9, rotate=0, opacity=1] (obs) at (-1.15,-1.95)
{
  Cost of the base policy $\mu$ ($\downarrow$ better) [cf.~Eq.~\eqref{eq:approximation_1}]
};

\node[inner sep=0pt,align=center, scale=0.9, rotate=0, opacity=1] (obs) at (-2.7,-4.8)
{
  $K=1$
};
\node[inner sep=0pt,align=center, scale=0.9, rotate=0, opacity=1] (obs) at (0.8,-4.8)
{
  $K=2$
};
\node[inner sep=0pt,align=center, scale=0.9, rotate=0, opacity=1] (obs) at (3.75,-4.8)
{
  $K=4$
};

\end{tikzpicture}

%% file: tikz/feature_extraction.tex
\begin{tikzpicture}
\pgfplotstableread{
1 4.96
2 13.23
4 41.57
6 73.55
8 107.24
10 140.38
12 172.69
}\zones
\pgfplotstableread{
1 4.96
2 13.23
4 66.90
6 118.88
8 149.77
10 245.29
12 393.78
}\nn

\pgfplotsset{/dummy/workaround/.style={/pgfplots/axis on top}}

\node[scale=1] (kth_cr) at (0,0.065)
{
\begin{tikzpicture}
  \begin{axis}
[
        width=10.25cm,
        height =3.5cm,
        axis y line=center,
        axis x line=bottom,
        scaled y ticks=false,
        yticklabel style={
        /pgf/number format/fixed,
        /pgf/number format/precision=5
      },
        xlabel style={below right},
        ylabel style={above left},
        axis line style={-{Latex[length=2mm]}},
        legend style={at={(0.695,1.03)}, cells={anchor=west}},
        legend columns=1,
        legend style={
          draw=none,
            /tikz/column 2/.style={
                column sep=5pt,
              }
              }
              ]
            \addplot[RoyalAzure,name path=l1, thick, mark=diamond] table [x index=0, y index=1, domain=0:1] {\zones};
            \addplot[Red,name path=l1, thick, mark=x] table [x index=0, y index=1, domain=0:1] {\nn};

              \legend{{Zone features ($|\mathcal{F}|=4$)}, Neural network features ($|\mathcal{F}|=4$)}
            \end{axis}
\node[inner sep=0pt,align=center, scale=1, rotate=0, opacity=1] (obs) at (3.75,2.1)
{
Cost of the base policy $\mu$ ($\downarrow$ better) [cf.~Eq.~\eqref{eq:approximation_1}]
};
\node[inner sep=0pt,align=center, scale=1, rotate=0, opacity=1] (obs) at (4.7,-0.8)
{
  Number of replicas $K$
};
\end{tikzpicture}
};
\end{tikzpicture}

%% file: tikz/security_alerts.tex
\begin{tikzpicture}
  \begin{axis}
[
        xmin=0.5,
        xmax=14,
        ymax=16400,
        ymin=-200000,
        boxplot/draw direction = y,
        ymode=log,
        width=11.5cm,
        height=3.5cm,
        axis y line=center,
        axis x line=bottom,
        scaled y ticks=false,
        yticklabel style={
        /pgf/number format/fixed,
        /pgf/number format/precision=5
      },
      xtick=\empty,
      xticklabels={},
        xlabel style={below right},
        ylabel style={above left},
        axis line style={-{Latex[length=2mm]}},
        legend style={at={(0.6,0.9)}},
        legend columns=2,
        legend style={
          draw=none,
            /tikz/column 2/.style={
                column sep=5pt,
              }
              }
              ]

            \addplot+[
    fill=Blue,draw=black,fill opacity=0.2,
    boxplot/box extend=0.5,
    x=1,
    boxplot prepared={
      median=836.31,
      upper quartile=1050,
      lower quartile=470,
      upper whisker=1400,
      lower whisker=200,
    },
    ] coordinates {};

\addplot+[
    fill=OliveGreen,draw=black,fill opacity=0.2,
    boxplot/box extend=0.5,
    boxplot prepared={
      median=743.31,
      upper quartile=1073,
      lower quartile=413,
      upper whisker=1270,
      lower whisker=237,
    }
    ] coordinates {};

\addplot+[
    fill=Black,draw=black,fill opacity=0.2,
    boxplot/box extend=0.5,
    boxplot prepared={
      median=756,
      upper quartile=1010,
      lower quartile=488,
      upper whisker=1270,
      lower whisker=242,
    }
    ] coordinates {};

\addplot+[
    fill=Dandelion,draw=black,fill opacity=0.2,
    boxplot/box extend=0.5,
    boxplot prepared={
      median=728.31,
      upper quartile=1231,
      lower quartile=410,
      upper whisker=1480,
      lower whisker=155,
    }
    ] coordinates {};
\addplot+[
    fill=DarkOrchid,draw=black,fill opacity=0.2,
    boxplot/box extend=0.5,
    boxplot prepared={
      median=916,
      upper quartile=1210,
      lower quartile=370,
      upper whisker=1599,
      lower whisker=178,
    }
    ] coordinates {};
\addplot+[
    fill=Maroon,draw=black,fill opacity=0.2,
    boxplot/box extend=0.5,
    boxplot prepared={
      median=715,
      upper quartile=990,
      lower quartile=410,
      upper whisker=1250,
      lower whisker=132,
    }
    ] coordinates {};
\addplot+[
    fill=Mulberry,draw=black,fill opacity=0.2,
    boxplot/box extend=0.5,
    boxplot prepared={
      median=989.31,
      upper quartile=1210,
      lower quartile=670,
      upper whisker=1700,
      lower whisker=202,
    }
    ] coordinates {};

\addplot+[
    fill=Red,draw=black,fill opacity=0.2,
    boxplot/box extend=0.5,
    boxplot prepared={
      median=4471,
      upper quartile=5210,
      lower quartile=3781,
      upper whisker=6100,
      lower whisker=2901,
    }
    ] coordinates {};

\addplot+[
    fill=Salmon,draw=black,fill opacity=0.2,
    boxplot/box extend=0.5,
    boxplot prepared={
      median=1410,
      upper quartile=1600,
      lower quartile=1100,
      upper whisker=1900,
      lower whisker=501,
    }
    ] coordinates {};

\addplot+[
    fill=White,draw=black,fill opacity=0.2,
    boxplot/box extend=0.5,
    boxplot prepared={
      median=999,
      upper quartile=1200,
      lower quartile=700,
      upper whisker=1450,
      lower whisker=240,
    }
    ] coordinates {};

\addplot+[
    fill=Yellow,draw=black,fill opacity=0.2,
    boxplot/box extend=0.5,
    boxplot prepared={
      median=756,
      upper quartile=899,
      lower quartile=400,
      upper whisker=1203,
      lower whisker=149,
    }
    ] coordinates {};      

\addplot+[
    fill=Violet,draw=black,fill opacity=0.2,
    boxplot/box extend=0.5,
    boxplot prepared={
      median=8500,
      upper quartile=12000,
      lower quartile=6000,
      upper whisker=15910,
      lower whisker=4501,
    }
    ] coordinates {};

\addplot+[
    fill=YellowOrange,draw=black,fill opacity=0.2,
    boxplot/box extend=0.5,
    boxplot prepared={
      median=1288,
      upper quartile=1510,
      lower quartile=1021,
      upper whisker=1641,
      lower whisker=301,
    }
    ] coordinates {};    
  \end{axis}

\node[inner sep=0pt,align=center, scale=0.9, rotate=0, opacity=1] (obs) at (2.05,1.9)
{
  Number of security alerts $z_k$
};

\node[inner sep=0pt,align=center, scale=0.8, opacity=1, rotate=70] (obs) at (0.05,-0.8)
{
  \textsc{no attack}
};
\node[inner sep=0pt,align=center, scale=0.8, opacity=1, rotate=70] (obs) at (0.7,-1)
{
  \textsc{cve-2010-0426}
};
\node[inner sep=0pt,align=center, scale=0.8, opacity=1, rotate=70] (obs) at (1.4,-1)
{
  \textsc{cve-2015-3306}
};

\node[inner sep=0pt,align=center, scale=0.8, opacity=1, rotate=70] (obs) at (2.1,-1)
{
  \textsc{cve-2015-5602}
};
\node[inner sep=0pt,align=center, scale=0.8, opacity=1, rotate=70] (obs) at (2.9,-0.95)
{
  \textsc{sql-injection}
};
\node[inner sep=0pt,align=center, scale=0.8, opacity=1, rotate=70] (obs) at (3.65,-1.1)
{
  \textsc{cve-2016-10033}
};
\node[inner sep=0pt,align=center, scale=0.8, opacity=1, rotate=70] (obs) at (4.5,-0.7)
{
  \textsc{ftp-brute}
};
\node[inner sep=0pt,align=center, scale=0.8, opacity=1, rotate=70] (obs) at (5.25,-0.7)
{
  \textsc{ping-scan}
};
\node[inner sep=0pt,align=center, scale=0.8, opacity=1, rotate=70] (obs) at (6.05,-0.7)
{
  \textsc{ssh-brute}
};
\node[inner sep=0pt,align=center, scale=0.8, opacity=1, rotate=70] (obs) at (6.8,-0.65)
{
  \textsc{sambacry}
};
\node[inner sep=0pt,align=center, scale=0.8, opacity=1, rotate=70] (obs) at (7.45,-0.8)
{
  \textsc{shellshock}
};
\node[inner sep=0pt,align=center, scale=0.8, opacity=1, rotate=70] (obs) at (8.25,-0.7)
{
  \textsc{tcp scan}
};
\node[inner sep=0pt,align=center, scale=0.8, opacity=1, rotate=70] (obs) at (8.8,-1)
{
  \textsc{telnet brute}
};
\end{tikzpicture}

%% file: tikz/adaption_eval.tex
\begin{tikzpicture}

\pgfplotstableread{
0 0.48
0.93 0.89
5.45 0.943
7.45 0.985
19.31 1
60 1
}\rollout

\pgfplotstableread{
0 0
10 0.006
20 0.0075
30 0.0084
40 0.0099
50 0.01
60 0.0139
}\ppo

\pgfplotstableread{
0 0.835
10 0.82
20 0.81
30 0.80
40 0.80
50 0.81
60 0.815
}\ppob

\pgfplotsset{/dummy/workaround/.style={/pgfplots/axis on top}}

\node[scale=1] (kth_cr) at (0,-3.3)
{
\begin{tikzpicture}
  \begin{axis}
[
        xmin=0,
        xmax=62,
        ymax=1.09,
        ymin=0,
        width=11.5cm,
        height=3.8cm,
        axis y line=center,
        axis x line=bottom,
        scaled y ticks=false,
        yticklabel style={
        /pgf/number format/fixed,
        /pgf/number format/precision=5
        },
        xlabel style={below right},
        ylabel style={above left},
        axis line style={-{Latex[length=2mm]}},
        smooth,
        legend style={at={(1.04,0.7)}},
        legend columns=1,
        legend style={
          scale=0.85,
          transform shape,
          draw=none,
          anchor=north east,
          align={left},
          nodes={anchor=west},
          /tikz/every odd column/.style={anchor=west},
            /tikz/column 2/.style={
                column sep=5pt,
              }
              }
              ]
            \addplot[Red,name path=l1, thick, mark=x, mark repeat=1] table [x index=0, y index=1, domain=0:95] {\rollout};
\addplot[RoyalAzure,name path=l1, thick, mark=diamond, mark repeat=1] table [x index=0, y index=1, domain=0:95] {\ppo};              
\addplot[OliveGreen,name path=l1, thick, mark=triangle, mark repeat=1] table [x index=0, y index=1, domain=0:95] {\ppob};              
\legend{Our method, PPO retraining with random initial policy \cite{ppo}, PPO retraining with initial policy computed for Scenario $1$ \cite{ppo}}
  \end{axis}
\node[inner sep=0pt,align=center, scale=1, rotate=0, opacity=1] (obs) at (4.835,-0.7)
{
  Adaptation time $t$ (seconds)
};
\node[inner sep=0pt,align=center, scale=1, rotate=0, opacity=1] (obs) at (4,2.4)
{
  Adaptation completion $A(\mu_t)$ [cf. Eq.~\eqref{eq:adaptation_metric}] ($\uparrow$ better)
};
\end{tikzpicture}
};

\end{tikzpicture}

%% file: tikz/recovery_frequency.tex
\begin{tikzpicture}
    
\pgfplotstableread{
1 1
2 0.1
}\ours

\pgfplotstableread{
1 1
2 0.1
}\ppo

\pgfplotstableread{
1 2.5
2 1.6
}\periodic

\pgfplotstableread{
1 2.5
2 0.3
}\base

\pgfplotsset{/dummy/workaround/.style={/pgfplots/axis on top}}

\node[scale=1] (kth_cr) at (0,0)
{
\begin{tikzpicture}
\begin{axis}[
   ybar=-0.55cm,
    title style={align=center},
    ticks=both,
    xmin=0.5,
    xmax=2.5,
    ymax=3.5,
    ymin=-0,
    xticklabels=\empty,
    axis x line = bottom,
    axis y line = left,
    axis line style={-|},
    xtick=data,
    ymajorgrids,
    legend style={at={(0.9, 1.2)}, legend columns=4, draw=none, /tikz/column 2/.style={
                column sep=10pt,
              }},
   x tick label style={align=center, yshift=-0.1cm},
   enlarge x limits=0.015,
    width=11.5cm,
    height=3.5cm,
    bar width=0.4cm,
group style={
       group size=3 by 1,
       horizontal sep=0pt,
       group sep=2000pt
    }
    ]
  \addplot+[draw=black, color=black, fill=Red,
postaction={pattern=crosshatch},
  nodes near coords,
  every node near coord/.append style={
      anchor=south,
      shift={(axis direction cs:0,0.1)}, 
      font=\small\bfseries, fill=white,
scale=0.95,
  }
] coordinates {(0.5, 1)};
\addplot+[draw=black, color=black, fill=Blue!80,
postaction={pattern=dots},
  nodes near coords,
  every node near coord/.append style={
      anchor=south,
      shift={(axis direction cs:0,0.1)}, 
      font=\small\bfseries, fill=white,
scale=0.95,
  }
] coordinates {(0.65, 1)};
\addplot+[draw=black, color=black, fill=OliveGreen!60,
postaction={pattern=horizontal lines},
  nodes near coords,
  every node near coord/.append style={
      anchor=south,
      shift={(axis direction cs:0,0.1)}, 
      font=\small\bfseries, fill=white,
scale=0.95,
}] coordinates {(0.8, 2.5)};

  \addplot+[draw=black, color=black, fill=Red,
postaction={pattern=crosshatch},
  nodes near coords,
  every node near coord/.append style={
      anchor=south,
      shift={(axis direction cs:0,0.1)}, 
      font=\small\bfseries, fill=white,
scale=0.95,
  }
] coordinates {(1.05, 1)};
\addplot+[draw=black, color=black, fill=Blue!80,
postaction={pattern=dots},
  nodes near coords,
  every node near coord/.append style={
      anchor=south,
      shift={(axis direction cs:0,0.1)}, 
      font=\small\bfseries, fill=white,
scale=0.95,
  }
  ] coordinates {(1.2, 1)};

\addplot+[draw=black, color=black, fill=OliveGreen!60,
postaction={pattern=horizontal lines},
  nodes near coords,
  every node near coord/.append style={
      anchor=south,
      shift={(axis direction cs:0,0.1)}, 
      font=\small\bfseries, fill=white,
scale=0.95,
}] coordinates {(1.35, 2.5)};

\addplot+[draw=black, color=black, fill=Red,
postaction={pattern=crosshatch},
  nodes near coords,
  every node near coord/.append style={
      anchor=south,
      shift={(axis direction cs:0,0.1)}, 
      font=\small\bfseries, fill=white,
scale=0.95,
  }
] coordinates {(1.65, 0.1)};
\addplot+[draw=black, color=black, fill=Blue!80,
postaction={pattern=dots},
  nodes near coords,
  every node near coord/.append style={
      anchor=south,
      shift={(axis direction cs:0,0.1)}, 
      font=\small\bfseries, fill=white,
scale=0.95,
  }] coordinates {(1.8, 0.1)};
\addplot+[draw=black, color=black, fill=OliveGreen!60,
postaction={pattern=horizontal lines},
  nodes near coords,
  every node near coord/.append style={
      anchor=south,
      shift={(axis direction cs:0,0.1)}, 
      font=\small\bfseries, fill=white,
scale=0.95,
}] coordinates {(1.95, 1.6)};

\addplot+[draw=black, color=black, fill=Red,
postaction={pattern=crosshatch},
  nodes near coords,
  every node near coord/.append style={
      anchor=south,
      shift={(axis direction cs:0,0.1)}, 
      font=\small\bfseries, fill=white,
scale=0.95,
  }
] coordinates {(2.2, 0.1)};

\addplot+[draw=black, color=black, fill=Blue!80,
postaction={pattern=dots},
  nodes near coords,
  every node near coord/.append style={
      anchor=south,
      shift={(axis direction cs:0,0.1)}, 
      font=\small\bfseries, fill=white,
scale=0.95,
}] coordinates {(2.35, 0.9)};

\addplot+[draw=black, color=black, fill=OliveGreen!60,
postaction={pattern=horizontal lines},
  nodes near coords,
  every node near coord/.append style={
      anchor=south,
      shift={(axis direction cs:0,0.1)}, 
      font=\small\bfseries, fill=white,
scale=0.95,
}] coordinates {(2.5, 1.6)};

    \legend{Our method,PPO$\quad$,Periodic$\quad$}
  \end{axis}

\node[inner sep=0pt,align=center, scale=1] (obs) at (1.5,-0.3)
{
\hyperref[testbed_scenario_1]{Scenario $1$}
};  
\node[inner sep=0pt,align=center, scale=1] (obs) at (3.75,-0.3)
{
\hyperref[testbed_scenario_1]{Scenario $2$}
};  

\node[inner sep=0pt,align=center, scale=1] (obs) at (6.25,-0.3)
{
\hyperref[testbed_scenario_1]{Scenario $1$}
};  
\node[inner sep=0pt,align=center, scale=1] (obs) at (8.45,-0.3)
{
\hyperref[testbed_scenario_1]{Scenario $2$}
};  

\node[inner sep=0pt,align=center, scale=1] (obs) at (2.4,-0.75)
{
Time-to-recovery ($\downarrow$ better)
};
\node[inner sep=0pt,align=center, scale=1] (obs) at (7.5,-0.75)
{
Recoveries per time step ($\downarrow$ better)
};
\end{tikzpicture}
};

\end{tikzpicture}

%% file: tikz/cage_network.tex
\begin{tikzpicture}[fill=white, >=stealth,
    node distance=3cm,
    database/.style={
      cylinder,
      cylinder uses custom fill,
      shape border rotate=90,
      aspect=0.25,
      draw}]

    \tikzset{
node distance = 9em and 4em,
sloped,
   box/.style = {%
    shape=rectangle,
    rounded corners,
    draw=blue!40,
    fill=blue!15,
    align=center,
    font=\fontsize{12}{12}\selectfont},
 arrow/.style = {%
    line width=0.1mm,
    -{Triangle[length=5mm,width=2mm]},
    shorten >=1mm, shorten <=1mm,
    font=\fontsize{8}{8}\selectfont},
}
\node[scale=0.8] (kth_cr) at (5.8,2)
{
  \begin{tikzpicture}

\node[inner sep=0pt,align=center, scale=0.8] (time) at (0.1,-5.5)
{
  \begin{tikzpicture}
\draw[-, color=black, fill=gray2] (-3.5,6.1) to (4.15,6.1) to (4.15,9.45) to (-3.5, 9.45) to (-3.5,6.1);

\node[scale=0.11](public_gw) at (0.05,8.95) {\router{}};
\draw[-, color=black] (0.05, 8.77) to (0.05, 8.35);
\draw[-, color=black] (0.05, 9.1) to (0.05, 9.6);
\draw[-, color=black] (0.05, 8.2) to (0.05, 7.9);
\draw[-, color=black] (-3.3, 9.6) to (4, 9.6);
\draw[-, color=black] (-2, 9.6) to (-2, 9.8);
\draw[-, color=black] (1, 9.6) to (1, 9.8);
\draw[-, color=black] (1.65, 9.6) to (1.65, 9.8);
\draw[-, color=black] (2.3, 9.6) to (2.3, 9.8);
\draw[-, color=black] (3, 9.6) to (3, 9.8);

\node[inner sep=0pt,align=center,rotate=90] (gpu1) at (0.1,8.4)
  {\scalebox{0.6}{
     \includegraphics{firewall.pdf}
   }
 };

\node[server, scale=0.65](s1) at (-2.22,8.5) {};
\node[server, scale=0.65](s1) at (-2.22,7.8) {};
\node[server, scale=0.65](s1) at (-2.22,7.1) {};

\node[rack switch, xshift=0.1cm,yshift=0.3cm, scale=0.6] at (0.3,7.5) (sw1){};
\draw[-, color=black] (-0.31, 7.8) to (-1.8, 7.8);
\draw[-, color=black] (-1.8, 8.6) to (-1.8, 7);
\draw[-, color=black] (-1.8, 8.5) to (-2.1, 8.5);
\draw[-, color=black] (-1.8, 7.8) to (-2.1, 7.8);
\draw[-, color=black] (-1.8, 7.1) to (-2.1, 7.1);

\node[inner sep=0pt,align=center, scale=1] (hacker) at (-1.88,6.4)
{\textsc{enterprise zone}};

\node[inner sep=0pt,align=center, scale=1] (hacker) at (2.35,6.4)
{\textsc{operational zone}};

\node[inner sep=0pt,align=center, scale=1] (hacker) at (2.25,9.24)
{\textsc{user zone}};

\node[inner sep=0pt,align=center,rotate=0] (gpu1) at (1.1,7.9)
  {\scalebox{0.6}{
     \includegraphics{firewall.pdf}
   }
 };

 \node[scale=0.11](public_gw) at (1.8,7.83) {\router{}};
 \node[rack switch, xshift=0.1cm,yshift=0.3cm, scale=0.6] at (3.1,7.48) (sw1){};

 \node[server, scale=0.65](s1) at (3,8.6) {};

\node[inner sep=0pt,align=center] (client1) at (3.7,7)
  {\scalebox{0.06}{
     \includegraphics{laptop3.pdf}
   }
 };
\node[inner sep=0pt,align=center] (client1) at (3,7)
  {\scalebox{0.06}{
     \includegraphics{laptop3.pdf}
   }
 };
\node[inner sep=0pt,align=center] (client1) at (2.3,7)
  {\scalebox{0.06}{
     \includegraphics{laptop3.pdf}
   }
 };

\draw[-, color=black] (0.6, 7.8) to (0.9, 7.8);
\draw[-, color=black] (1.17, 7.8) to (1.47, 7.8);
\draw[-, color=black] (2.13, 7.8) to (2.5, 7.8);

\draw[-, color=black] (1.83, 7.45) to (4.05, 7.45);

\draw[-, color=black] (2.25, 7.28) to (2.25, 7.45);
\draw[-, color=black] (2.95, 7.28) to (2.95, 7.45);
\draw[-, color=black] (3.65, 7.28) to (3.65, 7.45);
\draw[-, color=black] (2.95, 7.92) to (2.95, 8.34);

\node[inner sep=0pt,align=center] (hacker2) at (-1.95,10)
  {\scalebox{0.1}{
     \includegraphics{hacker.png}
   }
 };
\node[inner sep=0pt,align=center] (client1) at (1.05,10)
  {\scalebox{0.06}{
     \includegraphics{laptop3.pdf}
   }
 };

\node[inner sep=0pt,align=center] (client1) at (1.7,10)
  {\scalebox{0.06}{
     \includegraphics{laptop3.pdf}
   }
 };

\node[inner sep=0pt,align=center] (client1) at (2.35,10)
  {\scalebox{0.06}{
     \includegraphics{laptop3.pdf}
   }
 };


  \node[inner sep=0pt,align=center, scale=0.9] (hacker) at (-1.3,5.7)
  {observation $z_k$};
  \node[inner sep=0pt,align=center, scale=0.9] (hacker) at (2.3,5.7)
  {control $u_k$};
   \node[inner sep=0pt,align=center, scale=0.95] (policy_1) at (0.5,5.5)
  {$\mu$};
  \node[inner sep=0pt,align=center, scale=0.9] (hacker) at (0.5,5.18)
  {Security policy};
  \draw[-{Latex[width=1.4mm]}, color=black, line width=0.2mm, rounded corners] (-2.5,6.1) to (-2.5,5.5) to (0.3, 5.5);
\draw[-{Latex[width=1.4mm]}, color=black, line width=0.2mm, rounded corners] (0.7, 5.5) to (3.3, 5.5) to (3.3, 6.1);

\node[inner sep=0pt,align=center] (client2) at (3.05,10)
  {\scalebox{0.06}{
     \includegraphics{laptop3.pdf}
   }
 };

  \node[inner sep=0pt,align=center, scale=0.9] (hacker) at (-2,10.5)
  {Attacker};

  \node[inner sep=0pt,align=center, scale=0.9] (hacker) at (2,10.5)
  {Clients};

\node[inner sep=0pt,align=center, scale=0.45] (hacker) at (-2.4,7.83)
 {
$2$
};
\node[inner sep=0pt,align=center, scale=0.45] (hacker) at (-2.4,8.55)
 {
$1$
};
\node[inner sep=0pt,align=center, scale=0.45] (hacker) at (-2.4,7.1)
 {
$3$
};

\node[inner sep=0pt,align=center, scale=0.45] (hacker) at (2.3,6.8)
 {
$2$
};
\node[inner sep=0pt,align=center, scale=0.45] (hacker) at (3,6.8)
 {
$3$
};
\node[inner sep=0pt,align=center, scale=0.45] (hacker) at (3.7,6.8)
 {
$4$
};

\node[inner sep=0pt,align=center, scale=0.45] (hacker) at (2.8,8.6)
 {
$1$
};

\node[inner sep=0pt,align=center, scale=0.45] (hacker) at (1.02,9.8)
 {
$1$
};

\node[inner sep=0pt,align=center, scale=0.45] (hacker) at (1.68,9.8)
 {
$2$
};
\node[inner sep=0pt,align=center, scale=0.45] (hacker) at (2.33,9.8)
 {
$3$
};
\node[inner sep=0pt,align=center, scale=0.45] (hacker) at (3.01,9.8)
 {
$4$
};

    \end{tikzpicture}
  };

      \end{tikzpicture}
};

\end{tikzpicture}

%% file: tikz/adaption_eval_2.tex
\begin{tikzpicture}

\pgfplotstableread{
0 0
0.05 0.627
0.1 0.685
0.5 0.785
1 0.89
5 0.94
15 0.95
30 0.98
60 0.98
}\cpomcp

\pgfplotstableread{
0 0.61
0.01 0.67
0.95 0.77
2.39 0.89
6.41 0.98
8.29 0.985
14.80 1
60 1
}\rollout

\pgfplotstableread{
0 0
10 0.0001
20 0.0001
30 0.0001
40 0.0001
50 0.0001
60 0.0001
}\ppo

\pgfplotstableread{
0 0.68
10 0.65
20 0.64
30 0.64
40 0.66
50 0.665
60 0.675
}\ppob

\pgfplotsset{/dummy/workaround/.style={/pgfplots/axis on top}}

\node[scale=1] (kth_cr) at (0,-3.3)
{
\begin{tikzpicture}
  \begin{axis}
[
        xmin=0,
        xmax=62,
        ymax=1.09,
        ymin=0,
        width=11.5cm,
        height=3.8cm,
        axis y line=center,
        axis x line=bottom,
        scaled y ticks=false,
        yticklabel style={
        /pgf/number format/fixed,
        /pgf/number format/precision=5
        },
        xlabel style={below right},
        ylabel style={above left},
        axis line style={-{Latex[length=2mm]}},
        smooth,
        legend style={at={(0.99,0.52)}},
        legend columns=2,
        legend style={
          scale=0.75,
          transform shape,
          /tikz/every node/.style={anchor=west},
          draw=none,
            /tikz/column 2/.style={
                column sep=2pt,
              },
              /tikz/every even column/.style={column sep=2pt},
              }
              ]
            \addplot+[Red,name path=l1, thick, mark=x, mark repeat=1,id=plot1] table [x index=0, y index=1, domain=0:95] {\rollout};
\addplot[RoyalAzure,name path=l1, thick, mark=diamond, mark repeat=1] table [x index=0, y index=1, domain=0:95] {\ppo};              
\addplot[OliveGreen,name path=l1, thick, mark=triangle, mark repeat=1] table [x index=0, y index=1, domain=0:95] {\ppob};              
\addplot[Purple,name path=l1, thick, mark=oplus, mark repeat=1] table [x index=0, y index=1, domain=0:95] {\cpomcp};

  \end{axis}

  
\node[inner sep=0pt,align=center, scale=1, rotate=0, opacity=1] (obs) at (4.835,-0.7)
{
  Adaptation time $t$ (seconds)
};
\node[inner sep=0pt,align=center, scale=1, rotate=0, opacity=1] (obs) at (4,2.4)
{
  Adaptation completion $A(\mu_t)$ [cf. \eqqref{eq:adaptation_metric}] ($\uparrow$ better)
};
\node[inner sep=0pt,align=center, scale=0.9, rotate=0, opacity=1] (obs) at (0.9,0.9)
{
  \legendSymbol{plot1} Ours
};

\node[inner sep=0pt,align=center, scale=0.9, rotate=0, opacity=1] (obs) at (4.9,0.9)
{
  \legendSymboltwo{plot1} Cardiff retraining with random policy \cite{vyas2023automated}
};

\node[inner sep=0pt,align=center, scale=0.9, rotate=0, opacity=1] (obs) at (1.5,0.45)
{
  \legendSymbolthree{plot1} C-POMCP \cite{hammar2024optimaldefenderstrategiescage2}
};
\node[inner sep=0pt,align=center, scale=0.9, rotate=0, opacity=1] (obs) at (6.5,0.45)
{
  \legendSymbolfour{plot1} Cardiff retraining with policy for Scenario $1$ \cite{vyas2023automated}
};

\end{tikzpicture}
};

\end{tikzpicture}